\documentclass[longauth]{aaEC}

\usepackage{euclid}
\usepackage{graphicx}
\usepackage{natbib}
\usepackage{scalerel}
\usepackage{multirow}
\usepackage{booktabs}
\usepackage[version=4]{mhchem}
\usepackage[nolist,nohyperlinks]{acronym}
\usepackage[table]{xcolor}
\usepackage{txfonts}
\usepackage{soul} 
\usepackage[pdfencoding=auto,psdextra]{hyperref}
\hypersetup{
    colorlinks=true,
    linkcolor=blue,
    filecolor=magenta,      
    urlcolor=blue,
    citecolor=blue
}
\makeatletter
\renewcommand*\aa@pageof{, page \thepage{} of \pageref*{LastPage}}
\makeatother

\usepackage[utf8]{inputenc}

\usepackage[switch, modulo]{lineno}

\DeclareSIUnit \kelvin{K}
\DeclareSIUnit \erg{erg}
\DeclareSIUnit \grv{grooves}

\begin{document}

\title{\Euclid preparation}
\subtitle{TBD. Overview of \Euclid infrared detector performance from ground tests}

\acrodef{ASIC}{application specific integrated circuit}
\acrodef{BFE}{brighter-fatter effect}
\acrodef{CaLA}{camera-lens assembly}
\acrodef{CCD}{charge-coupled device}
\acrodef{CoLA}{corrector-lens assembly}
\acrodef{CDS}{Correlated Double Sampling}
\acrodef{CFC}{cryo-flex cable}
\acrodef{CGH}{computer-generated holograms}
\acrodef{CNES}{Centre National d'Etude Spacial}
\acrodef{CTE}{coefficient of thermal expansion}
\acrodef{DCU}{Detector Control Unit}
\acrodef{DPU}{Data Processing Unit}
\acrodef{DS}{Detector System}
\acrodef{EDS}{\Euclid Deep Survey}
\acrodef{EE}{encircled energy}
\acrodef{ESA}{European Space Agency}
\acrodef{EWS}{\Euclid Wide Survey}
\acrodef{FDIR}{Fault Detection, Isolation and Recovery}
\acrodef{FOM}{figure of merit}
\acrodef{FOV}{field of view}
\acrodef{FPA}{focal plane array}
\acrodef{FWA}{filter-wheel assembly}
\acrodef{FWC}{full-well capacity}
\acrodef{FWHM}{full width at half maximum}
\acrodef{GWA}{grism-wheel assembly}
\acrodef{H2RG}{HAWAII-2RG}
\acrodef{IAD}{ion-assisted deposition}
\acrodef{ICU}{Instrument Control Unit}
\acrodef{IPC}{inter-pixel capacitance}
\acrodef{LED}{light-emitting diode}
\acrodef{MACC}{Multiple Accumulated}
\acrodef{MMU}{Mass Memory Unit}
\acrodef{NA}{numerical aperture}
\acrodef{NASA}{National Aeronautic and Space Administration}
\acrodef{MZ-CGH}{multi-zonal computer-generated hologram}
\acrodef{NI-CU}{NISP calibration unit}
\acrodef{NI-OA}{near-infrared optical assembly}
\acrodef{NI-GWA}{NISP Grism Wheel Assembly}
\acrodef{PARMS}{plasma-assisted reactive magnetron sputtering}
\acrodef{PLM}{payload module}
\acrodef{PTFE}{polytetrafluoroethylene}
\acrodef{PV}{performance verification}
\acrodef{PWM}{Pulse-Width Modulation}
\acrodef{PSF}{point spread function}
\acrodef{QE}{quantum efficiency}
\acrodef{ROIC}{readout-integrated chip}
\acrodef{ROS}{reference observing sequence}
\acrodef{SCA}{sensor chip array}
\acrodef{SCE}{sensor chip electronic}
\acrodef{SCS}{sensor chip system}
\acrodef{SGS}{science ground segment}
\acrodef{SHS}{Shack-Hartmann sensor}
\acrodef{SNR}[SNR]{signal-to-noise ratio}
\acrodef{SED}{spectral energy distribution}
\acrodef{SiC}{silicon carbide}
\acrodef{SVM}{service module}
\acrodef{VIS}{visible imager}
\acrodef{WD}{white dwarf}
\acrodef{WFE}{wavefront error}
\acrodef{ZP}{zero point}

\newcommand{\orcid}[1]{} 
\author{Euclid Collaboration: B.~Kubik\orcid{0009-0006-5823-4880}\thanks{\email{bkubik@ipnl.in2p3.fr}}\inst{\ref{aff1}}
\and R.~Barbier\inst{\ref{aff1}}
\and J.~Clemens\inst{\ref{aff2}}
\and S.~Ferriol\inst{\ref{aff1}}
\and A.~Secroun\orcid{0000-0003-0505-3710}\inst{\ref{aff2}}
\and G.~Smadja\inst{\ref{aff1}}
\and W.~Gillard\orcid{0000-0003-4744-9748}\inst{\ref{aff2}}
\and N.~Fourmanoit\orcid{0009-0005-6816-6925}\inst{\ref{aff2}}
\and A.~Ealet\orcid{0000-0003-3070-014X}\inst{\ref{aff1}}
\and S.~Conseil\orcid{0000-0002-3657-4191}\inst{\ref{aff1}}
\and J.~Zoubian\inst{\ref{aff2}}
\and R.~Kohley\inst{\ref{aff3}}
\and J.-C.~Salvignol\inst{\ref{aff4}}
\and L.~Conversi\orcid{0000-0002-6710-8476}\inst{\ref{aff5},\ref{aff3}}
\and T.~Maciaszek\inst{\ref{aff6}}
\and H.~Cho\inst{\ref{aff7}}
\and W.~Holmes\inst{\ref{aff7}}
\and M.~Seiffert\orcid{0000-0002-7536-9393}\inst{\ref{aff7}}
\and A.~Waczynski\inst{\ref{aff8}}
\and S.~Wachter\inst{\ref{aff9}}
\and K.~Jahnke\orcid{0000-0003-3804-2137}\inst{\ref{aff10}}
\and F.~Grupp\inst{\ref{aff11},\ref{aff12}}
\and C.~Bonoli\inst{\ref{aff13}}
\and L.~Corcione\orcid{0000-0002-6497-5881}\inst{\ref{aff14}}
\and S.~Dusini\orcid{0000-0002-1128-0664}\inst{\ref{aff15}}
\and E.~Medinaceli\orcid{0000-0002-4040-7783}\inst{\ref{aff16}}
\and R.~Laureijs\inst{\ref{aff17},\ref{aff4}}
\and G.~D.~Racca\orcid{0000-0002-9883-8981}\inst{\ref{aff4},\ref{aff18}}
\and A.~Bonnefoi\inst{\ref{aff19}}
\and M.~Carle\inst{\ref{aff19}}
\and A.~Costille\inst{\ref{aff19}}
\and F.~Ducret\inst{\ref{aff19}}
\and J-L.~Gimenez\inst{\ref{aff19}}
\and D.~Le~Mignant\orcid{0000-0002-5339-5515}\inst{\ref{aff19}}
\and L.~Martin\inst{\ref{aff19}}
\and L.~Caillat\inst{\ref{aff2}}
\and L.~Valenziano\orcid{0000-0002-1170-0104}\inst{\ref{aff16},\ref{aff20}}
\and N.~Auricchio\orcid{0000-0003-4444-8651}\inst{\ref{aff16}}
\and P.~Battaglia\orcid{0000-0002-7337-5909}\inst{\ref{aff16}}
\and A.~Derosa\inst{\ref{aff16}}
\and R.~Farinelli\inst{\ref{aff16}}
\and F.~Cogato\orcid{0000-0003-4632-6113}\inst{\ref{aff21},\ref{aff16}}
\and G.~Morgante\inst{\ref{aff16}}
\and M.~Trifoglio\orcid{0000-0002-2505-3630}\inst{\ref{aff16}}
\and V.~Capobianco\orcid{0000-0002-3309-7692}\inst{\ref{aff14}}
\and S.~Ligori\orcid{0000-0003-4172-4606}\inst{\ref{aff14}}
\and E.~Borsato\inst{\ref{aff22},\ref{aff15}}
\and C.~Sirignano\orcid{0000-0002-0995-7146}\inst{\ref{aff22},\ref{aff15}}
\and L.~Stanco\orcid{0000-0002-9706-5104}\inst{\ref{aff15}}
\and S.~Ventura\inst{\ref{aff15}}
\and R.~Toledo-Moreo\orcid{0000-0002-2997-4859}\inst{\ref{aff23}}
\and L.~Patrizii\inst{\ref{aff24}}
\and Y.~Copin\orcid{0000-0002-5317-7518}\inst{\ref{aff1}}
\and R.~Foltz\inst{\ref{aff8}}
\and E.~Prieto\inst{\ref{aff19}}
\and N.~Aghanim\orcid{0000-0002-6688-8992}\inst{\ref{aff25}}
\and B.~Altieri\orcid{0000-0003-3936-0284}\inst{\ref{aff3}}
\and S.~Andreon\orcid{0000-0002-2041-8784}\inst{\ref{aff26}}
\and C.~Baccigalupi\orcid{0000-0002-8211-1630}\inst{\ref{aff27},\ref{aff28},\ref{aff29},\ref{aff30}}
\and M.~Baldi\orcid{0000-0003-4145-1943}\inst{\ref{aff31},\ref{aff16},\ref{aff24}}
\and A.~Balestra\orcid{0000-0002-6967-261X}\inst{\ref{aff13}}
\and S.~Bardelli\orcid{0000-0002-8900-0298}\inst{\ref{aff16}}
\and F.~Bernardeau\orcid{0009-0007-3015-2581}\inst{\ref{aff32},\ref{aff33}}
\and A.~Biviano\orcid{0000-0002-0857-0732}\inst{\ref{aff28},\ref{aff27}}
\and A.~Bonchi\orcid{0000-0002-2667-5482}\inst{\ref{aff34}}
\and E.~Branchini\orcid{0000-0002-0808-6908}\inst{\ref{aff35},\ref{aff36},\ref{aff26}}
\and M.~Brescia\orcid{0000-0001-9506-5680}\inst{\ref{aff37},\ref{aff38}}
\and J.~Brinchmann\orcid{0000-0003-4359-8797}\inst{\ref{aff39},\ref{aff40}}
\and S.~Camera\orcid{0000-0003-3399-3574}\inst{\ref{aff41},\ref{aff42},\ref{aff14}}
\and G.~Ca\~nas-Herrera\orcid{0000-0003-2796-2149}\inst{\ref{aff4},\ref{aff43},\ref{aff18}}
\and C.~Carbone\orcid{0000-0003-0125-3563}\inst{\ref{aff44}}
\and J.~Carretero\orcid{0000-0002-3130-0204}\inst{\ref{aff45},\ref{aff46}}
\and S.~Casas\orcid{0000-0002-4751-5138}\inst{\ref{aff47}}
\and F.~J.~Castander\orcid{0000-0001-7316-4573}\inst{\ref{aff48},\ref{aff49}}
\and M.~Castellano\orcid{0000-0001-9875-8263}\inst{\ref{aff50}}
\and G.~Castignani\orcid{0000-0001-6831-0687}\inst{\ref{aff16}}
\and S.~Cavuoti\orcid{0000-0002-3787-4196}\inst{\ref{aff38},\ref{aff51}}
\and K.~C.~Chambers\orcid{0000-0001-6965-7789}\inst{\ref{aff52}}
\and A.~Cimatti\inst{\ref{aff53}}
\and C.~Colodro-Conde\inst{\ref{aff54}}
\and G.~Congedo\orcid{0000-0003-2508-0046}\inst{\ref{aff55}}
\and C.~J.~Conselice\orcid{0000-0003-1949-7638}\inst{\ref{aff56}}
\and F.~Courbin\orcid{0000-0003-0758-6510}\inst{\ref{aff57},\ref{aff58}}
\and H.~M.~Courtois\orcid{0000-0003-0509-1776}\inst{\ref{aff59}}
\and A.~Da~Silva\orcid{0000-0002-6385-1609}\inst{\ref{aff60},\ref{aff61}}
\and R.~da~Silva\orcid{0000-0003-4788-677X}\inst{\ref{aff50},\ref{aff34}}
\and H.~Degaudenzi\orcid{0000-0002-5887-6799}\inst{\ref{aff62}}
\and G.~De~Lucia\orcid{0000-0002-6220-9104}\inst{\ref{aff28}}
\and A.~M.~Di~Giorgio\orcid{0000-0002-4767-2360}\inst{\ref{aff63}}
\and H.~Dole\orcid{0000-0002-9767-3839}\inst{\ref{aff25}}
\and M.~Douspis\orcid{0000-0003-4203-3954}\inst{\ref{aff25}}
\and F.~Dubath\orcid{0000-0002-6533-2810}\inst{\ref{aff62}}
\and C.~A.~J.~Duncan\orcid{0009-0003-3573-0791}\inst{\ref{aff55},\ref{aff56}}
\and X.~Dupac\inst{\ref{aff3}}
\and S.~Escoffier\orcid{0000-0002-2847-7498}\inst{\ref{aff2}}
\and M.~Farina\orcid{0000-0002-3089-7846}\inst{\ref{aff63}}
\and F.~Faustini\orcid{0000-0001-6274-5145}\inst{\ref{aff50},\ref{aff34}}
\and F.~Finelli\orcid{0000-0002-6694-3269}\inst{\ref{aff16},\ref{aff20}}
\and S.~Fotopoulou\orcid{0000-0002-9686-254X}\inst{\ref{aff64}}
\and M.~Frailis\orcid{0000-0002-7400-2135}\inst{\ref{aff28}}
\and E.~Franceschi\orcid{0000-0002-0585-6591}\inst{\ref{aff16}}
\and M.~Fumana\orcid{0000-0001-6787-5950}\inst{\ref{aff44}}
\and S.~Galeotta\orcid{0000-0002-3748-5115}\inst{\ref{aff28}}
\and B.~Gillis\orcid{0000-0002-4478-1270}\inst{\ref{aff55}}
\and C.~Giocoli\orcid{0000-0002-9590-7961}\inst{\ref{aff16},\ref{aff24}}
\and J.~Gracia-Carpio\inst{\ref{aff11}}
\and B.~R.~Granett\orcid{0000-0003-2694-9284}\inst{\ref{aff26}}
\and A.~Grazian\orcid{0000-0002-5688-0663}\inst{\ref{aff13}}
\and L.~Guzzo\orcid{0000-0001-8264-5192}\inst{\ref{aff65},\ref{aff26},\ref{aff66}}
\and S.~V.~H.~Haugan\orcid{0000-0001-9648-7260}\inst{\ref{aff67}}
\and J.~Hoar\inst{\ref{aff3}}
\and H.~Hoekstra\orcid{0000-0002-0641-3231}\inst{\ref{aff18}}
\and I.~M.~Hook\orcid{0000-0002-2960-978X}\inst{\ref{aff68}}
\and F.~Hormuth\inst{\ref{aff69}}
\and A.~Hornstrup\orcid{0000-0002-3363-0936}\inst{\ref{aff70},\ref{aff71}}
\and P.~Hudelot\inst{\ref{aff33}}
\and M.~Jhabvala\inst{\ref{aff8}}
\and E.~Keih\"anen\orcid{0000-0003-1804-7715}\inst{\ref{aff72}}
\and S.~Kermiche\orcid{0000-0002-0302-5735}\inst{\ref{aff2}}
\and A.~Kiessling\orcid{0000-0002-2590-1273}\inst{\ref{aff7}}
\and M.~K\"ummel\orcid{0000-0003-2791-2117}\inst{\ref{aff12}}
\and M.~Kunz\orcid{0000-0002-3052-7394}\inst{\ref{aff73}}
\and H.~Kurki-Suonio\orcid{0000-0002-4618-3063}\inst{\ref{aff74},\ref{aff75}}
\and Q.~Le~Boulc'h\inst{\ref{aff76}}
\and A.~M.~C.~Le~Brun\orcid{0000-0002-0936-4594}\inst{\ref{aff77}}
\and P.~Liebing\inst{\ref{aff78}}
\and P.~B.~Lilje\orcid{0000-0003-4324-7794}\inst{\ref{aff67}}
\and V.~Lindholm\orcid{0000-0003-2317-5471}\inst{\ref{aff74},\ref{aff75}}
\and I.~Lloro\orcid{0000-0001-5966-1434}\inst{\ref{aff79}}
\and G.~Mainetti\orcid{0000-0003-2384-2377}\inst{\ref{aff76}}
\and D.~Maino\inst{\ref{aff65},\ref{aff44},\ref{aff66}}
\and E.~Maiorano\orcid{0000-0003-2593-4355}\inst{\ref{aff16}}
\and O.~Mansutti\orcid{0000-0001-5758-4658}\inst{\ref{aff28}}
\and S.~Marcin\inst{\ref{aff80}}
\and O.~Marggraf\orcid{0000-0001-7242-3852}\inst{\ref{aff81}}
\and M.~Martinelli\orcid{0000-0002-6943-7732}\inst{\ref{aff50},\ref{aff82}}
\and N.~Martinet\orcid{0000-0003-2786-7790}\inst{\ref{aff19}}
\and F.~Marulli\orcid{0000-0002-8850-0303}\inst{\ref{aff21},\ref{aff16},\ref{aff24}}
\and R.~Massey\orcid{0000-0002-6085-3780}\inst{\ref{aff83}}
\and S.~Maurogordato\inst{\ref{aff84}}
\and H.~J.~McCracken\orcid{0000-0002-9489-7765}\inst{\ref{aff33}}
\and S.~Mei\orcid{0000-0002-2849-559X}\inst{\ref{aff85},\ref{aff86}}
\and M.~Melchior\inst{\ref{aff87}}
\and Y.~Mellier\inst{\ref{aff88},\ref{aff33}}
\and M.~Meneghetti\orcid{0000-0003-1225-7084}\inst{\ref{aff16},\ref{aff24}}
\and E.~Merlin\orcid{0000-0001-6870-8900}\inst{\ref{aff50}}
\and G.~Meylan\inst{\ref{aff89}}
\and A.~Mora\orcid{0000-0002-1922-8529}\inst{\ref{aff90}}
\and M.~Moresco\orcid{0000-0002-7616-7136}\inst{\ref{aff21},\ref{aff16}}
\and P.~W.~Morris\orcid{0000-0002-5186-4381}\inst{\ref{aff91}}
\and L.~Moscardini\orcid{0000-0002-3473-6716}\inst{\ref{aff21},\ref{aff16},\ref{aff24}}
\and R.~Nakajima\orcid{0009-0009-1213-7040}\inst{\ref{aff81}}
\and C.~Neissner\orcid{0000-0001-8524-4968}\inst{\ref{aff92},\ref{aff46}}
\and R.~C.~Nichol\orcid{0000-0003-0939-6518}\inst{\ref{aff93}}
\and S.-M.~Niemi\orcid{0009-0005-0247-0086}\inst{\ref{aff4}}
\and C.~Padilla\orcid{0000-0001-7951-0166}\inst{\ref{aff92}}
\and S.~Paltani\orcid{0000-0002-8108-9179}\inst{\ref{aff62}}
\and F.~Pasian\orcid{0000-0002-4869-3227}\inst{\ref{aff28}}
\and K.~Pedersen\inst{\ref{aff94}}
\and W.~J.~Percival\orcid{0000-0002-0644-5727}\inst{\ref{aff95},\ref{aff96},\ref{aff97}}
\and V.~Pettorino\inst{\ref{aff4}}
\and S.~Pires\orcid{0000-0002-0249-2104}\inst{\ref{aff98}}
\and G.~Polenta\orcid{0000-0003-4067-9196}\inst{\ref{aff34}}
\and M.~Poncet\inst{\ref{aff6}}
\and L.~A.~Popa\inst{\ref{aff99}}
\and L.~Pozzetti\orcid{0000-0001-7085-0412}\inst{\ref{aff16}}
\and F.~Raison\orcid{0000-0002-7819-6918}\inst{\ref{aff11}}
\and R.~Rebolo\orcid{0000-0003-3767-7085}\inst{\ref{aff54},\ref{aff100},\ref{aff101}}
\and A.~Renzi\orcid{0000-0001-9856-1970}\inst{\ref{aff22},\ref{aff15}}
\and J.~Rhodes\orcid{0000-0002-4485-8549}\inst{\ref{aff7}}
\and G.~Riccio\inst{\ref{aff38}}
\and E.~Romelli\orcid{0000-0003-3069-9222}\inst{\ref{aff28}}
\and M.~Roncarelli\orcid{0000-0001-9587-7822}\inst{\ref{aff16}}
\and E.~Rossetti\orcid{0000-0003-0238-4047}\inst{\ref{aff31}}
\and R.~Saglia\orcid{0000-0003-0378-7032}\inst{\ref{aff12},\ref{aff11}}
\and Z.~Sakr\orcid{0000-0002-4823-3757}\inst{\ref{aff102},\ref{aff103},\ref{aff104}}
\and D.~Sapone\orcid{0000-0001-7089-4503}\inst{\ref{aff105}}
\and B.~Sartoris\orcid{0000-0003-1337-5269}\inst{\ref{aff12},\ref{aff28}}
\and J.~A.~Schewtschenko\orcid{0000-0002-4913-6393}\inst{\ref{aff55}}
\and M.~Schirmer\orcid{0000-0003-2568-9994}\inst{\ref{aff10}}
\and P.~Schneider\orcid{0000-0001-8561-2679}\inst{\ref{aff81}}
\and T.~Schrabback\orcid{0000-0002-6987-7834}\inst{\ref{aff106}}
\and M.~Scodeggio\inst{\ref{aff44}}
\and E.~Sefusatti\orcid{0000-0003-0473-1567}\inst{\ref{aff28},\ref{aff27},\ref{aff29}}
\and G.~Seidel\orcid{0000-0003-2907-353X}\inst{\ref{aff10}}
\and S.~Serrano\orcid{0000-0002-0211-2861}\inst{\ref{aff49},\ref{aff107},\ref{aff48}}
\and P.~Simon\inst{\ref{aff81}}
\and G.~Sirri\orcid{0000-0003-2626-2853}\inst{\ref{aff24}}
\and J.~Steinwagner\orcid{0000-0001-7443-1047}\inst{\ref{aff11}}
\and P.~Tallada-Cresp\'{i}\orcid{0000-0002-1336-8328}\inst{\ref{aff45},\ref{aff46}}
\and D.~Tavagnacco\orcid{0000-0001-7475-9894}\inst{\ref{aff28}}
\and A.~N.~Taylor\inst{\ref{aff55}}
\and H.~I.~Teplitz\orcid{0000-0002-7064-5424}\inst{\ref{aff108}}
\and I.~Tereno\orcid{0000-0002-4537-6218}\inst{\ref{aff60},\ref{aff109}}
\and S.~Toft\orcid{0000-0003-3631-7176}\inst{\ref{aff110},\ref{aff111}}
\and F.~Torradeflot\orcid{0000-0003-1160-1517}\inst{\ref{aff46},\ref{aff45}}
\and A.~Tsyganov\inst{\ref{aff112}}
\and I.~Tutusaus\orcid{0000-0002-3199-0399}\inst{\ref{aff103}}
\and J.~Valiviita\orcid{0000-0001-6225-3693}\inst{\ref{aff74},\ref{aff75}}
\and T.~Vassallo\orcid{0000-0001-6512-6358}\inst{\ref{aff12},\ref{aff28}}
\and G.~Verdoes~Kleijn\orcid{0000-0001-5803-2580}\inst{\ref{aff17}}
\and A.~Veropalumbo\orcid{0000-0003-2387-1194}\inst{\ref{aff26},\ref{aff36},\ref{aff35}}
\and Y.~Wang\orcid{0000-0002-4749-2984}\inst{\ref{aff108}}
\and J.~Weller\orcid{0000-0002-8282-2010}\inst{\ref{aff12},\ref{aff11}}
\and A.~Zacchei\orcid{0000-0003-0396-1192}\inst{\ref{aff28},\ref{aff27}}
\and G.~Zamorani\orcid{0000-0002-2318-301X}\inst{\ref{aff16}}
\and F.~M.~Zerbi\inst{\ref{aff26}}
\and E.~Zucca\orcid{0000-0002-5845-8132}\inst{\ref{aff16}}
\and V.~Allevato\orcid{0000-0001-7232-5152}\inst{\ref{aff38}}
\and M.~Ballardini\orcid{0000-0003-4481-3559}\inst{\ref{aff113},\ref{aff114},\ref{aff16}}
\and M.~Bolzonella\orcid{0000-0003-3278-4607}\inst{\ref{aff16}}
\and E.~Bozzo\orcid{0000-0002-8201-1525}\inst{\ref{aff62}}
\and C.~Burigana\orcid{0000-0002-3005-5796}\inst{\ref{aff115},\ref{aff20}}
\and R.~Cabanac\orcid{0000-0001-6679-2600}\inst{\ref{aff103}}
\and A.~Cappi\inst{\ref{aff16},\ref{aff84}}
\and P.~Casenove\orcid{0009-0006-6736-1670}\inst{\ref{aff6}}
\and D.~Di~Ferdinando\inst{\ref{aff24}}
\and J.~A.~Escartin~Vigo\inst{\ref{aff11}}
\and L.~Gabarra\orcid{0000-0002-8486-8856}\inst{\ref{aff116}}
\and W.~G.~Hartley\inst{\ref{aff62}}
\and J.~Mart\'{i}n-Fleitas\orcid{0000-0002-8594-569X}\inst{\ref{aff90}}
\and S.~Matthew\orcid{0000-0001-8448-1697}\inst{\ref{aff55}}
\and N.~Mauri\orcid{0000-0001-8196-1548}\inst{\ref{aff53},\ref{aff24}}
\and R.~B.~Metcalf\orcid{0000-0003-3167-2574}\inst{\ref{aff21},\ref{aff16}}
\and A.~Pezzotta\orcid{0000-0003-0726-2268}\inst{\ref{aff117},\ref{aff11}}
\and M.~P\"ontinen\orcid{0000-0001-5442-2530}\inst{\ref{aff74}}
\and C.~Porciani\orcid{0000-0002-7797-2508}\inst{\ref{aff81}}
\and I.~Risso\orcid{0000-0003-2525-7761}\inst{\ref{aff118}}
\and V.~Scottez\inst{\ref{aff88},\ref{aff119}}
\and M.~Sereno\orcid{0000-0003-0302-0325}\inst{\ref{aff16},\ref{aff24}}
\and M.~Tenti\orcid{0000-0002-4254-5901}\inst{\ref{aff24}}
\and M.~Viel\orcid{0000-0002-2642-5707}\inst{\ref{aff27},\ref{aff28},\ref{aff30},\ref{aff29},\ref{aff120}}
\and M.~Wiesmann\orcid{0009-0000-8199-5860}\inst{\ref{aff67}}
\and Y.~Akrami\orcid{0000-0002-2407-7956}\inst{\ref{aff121},\ref{aff122}}
\and I.~T.~Andika\orcid{0000-0001-6102-9526}\inst{\ref{aff123},\ref{aff124}}
\and S.~Anselmi\orcid{0000-0002-3579-9583}\inst{\ref{aff15},\ref{aff22},\ref{aff125}}
\and M.~Archidiacono\orcid{0000-0003-4952-9012}\inst{\ref{aff65},\ref{aff66}}
\and F.~Atrio-Barandela\orcid{0000-0002-2130-2513}\inst{\ref{aff126}}
\and D.~Bertacca\orcid{0000-0002-2490-7139}\inst{\ref{aff22},\ref{aff13},\ref{aff15}}
\and M.~Bethermin\orcid{0000-0002-3915-2015}\inst{\ref{aff127}}
\and A.~Blanchard\orcid{0000-0001-8555-9003}\inst{\ref{aff103}}
\and L.~Blot\orcid{0000-0002-9622-7167}\inst{\ref{aff128},\ref{aff77}}
\and S.~Borgani\orcid{0000-0001-6151-6439}\inst{\ref{aff129},\ref{aff27},\ref{aff28},\ref{aff29},\ref{aff120}}
\and M.~L.~Brown\orcid{0000-0002-0370-8077}\inst{\ref{aff56}}
\and S.~Bruton\orcid{0000-0002-6503-5218}\inst{\ref{aff91}}
\and A.~Calabro\orcid{0000-0003-2536-1614}\inst{\ref{aff50}}
\and B.~Camacho~Quevedo\orcid{0000-0002-8789-4232}\inst{\ref{aff49},\ref{aff48}}
\and F.~Caro\inst{\ref{aff50}}
\and C.~S.~Carvalho\inst{\ref{aff109}}
\and T.~Castro\orcid{0000-0002-6292-3228}\inst{\ref{aff28},\ref{aff29},\ref{aff27},\ref{aff120}}
\and Y.~Charles\inst{\ref{aff19}}
\and R.~Chary\orcid{0000-0001-7583-0621}\inst{\ref{aff108},\ref{aff130}}
\and A.~R.~Cooray\orcid{0000-0002-3892-0190}\inst{\ref{aff131}}
\and O.~Cucciati\orcid{0000-0002-9336-7551}\inst{\ref{aff16}}
\and S.~Davini\orcid{0000-0003-3269-1718}\inst{\ref{aff36}}
\and F.~De~Paolis\orcid{0000-0001-6460-7563}\inst{\ref{aff132},\ref{aff133},\ref{aff134}}
\and G.~Desprez\orcid{0000-0001-8325-1742}\inst{\ref{aff17}}
\and A.~D\'iaz-S\'anchez\orcid{0000-0003-0748-4768}\inst{\ref{aff135}}
\and S.~Di~Domizio\orcid{0000-0003-2863-5895}\inst{\ref{aff35},\ref{aff36}}
\and J.~M.~Diego\orcid{0000-0001-9065-3926}\inst{\ref{aff136}}
\and P.~Dimauro\orcid{0000-0001-7399-2854}\inst{\ref{aff50},\ref{aff137}}
\and A.~Enia\orcid{0000-0002-0200-2857}\inst{\ref{aff31},\ref{aff16}}
\and Y.~Fang\inst{\ref{aff12}}
\and A.~M.~N.~Ferguson\inst{\ref{aff55}}
\and A.~G.~Ferrari\orcid{0009-0005-5266-4110}\inst{\ref{aff24}}
\and A.~Finoguenov\orcid{0000-0002-4606-5403}\inst{\ref{aff74}}
\and A.~Fontana\orcid{0000-0003-3820-2823}\inst{\ref{aff50}}
\and A.~Franco\orcid{0000-0002-4761-366X}\inst{\ref{aff133},\ref{aff132},\ref{aff134}}
\and K.~Ganga\orcid{0000-0001-8159-8208}\inst{\ref{aff85}}
\and J.~Garc\'ia-Bellido\orcid{0000-0002-9370-8360}\inst{\ref{aff121}}
\and T.~Gasparetto\orcid{0000-0002-7913-4866}\inst{\ref{aff28}}
\and V.~Gautard\inst{\ref{aff138}}
\and E.~Gaztanaga\orcid{0000-0001-9632-0815}\inst{\ref{aff48},\ref{aff49},\ref{aff139}}
\and F.~Giacomini\orcid{0000-0002-3129-2814}\inst{\ref{aff24}}
\and F.~Gianotti\orcid{0000-0003-4666-119X}\inst{\ref{aff16}}
\and G.~Gozaliasl\orcid{0000-0002-0236-919X}\inst{\ref{aff140},\ref{aff74}}
\and M.~Guidi\orcid{0000-0001-9408-1101}\inst{\ref{aff31},\ref{aff16}}
\and C.~M.~Gutierrez\orcid{0000-0001-7854-783X}\inst{\ref{aff141}}
\and A.~Hall\orcid{0000-0002-3139-8651}\inst{\ref{aff55}}
\and H.~Hildebrandt\orcid{0000-0002-9814-3338}\inst{\ref{aff142}}
\and J.~Hjorth\orcid{0000-0002-4571-2306}\inst{\ref{aff94}}
\and J.~J.~E.~Kajava\orcid{0000-0002-3010-8333}\inst{\ref{aff143},\ref{aff144}}
\and Y.~Kang\orcid{0009-0000-8588-7250}\inst{\ref{aff62}}
\and V.~Kansal\orcid{0000-0002-4008-6078}\inst{\ref{aff145},\ref{aff146}}
\and D.~Karagiannis\orcid{0000-0002-4927-0816}\inst{\ref{aff113},\ref{aff147}}
\and K.~Kiiveri\inst{\ref{aff72}}
\and C.~C.~Kirkpatrick\inst{\ref{aff72}}
\and S.~Kruk\orcid{0000-0001-8010-8879}\inst{\ref{aff3}}
\and J.~Le~Graet\orcid{0000-0001-6523-7971}\inst{\ref{aff2}}
\and L.~Legrand\orcid{0000-0003-0610-5252}\inst{\ref{aff148},\ref{aff149}}
\and M.~Lembo\orcid{0000-0002-5271-5070}\inst{\ref{aff113},\ref{aff114}}
\and F.~Lepori\orcid{0009-0000-5061-7138}\inst{\ref{aff150}}
\and G.~Leroy\orcid{0009-0004-2523-4425}\inst{\ref{aff151},\ref{aff83}}
\and G.~F.~Lesci\orcid{0000-0002-4607-2830}\inst{\ref{aff21},\ref{aff16}}
\and J.~Lesgourgues\orcid{0000-0001-7627-353X}\inst{\ref{aff47}}
\and L.~Leuzzi\orcid{0009-0006-4479-7017}\inst{\ref{aff21},\ref{aff16}}
\and T.~I.~Liaudat\orcid{0000-0002-9104-314X}\inst{\ref{aff152}}
\and A.~Loureiro\orcid{0000-0002-4371-0876}\inst{\ref{aff153},\ref{aff154}}
\and J.~Macias-Perez\orcid{0000-0002-5385-2763}\inst{\ref{aff155}}
\and G.~Maggio\orcid{0000-0003-4020-4836}\inst{\ref{aff28}}
\and M.~Magliocchetti\orcid{0000-0001-9158-4838}\inst{\ref{aff63}}
\and C.~Mancini\orcid{0000-0002-4297-0561}\inst{\ref{aff44}}
\and F.~Mannucci\orcid{0000-0002-4803-2381}\inst{\ref{aff156}}
\and R.~Maoli\orcid{0000-0002-6065-3025}\inst{\ref{aff157},\ref{aff50}}
\and C.~J.~A.~P.~Martins\orcid{0000-0002-4886-9261}\inst{\ref{aff158},\ref{aff39}}
\and L.~Maurin\orcid{0000-0002-8406-0857}\inst{\ref{aff25}}
\and M.~Miluzio\inst{\ref{aff3},\ref{aff159}}
\and P.~Monaco\orcid{0000-0003-2083-7564}\inst{\ref{aff129},\ref{aff28},\ref{aff29},\ref{aff27}}
\and A.~Montoro\orcid{0000-0003-4730-8590}\inst{\ref{aff48},\ref{aff49}}
\and C.~Moretti\orcid{0000-0003-3314-8936}\inst{\ref{aff30},\ref{aff120},\ref{aff28},\ref{aff27},\ref{aff29}}
\and C.~Murray\inst{\ref{aff85}}
\and S.~Nadathur\orcid{0000-0001-9070-3102}\inst{\ref{aff139}}
\and K.~Naidoo\orcid{0000-0002-9182-1802}\inst{\ref{aff139}}
\and A.~Navarro-Alsina\orcid{0000-0002-3173-2592}\inst{\ref{aff81}}
\and F.~Passalacqua\orcid{0000-0002-8606-4093}\inst{\ref{aff22},\ref{aff15}}
\and K.~Paterson\orcid{0000-0001-8340-3486}\inst{\ref{aff10}}
\and A.~Pisani\orcid{0000-0002-6146-4437}\inst{\ref{aff2}}
\and D.~Potter\orcid{0000-0002-0757-5195}\inst{\ref{aff150}}
\and S.~Quai\orcid{0000-0002-0449-8163}\inst{\ref{aff21},\ref{aff16}}
\and M.~Radovich\orcid{0000-0002-3585-866X}\inst{\ref{aff13}}
\and P.-F.~Rocci\inst{\ref{aff25}}
\and S.~Sacquegna\orcid{0000-0002-8433-6630}\inst{\ref{aff132},\ref{aff133},\ref{aff134}}
\and M.~Sahl\'en\orcid{0000-0003-0973-4804}\inst{\ref{aff160}}
\and D.~B.~Sanders\orcid{0000-0002-1233-9998}\inst{\ref{aff52}}
\and E.~Sarpa\orcid{0000-0002-1256-655X}\inst{\ref{aff30},\ref{aff120},\ref{aff29}}
\and A.~Schneider\orcid{0000-0001-7055-8104}\inst{\ref{aff150}}
\and D.~Sciotti\orcid{0009-0008-4519-2620}\inst{\ref{aff50},\ref{aff82}}
\and E.~Sellentin\inst{\ref{aff161},\ref{aff18}}
\and G.~Setnikar\orcid{0009-0000-0136-3397}\inst{\ref{aff1}}
\and L.~C.~Smith\orcid{0000-0002-3259-2771}\inst{\ref{aff162}}
\and K.~Tanidis\orcid{0000-0001-9843-5130}\inst{\ref{aff116}}
\and C.~Tao\orcid{0000-0001-7961-8177}\inst{\ref{aff2}}
\and G.~Testera\inst{\ref{aff36}}
\and R.~Teyssier\orcid{0000-0001-7689-0933}\inst{\ref{aff163}}
\and S.~Tosi\orcid{0000-0002-7275-9193}\inst{\ref{aff35},\ref{aff36},\ref{aff26}}
\and A.~Troja\orcid{0000-0003-0239-4595}\inst{\ref{aff22},\ref{aff15}}
\and M.~Tucci\inst{\ref{aff62}}
\and C.~Valieri\inst{\ref{aff24}}
\and A.~Venhola\orcid{0000-0001-6071-4564}\inst{\ref{aff164}}
\and D.~Vergani\orcid{0000-0003-0898-2216}\inst{\ref{aff16}}
\and G.~Verza\orcid{0000-0002-1886-8348}\inst{\ref{aff165}}
\and J.~R.~Weaver\orcid{0000-0003-1614-196X}\inst{\ref{aff166}}
\and L.~Zalesky\orcid{0000-0001-5680-2326}\inst{\ref{aff52}}}
										   
\institute{Universit\'e Claude Bernard Lyon 1, CNRS/IN2P3, IP2I Lyon, UMR 5822, Villeurbanne, F-69100, France\label{aff1}
\and
Aix-Marseille Universit\'e, CNRS/IN2P3, CPPM, Marseille, France\label{aff2}
\and
ESAC/ESA, Camino Bajo del Castillo, s/n., Urb. Villafranca del Castillo, 28692 Villanueva de la Ca\~nada, Madrid, Spain\label{aff3}
\and
European Space Agency/ESTEC, Keplerlaan 1, 2201 AZ Noordwijk, The Netherlands\label{aff4}
\and
European Space Agency/ESRIN, Largo Galileo Galilei 1, 00044 Frascati, Roma, Italy\label{aff5}
\and
Centre National d'Etudes Spatiales -- Centre spatial de Toulouse, 18 avenue Edouard Belin, 31401 Toulouse Cedex 9, France\label{aff6}
\and
Jet Propulsion Laboratory, California Institute of Technology, 4800 Oak Grove Drive, Pasadena, CA, 91109, USA\label{aff7}
\and
NASA Goddard Space Flight Center, Greenbelt, MD 20771, USA\label{aff8}
\and
Carnegie Observatories, Pasadena, CA 91101, USA\label{aff9}
\and
Max-Planck-Institut f\"ur Astronomie, K\"onigstuhl 17, 69117 Heidelberg, Germany\label{aff10}
\and
Max Planck Institute for Extraterrestrial Physics, Giessenbachstr. 1, 85748 Garching, Germany\label{aff11}
\and
Universit\"ats-Sternwarte M\"unchen, Fakult\"at f\"ur Physik, Ludwig-Maximilians-Universit\"at M\"unchen, Scheinerstrasse 1, 81679 M\"unchen, Germany\label{aff12}
\and
INAF-Osservatorio Astronomico di Padova, Via dell'Osservatorio 5, 35122 Padova, Italy\label{aff13}
\and
INAF-Osservatorio Astrofisico di Torino, Via Osservatorio 20, 10025 Pino Torinese (TO), Italy\label{aff14}
\and
INFN-Padova, Via Marzolo 8, 35131 Padova, Italy\label{aff15}
\and
INAF-Osservatorio di Astrofisica e Scienza dello Spazio di Bologna, Via Piero Gobetti 93/3, 40129 Bologna, Italy\label{aff16}
\and
Kapteyn Astronomical Institute, University of Groningen, PO Box 800, 9700 AV Groningen, The Netherlands\label{aff17}
\and
Leiden Observatory, Leiden University, Einsteinweg 55, 2333 CC Leiden, The Netherlands\label{aff18}
\and
Aix-Marseille Universit\'e, CNRS, CNES, LAM, Marseille, France\label{aff19}
\and
INFN-Bologna, Via Irnerio 46, 40126 Bologna, Italy\label{aff20}
\and
Dipartimento di Fisica e Astronomia "Augusto Righi" - Alma Mater Studiorum Universit\`a di Bologna, via Piero Gobetti 93/2, 40129 Bologna, Italy\label{aff21}
\and
Dipartimento di Fisica e Astronomia "G. Galilei", Universit\`a di Padova, Via Marzolo 8, 35131 Padova, Italy\label{aff22}
\and
Universidad Polit\'ecnica de Cartagena, Departamento de Electr\'onica y Tecnolog\'ia de Computadoras,  Plaza del Hospital 1, 30202 Cartagena, Spain\label{aff23}
\and
INFN-Sezione di Bologna, Viale Berti Pichat 6/2, 40127 Bologna, Italy\label{aff24}
\and
Universit\'e Paris-Saclay, CNRS, Institut d'astrophysique spatiale, 91405, Orsay, France\label{aff25}
\and
INAF-Osservatorio Astronomico di Brera, Via Brera 28, 20122 Milano, Italy\label{aff26}
\and
IFPU, Institute for Fundamental Physics of the Universe, via Beirut 2, 34151 Trieste, Italy\label{aff27}
\and
INAF-Osservatorio Astronomico di Trieste, Via G. B. Tiepolo 11, 34143 Trieste, Italy\label{aff28}
\and
INFN, Sezione di Trieste, Via Valerio 2, 34127 Trieste TS, Italy\label{aff29}
\and
SISSA, International School for Advanced Studies, Via Bonomea 265, 34136 Trieste TS, Italy\label{aff30}
\and
Dipartimento di Fisica e Astronomia, Universit\`a di Bologna, Via Gobetti 93/2, 40129 Bologna, Italy\label{aff31}
\and
Institut de Physique Th\'eorique, CEA, CNRS, Universit\'e Paris-Saclay 91191 Gif-sur-Yvette Cedex, France\label{aff32}
\and
Institut d'Astrophysique de Paris, UMR 7095, CNRS, and Sorbonne Universit\'e, 98 bis boulevard Arago, 75014 Paris, France\label{aff33}
\and
Space Science Data Center, Italian Space Agency, via del Politecnico snc, 00133 Roma, Italy\label{aff34}
\and
Dipartimento di Fisica, Universit\`a di Genova, Via Dodecaneso 33, 16146, Genova, Italy\label{aff35}
\and
INFN-Sezione di Genova, Via Dodecaneso 33, 16146, Genova, Italy\label{aff36}
\and
Department of Physics "E. Pancini", University Federico II, Via Cinthia 6, 80126, Napoli, Italy\label{aff37}
\and
INAF-Osservatorio Astronomico di Capodimonte, Via Moiariello 16, 80131 Napoli, Italy\label{aff38}
\and
Instituto de Astrof\'isica e Ci\^encias do Espa\c{c}o, Universidade do Porto, CAUP, Rua das Estrelas, PT4150-762 Porto, Portugal\label{aff39}
\and
Faculdade de Ci\^encias da Universidade do Porto, Rua do Campo de Alegre, 4150-007 Porto, Portugal\label{aff40}
\and
Dipartimento di Fisica, Universit\`a degli Studi di Torino, Via P. Giuria 1, 10125 Torino, Italy\label{aff41}
\and
INFN-Sezione di Torino, Via P. Giuria 1, 10125 Torino, Italy\label{aff42}
\and
Institute Lorentz, Leiden University, Niels Bohrweg 2, 2333 CA Leiden, The Netherlands\label{aff43}
\and
INAF-IASF Milano, Via Alfonso Corti 12, 20133 Milano, Italy\label{aff44}
\and
Centro de Investigaciones Energ\'eticas, Medioambientales y Tecnol\'ogicas (CIEMAT), Avenida Complutense 40, 28040 Madrid, Spain\label{aff45}
\and
Port d'Informaci\'{o} Cient\'{i}fica, Campus UAB, C. Albareda s/n, 08193 Bellaterra (Barcelona), Spain\label{aff46}
\and
Institute for Theoretical Particle Physics and Cosmology (TTK), RWTH Aachen University, 52056 Aachen, Germany\label{aff47}
\and
Institute of Space Sciences (ICE, CSIC), Campus UAB, Carrer de Can Magrans, s/n, 08193 Barcelona, Spain\label{aff48}
\and
Institut d'Estudis Espacials de Catalunya (IEEC),  Edifici RDIT, Campus UPC, 08860 Castelldefels, Barcelona, Spain\label{aff49}
\and
INAF-Osservatorio Astronomico di Roma, Via Frascati 33, 00078 Monteporzio Catone, Italy\label{aff50}
\and
INFN section of Naples, Via Cinthia 6, 80126, Napoli, Italy\label{aff51}
\and
Institute for Astronomy, University of Hawaii, 2680 Woodlawn Drive, Honolulu, HI 96822, USA\label{aff52}
\and
Dipartimento di Fisica e Astronomia "Augusto Righi" - Alma Mater Studiorum Universit\`a di Bologna, Viale Berti Pichat 6/2, 40127 Bologna, Italy\label{aff53}
\and
Instituto de Astrof\'{\i}sica de Canarias, V\'{\i}a L\'actea, 38205 La Laguna, Tenerife, Spain\label{aff54}
\and
Institute for Astronomy, University of Edinburgh, Royal Observatory, Blackford Hill, Edinburgh EH9 3HJ, UK\label{aff55}
\and
Jodrell Bank Centre for Astrophysics, Department of Physics and Astronomy, University of Manchester, Oxford Road, Manchester M13 9PL, UK\label{aff56}
\and
Institut de Ci\`{e}ncies del Cosmos (ICCUB), Universitat de Barcelona (IEEC-UB), Mart\'{i} i Franqu\`{e}s 1, 08028 Barcelona, Spain\label{aff57}
\and
Instituci\'o Catalana de Recerca i Estudis Avan\c{c}ats (ICREA), Passeig de Llu\'{\i}s Companys 23, 08010 Barcelona, Spain\label{aff58}
\and
UCB Lyon 1, CNRS/IN2P3, IUF, IP2I Lyon, 4 rue Enrico Fermi, 69622 Villeurbanne, France\label{aff59}
\and
Departamento de F\'isica, Faculdade de Ci\^encias, Universidade de Lisboa, Edif\'icio C8, Campo Grande, PT1749-016 Lisboa, Portugal\label{aff60}
\and
Instituto de Astrof\'isica e Ci\^encias do Espa\c{c}o, Faculdade de Ci\^encias, Universidade de Lisboa, Campo Grande, 1749-016 Lisboa, Portugal\label{aff61}
\and
Department of Astronomy, University of Geneva, ch. d'Ecogia 16, 1290 Versoix, Switzerland\label{aff62}
\and
INAF-Istituto di Astrofisica e Planetologia Spaziali, via del Fosso del Cavaliere, 100, 00100 Roma, Italy\label{aff63}
\and
School of Physics, HH Wills Physics Laboratory, University of Bristol, Tyndall Avenue, Bristol, BS8 1TL, UK\label{aff64}
\and
Dipartimento di Fisica "Aldo Pontremoli", Universit\`a degli Studi di Milano, Via Celoria 16, 20133 Milano, Italy\label{aff65}
\and
INFN-Sezione di Milano, Via Celoria 16, 20133 Milano, Italy\label{aff66}
\and
Institute of Theoretical Astrophysics, University of Oslo, P.O. Box 1029 Blindern, 0315 Oslo, Norway\label{aff67}
\and
Department of Physics, Lancaster University, Lancaster, LA1 4YB, UK\label{aff68}
\and
Felix Hormuth Engineering, Goethestr. 17, 69181 Leimen, Germany\label{aff69}
\and
Technical University of Denmark, Elektrovej 327, 2800 Kgs. Lyngby, Denmark\label{aff70}
\and
Cosmic Dawn Center (DAWN), Denmark\label{aff71}
\and
Department of Physics and Helsinki Institute of Physics, Gustaf H\"allstr\"omin katu 2, 00014 University of Helsinki, Finland\label{aff72}
\and
Universit\'e de Gen\`eve, D\'epartement de Physique Th\'eorique and Centre for Astroparticle Physics, 24 quai Ernest-Ansermet, CH-1211 Gen\`eve 4, Switzerland\label{aff73}
\and
Department of Physics, P.O. Box 64, 00014 University of Helsinki, Finland\label{aff74}
\and
Helsinki Institute of Physics, Gustaf H{\"a}llstr{\"o}min katu 2, University of Helsinki, Helsinki, Finland\label{aff75}
\and
Centre de Calcul de l'IN2P3/CNRS, 21 avenue Pierre de Coubertin 69627 Villeurbanne Cedex, France\label{aff76}
\and
Laboratoire d'etude de l'Univers et des phenomenes eXtremes, Observatoire de Paris, Universit\'e PSL, Sorbonne Universit\'e, CNRS, 92190 Meudon, France\label{aff77}
\and
Mullard Space Science Laboratory, University College London, Holmbury St Mary, Dorking, Surrey RH5 6NT, UK\label{aff78}
\and
SKA Observatory, Jodrell Bank, Lower Withington, Macclesfield, Cheshire SK11 9FT, UK\label{aff79}
\and
University of Applied Sciences and Arts of Northwestern Switzerland, School of Computer Science, 5210 Windisch, Switzerland\label{aff80}
\and
Universit\"at Bonn, Argelander-Institut f\"ur Astronomie, Auf dem H\"ugel 71, 53121 Bonn, Germany\label{aff81}
\and
INFN-Sezione di Roma, Piazzale Aldo Moro, 2 - c/o Dipartimento di Fisica, Edificio G. Marconi, 00185 Roma, Italy\label{aff82}
\and
Department of Physics, Institute for Computational Cosmology, Durham University, South Road, Durham, DH1 3LE, UK\label{aff83}
\and
Universit\'e C\^{o}te d'Azur, Observatoire de la C\^{o}te d'Azur, CNRS, Laboratoire Lagrange, Bd de l'Observatoire, CS 34229, 06304 Nice cedex 4, France\label{aff84}
\and
Universit\'e Paris Cit\'e, CNRS, Astroparticule et Cosmologie, 75013 Paris, France\label{aff85}
\and
CNRS-UCB International Research Laboratory, Centre Pierre Bin\'etruy, IRL2007, CPB-IN2P3, Berkeley, USA\label{aff86}
\and
University of Applied Sciences and Arts of Northwestern Switzerland, School of Engineering, 5210 Windisch, Switzerland\label{aff87}
\and
Institut d'Astrophysique de Paris, 98bis Boulevard Arago, 75014, Paris, France\label{aff88}
\and
Institute of Physics, Laboratory of Astrophysics, Ecole Polytechnique F\'ed\'erale de Lausanne (EPFL), Observatoire de Sauverny, 1290 Versoix, Switzerland\label{aff89}
\and
Aurora Technology for European Space Agency (ESA), Camino bajo del Castillo, s/n, Urbanizacion Villafranca del Castillo, Villanueva de la Ca\~nada, 28692 Madrid, Spain\label{aff90}
\and
California Institute of Technology, 1200 E California Blvd, Pasadena, CA 91125, USA\label{aff91}
\and
Institut de F\'{i}sica d'Altes Energies (IFAE), The Barcelona Institute of Science and Technology, Campus UAB, 08193 Bellaterra (Barcelona), Spain\label{aff92}
\and
School of Mathematics and Physics, University of Surrey, Guildford, Surrey, GU2 7XH, UK\label{aff93}
\and
DARK, Niels Bohr Institute, University of Copenhagen, Jagtvej 155, 2200 Copenhagen, Denmark\label{aff94}
\and
Waterloo Centre for Astrophysics, University of Waterloo, Waterloo, Ontario N2L 3G1, Canada\label{aff95}
\and
Department of Physics and Astronomy, University of Waterloo, Waterloo, Ontario N2L 3G1, Canada\label{aff96}
\and
Perimeter Institute for Theoretical Physics, Waterloo, Ontario N2L 2Y5, Canada\label{aff97}
\and
Universit\'e Paris-Saclay, Universit\'e Paris Cit\'e, CEA, CNRS, AIM, 91191, Gif-sur-Yvette, France\label{aff98}
\and
Institute of Space Science, Str. Atomistilor, nr. 409 M\u{a}gurele, Ilfov, 077125, Romania\label{aff99}
\and
Consejo Superior de Investigaciones Cientificas, Calle Serrano 117, 28006 Madrid, Spain\label{aff100}
\and
Universidad de La Laguna, Departamento de Astrof\'{\i}sica, 38206 La Laguna, Tenerife, Spain\label{aff101}
\and
Institut f\"ur Theoretische Physik, University of Heidelberg, Philosophenweg 16, 69120 Heidelberg, Germany\label{aff102}
\and
Institut de Recherche en Astrophysique et Plan\'etologie (IRAP), Universit\'e de Toulouse, CNRS, UPS, CNES, 14 Av. Edouard Belin, 31400 Toulouse, France\label{aff103}
\and
Universit\'e St Joseph; Faculty of Sciences, Beirut, Lebanon\label{aff104}
\and
Departamento de F\'isica, FCFM, Universidad de Chile, Blanco Encalada 2008, Santiago, Chile\label{aff105}
\and
Universit\"at Innsbruck, Institut f\"ur Astro- und Teilchenphysik, Technikerstr. 25/8, 6020 Innsbruck, Austria\label{aff106}
\and
Satlantis, University Science Park, Sede Bld 48940, Leioa-Bilbao, Spain\label{aff107}
\and
Infrared Processing and Analysis Center, California Institute of Technology, Pasadena, CA 91125, USA\label{aff108}
\and
Instituto de Astrof\'isica e Ci\^encias do Espa\c{c}o, Faculdade de Ci\^encias, Universidade de Lisboa, Tapada da Ajuda, 1349-018 Lisboa, Portugal\label{aff109}
\and
Cosmic Dawn Center (DAWN)\label{aff110}
\and
Niels Bohr Institute, University of Copenhagen, Jagtvej 128, 2200 Copenhagen, Denmark\label{aff111}
\and
Centre for Information Technology, University of Groningen, P.O. Box 11044, 9700 CA Groningen, The Netherlands\label{aff112}
\and
Dipartimento di Fisica e Scienze della Terra, Universit\`a degli Studi di Ferrara, Via Giuseppe Saragat 1, 44122 Ferrara, Italy\label{aff113}
\and
Istituto Nazionale di Fisica Nucleare, Sezione di Ferrara, Via Giuseppe Saragat 1, 44122 Ferrara, Italy\label{aff114}
\and
INAF, Istituto di Radioastronomia, Via Piero Gobetti 101, 40129 Bologna, Italy\label{aff115}
\and
Department of Physics, Oxford University, Keble Road, Oxford OX1 3RH, UK\label{aff116}
\and
INAF - Osservatorio Astronomico di Brera, via Emilio Bianchi 46, 23807 Merate, Italy\label{aff117}
\and
INAF-Osservatorio Astronomico di Brera, Via Brera 28, 20122 Milano, Italy, and INFN-Sezione di Genova, Via Dodecaneso 33, 16146, Genova, Italy\label{aff118}
\and
ICL, Junia, Universit\'e Catholique de Lille, LITL, 59000 Lille, France\label{aff119}
\and
ICSC - Centro Nazionale di Ricerca in High Performance Computing, Big Data e Quantum Computing, Via Magnanelli 2, Bologna, Italy\label{aff120}
\and
Instituto de F\'isica Te\'orica UAM-CSIC, Campus de Cantoblanco, 28049 Madrid, Spain\label{aff121}
\and
CERCA/ISO, Department of Physics, Case Western Reserve University, 10900 Euclid Avenue, Cleveland, OH 44106, USA\label{aff122}
\and
Technical University of Munich, TUM School of Natural Sciences, Physics Department, James-Franck-Str.~1, 85748 Garching, Germany\label{aff123}
\and
Max-Planck-Institut f\"ur Astrophysik, Karl-Schwarzschild-Str.~1, 85748 Garching, Germany\label{aff124}
\and
Laboratoire Univers et Th\'eorie, Observatoire de Paris, Universit\'e PSL, Universit\'e Paris Cit\'e, CNRS, 92190 Meudon, France\label{aff125}
\and
Departamento de F{\'\i}sica Fundamental. Universidad de Salamanca. Plaza de la Merced s/n. 37008 Salamanca, Spain\label{aff126}
\and
Universit\'e de Strasbourg, CNRS, Observatoire astronomique de Strasbourg, UMR 7550, 67000 Strasbourg, France\label{aff127}
\and
Center for Data-Driven Discovery, Kavli IPMU (WPI), UTIAS, The University of Tokyo, Kashiwa, Chiba 277-8583, Japan\label{aff128}
\and
Dipartimento di Fisica - Sezione di Astronomia, Universit\`a di Trieste, Via Tiepolo 11, 34131 Trieste, Italy\label{aff129}
\and
University of California, Los Angeles, CA 90095-1562, USA\label{aff130}
\and
Department of Physics \& Astronomy, University of California Irvine, Irvine CA 92697, USA\label{aff131}
\and
Department of Mathematics and Physics E. De Giorgi, University of Salento, Via per Arnesano, CP-I93, 73100, Lecce, Italy\label{aff132}
\and
INFN, Sezione di Lecce, Via per Arnesano, CP-193, 73100, Lecce, Italy\label{aff133}
\and
INAF-Sezione di Lecce, c/o Dipartimento Matematica e Fisica, Via per Arnesano, 73100, Lecce, Italy\label{aff134}
\and
Departamento F\'isica Aplicada, Universidad Polit\'ecnica de Cartagena, Campus Muralla del Mar, 30202 Cartagena, Murcia, Spain\label{aff135}
\and
Instituto de F\'isica de Cantabria, Edificio Juan Jord\'a, Avenida de los Castros, 39005 Santander, Spain\label{aff136}
\and
Observatorio Nacional, Rua General Jose Cristino, 77-Bairro Imperial de Sao Cristovao, Rio de Janeiro, 20921-400, Brazil\label{aff137}
\and
CEA Saclay, DFR/IRFU, Service d'Astrophysique, Bat. 709, 91191 Gif-sur-Yvette, France\label{aff138}
\and
Institute of Cosmology and Gravitation, University of Portsmouth, Portsmouth PO1 3FX, UK\label{aff139}
\and
Department of Computer Science, Aalto University, PO Box 15400, Espoo, FI-00 076, Finland\label{aff140}
\and
Instituto de Astrof\'\i sica de Canarias, c/ Via Lactea s/n, La Laguna 38200, Spain. Departamento de Astrof\'\i sica de la Universidad de La Laguna, Avda. Francisco Sanchez, La Laguna, 38200, Spain\label{aff141}
\and
Ruhr University Bochum, Faculty of Physics and Astronomy, Astronomical Institute (AIRUB), German Centre for Cosmological Lensing (GCCL), 44780 Bochum, Germany\label{aff142}
\and
Department of Physics and Astronomy, Vesilinnantie 5, 20014 University of Turku, Finland\label{aff143}
\and
Serco for European Space Agency (ESA), Camino bajo del Castillo, s/n, Urbanizacion Villafranca del Castillo, Villanueva de la Ca\~nada, 28692 Madrid, Spain\label{aff144}
\and
ARC Centre of Excellence for Dark Matter Particle Physics, Melbourne, Australia\label{aff145}
\and
Centre for Astrophysics \& Supercomputing, Swinburne University of Technology,  Hawthorn, Victoria 3122, Australia\label{aff146}
\and
Department of Physics and Astronomy, University of the Western Cape, Bellville, Cape Town, 7535, South Africa\label{aff147}
\and
DAMTP, Centre for Mathematical Sciences, Wilberforce Road, Cambridge CB3 0WA, UK\label{aff148}
\and
Kavli Institute for Cosmology Cambridge, Madingley Road, Cambridge, CB3 0HA, UK\label{aff149}
\and
Department of Astrophysics, University of Zurich, Winterthurerstrasse 190, 8057 Zurich, Switzerland\label{aff150}
\and
Department of Physics, Centre for Extragalactic Astronomy, Durham University, South Road, Durham, DH1 3LE, UK\label{aff151}
\and
IRFU, CEA, Universit\'e Paris-Saclay 91191 Gif-sur-Yvette Cedex, France\label{aff152}
\and
Oskar Klein Centre for Cosmoparticle Physics, Department of Physics, Stockholm University, Stockholm, SE-106 91, Sweden\label{aff153}
\and
Astrophysics Group, Blackett Laboratory, Imperial College London, London SW7 2AZ, UK\label{aff154}
\and
Univ. Grenoble Alpes, CNRS, Grenoble INP, LPSC-IN2P3, 53, Avenue des Martyrs, 38000, Grenoble, France\label{aff155}
\and
INAF-Osservatorio Astrofisico di Arcetri, Largo E. Fermi 5, 50125, Firenze, Italy\label{aff156}
\and
Dipartimento di Fisica, Sapienza Universit\`a di Roma, Piazzale Aldo Moro 2, 00185 Roma, Italy\label{aff157}
\and
Centro de Astrof\'{\i}sica da Universidade do Porto, Rua das Estrelas, 4150-762 Porto, Portugal\label{aff158}
\and
HE Space for European Space Agency (ESA), Camino bajo del Castillo, s/n, Urbanizacion Villafranca del Castillo, Villanueva de la Ca\~nada, 28692 Madrid, Spain\label{aff159}
\and
Theoretical astrophysics, Department of Physics and Astronomy, Uppsala University, Box 516, 751 37 Uppsala, Sweden\label{aff160}
\and
Mathematical Institute, University of Leiden, Einsteinweg 55, 2333 CA Leiden, The Netherlands\label{aff161}
\and
Institute of Astronomy, University of Cambridge, Madingley Road, Cambridge CB3 0HA, UK\label{aff162}
\and
Department of Astrophysical Sciences, Peyton Hall, Princeton University, Princeton, NJ 08544, USA\label{aff163}
\and
Space physics and astronomy research unit, University of Oulu, Pentti Kaiteran katu 1, FI-90014 Oulu, Finland\label{aff164}
\and
Center for Computational Astrophysics, Flatiron Institute, 162 5th Avenue, 10010, New York, NY, USA\label{aff165}
\and
Department of Astronomy, University of Massachusetts, Amherst, MA 01003, USA\label{aff166}}    

\date{Received Now; accepted Never}
 
\abstract{The paper describes the objectives, design and findings of the pre-launch ground characterisation campaigns of the \Euclid infrared detectors. The aim of the ground characterisations is to evaluate the performance of the detectors, to calibrate the pixel response, and to derive the pixel response correction methods. The detectors have been tested and characterised in the facilities set up for this purpose. The pixel properties, including baseline, bad pixels, quantum efficiency, inter pixel capacitance, quantum efficiency, dark current, readout noise, conversion gain, response nonlinearity, and image persistence were measured and characterised for each pixel. We describe in detail the test flow definition that allows us to derive the pixel properties and we present the data acquisition and data quality check software implemented for this purpose. We also outline the measurement protocols of all the pixel properties presented and we provide a comprehensive overview of the performance of the \Euclid infrared detectors as derived after tuning the operating parameters of the detectors. 
The main conclusion of this work is that the performance of the infrared detectors \Euclid meets the requirements. Pixels classified as non-functioning accounted for less than 0.2\% of all science pixels. IPC coupling is minimal and crosstalk between adjacent pixels is less than 1\% between adjacent pixels. 95\% of the pixels show a QE greater than 80\% across the entire spectral range of the Euclid mission. The conversion gain is approximately 0.52\,ADU/e$^-$, with a variation less than 1\% between channels of the same detector. The reset noise is approximately equal to 23 ADU after reference pixels correction. The readout noise of a single frame is approximately \SI{13}{e^-} while the signal estimator noise is measured at 7\,e$^-$ in photometric mode and 9\,e$^-$ in spectroscopic acquisition mode. The deviation from linear response at signal levels up to 80\,ke$^-$ is less than 5\% for 95\% of the pixels. Median persistence amplitudes are less than 0.3\% of the signal, though persistence exhibits significant spatial variation and differences between detectors.
}


   \authorrunning{Euclid Collaboration: B.\ Kubik et al.}
   \titlerunning{\Euclid: NISP detectors performance on ground}
   \maketitle

\section{Introduction\label{sec:intro}}
The \Euclid mission, led by the European Space Agency (ESA) in collaboration with NASA, represents a cornerstone in our quest to understand the nature of dark energy and dark matter \citep{Laureijs11,EuclidSkyOverview}. 
With its wide field of view of \SI{0.5}{deg^2}, the \Euclid telescope will scan \SI{14000}{deg^2} of the extragalactic sky from the second Sun–Earth Lagrange point \citep{Scaramella-EP1}.

Equipped with two cutting-edge instruments -- the visible imager VIS \citep{EuclidSkyVIS} and the Near Infrared Spectrometer and Photometer \citep[NISP;][]{EuclidSkyNISP} -- the mission aims to map the geometry of the Universe with unprecedented precision. Central to the success of NISP is its reliance on HAWAII-2RG (H2RG)\footnote{HAWAII is an acronym for HgCdTe Astronomical Wide Area Infrared Imager. It includes a family of HxRG detectors, where H stands for HAWAII, the number x=1,\,2,\,4 denotes 1024\,$\times$\,1024, 2048\,$\times$\,2048 or 4096\,$\times$\,4096 pixels, R stands for reference pixels, and G for guide window capability.} detector arrays whose characterization and calibration are crucial for ensuring scientific accuracy.

Achieving the ambitious observational goals of the \Euclid telescope depends on reduced instrument systematics and precise measurements from the detectors. To maximize the performance of the signal detection chain, we must generally maximize the observed signal and minimize any sources of noise. To do this, we need detectors with high quantum efficiency (QE), low dark current and low readout noise. Likewise, we want devices with low persistence, because that false current from the previous exposures would also contribute to noise and is difficult to model and remove. At the same time the capacitive couplings between pixels should be minimal because they affect the instrument's point spread function (PSF). 

With the increasing demand for low-signal observations and the growing complexity of scientific requirements, achieving greater precision in detector performance measurements is becoming more critical. The ground characterisation performed in 2019 by the NISP detector team focuses on calibrating pixel responses and creating pixel maps that account for both systematic errors and correction functions. These functions correct signal values modified by effects present in the detector ensuring the integrity of the scientific data.

This paper presents the methodology of characterisation and summarizes the key properties of the flight detectors. In Sect.~\ref{sec:euclid_detectors} we present the architecture of the detector system, the detector readout principle, the common-mode correction approach, as well as the signal estimation algorithm employed during the \Euclid mission and in the generation of certain calibration products. In Sect.~\ref{sec:ground_charac} we outline the objectives of the characterisation campaign, detail and motivate the steps of the test-flow and describe the software developed for automated data acquisition and verification. Section~\ref{sec:tuning} presents the procedure of tuning the fundamental parameters of the detector system. The main results are presented in Sect.~\ref{sec:properties}, which provides a comprehensive overview of all detector properties measured during ground tests. For each property, we begin by explaining its physical significance and role in the overall performance of the detector system. Next, the specific measurement protocols used to evaluate each property are discussed, outlining the steps taken to ensure accuracy and consistency. The measurement results are then presented, highlighting their potential implications for detector performance. Section~\ref{sec:conclusions} summarizes the obtained results. Furthermore, a summary table of the main detector properties is given in Appendix~\ref{app:det_prop}. Appendix~\ref{app:ortho} presents the formalism of orthogonal polynomials employed in the nonlinearity characterisation.

\section{\Euclid IR detectors\label{sec:euclid_detectors}}
This section provides an overview of the IR detectors as utilised onboard the satellite. It details the architecture of the Sensor Chip System (SCS), the acquisition modes, the correction of common-mode signals and the signal estimation algorithms implemented within the onboard electronics.

\subsection{SCS architecture -- detailed description}

\begin{figure}
\centering
\includegraphics[width=\linewidth]{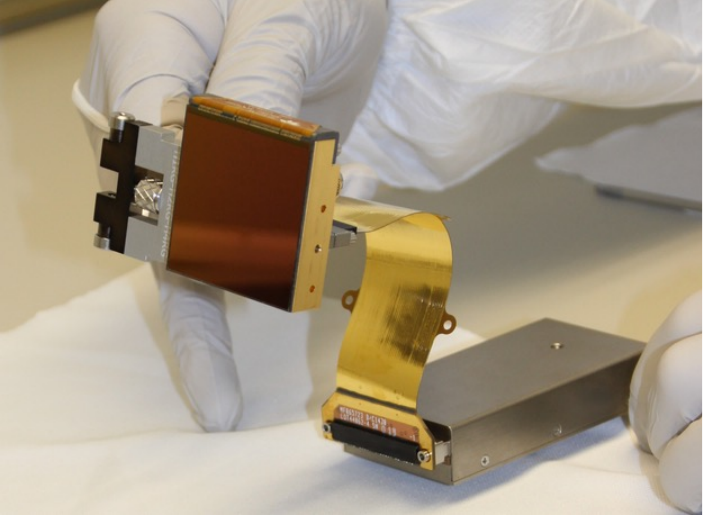}~~
\caption{\label{fig:scs} Photo of a \Euclid-like engineering grade sensor chip system taken in the clean room of the CNRS-IN2P3 Center for Particle Physics in Marseille (CPPM). The sensitive surface is 3.6\,cm\,$\times$\,3.6\,cm. Credit: CPPM-CNRS.
} 
\end{figure}

The NISP focal plane, described in detail in \citet{Maciaszek2022}, is composed of a $4\times 4$ mosaic of HgCdTe near-infrared detectors manufactured by Teledyne Imaging Sensors for the \Euclid mission. Each SCS is composed of three parts: the H2RG array, also called the sensor chip assembly (SCA), sensor chip electronics (SCE) and cryo-flex cable (CFC) as seen in Fig.~\ref{fig:scs}.

Each of the 16 SCAs consists of an area of 2040\,$\times$\,2040 science pixels surrounded by a 4-pixel wide border of reference pixels on all sides. The pixel pitch is 18\,\micron\ in both directions and the cut-off wavelength is 2.3\,\micron. Each pixel is made of HgCdTe semiconducting material and is connected to a readout integrated circuit (ROIC) via indium bumps. 

The SCE is an application specific integrated circuit (ASIC) including a system for image digitisation, enhancement, control and retrieval (SIDECAR) functioning at \SI{135}{K} during the ground tests. The main functionalities of the SCE are the overall ROIC control and sequencing, generation of biases to polarise the photosensitive volume and Analogue-to-Digital conversion (ADC) which provides signal in Analogue Digital Units (ADU). 
\Euclid's SCEs work in single ended analogue readout, allowing for baseline\footnote{Baseline is defined as the average value of the first 16 frames after a pixel reset. See Sects.~\ref{sec:baseline_tunning} and \ref{sec:baseline} for more details.} adjustment with tunable voltage register. The SIDECAR presents a digital interface to instrument electronics through low voltage differential signal (LVDS) communication. A detailed description of the SIDECAR ASIC architecture and functionalities can be found in \citet{LooseSPIE2005,Loose2007} and \citet{BeleticSPIE2008} .


The ROIC, or multiplexer, contains digital circuits and switches to address and readout signal voltages in the detector array. The operation of the ROIC is controlled by firmware loaded in the ASIC. This firmware was developed specifically for Euclid. The ROIC is configured to read 32 channels, each 64\,$\times$\,2048, in parallel and in buffered mode as shown in Fig. \ref{fig:frame_geometry}. Pixels within each channel are addressed in ‘slow’ mode readout at a rate of 100 kHz. After all pixels are read (or reset) there are 224 buffer lines for the ROIC and ASIC to perform housekeeping tasks and start a new frame. This yields a total frame time $t_{\rm fr} = 1.45408\,\mathrm{s}$. 
The NISP focal plane is designed for SCA operation at temperatures from 100 K to 85 K and has been tested over this temperature range. 
The full description of the detection physics is beyond the scope of this paper, the reader can refer to \citet{Mosby2020JATIS} and references therein for a recent and complete description of the HxRG sensors.

The cryo-flex cable connects the SCA to the SCE with a thermal conductance of \SI{0.85}{mW.K^{-1}} \citep{Holmes2019} keeping the two parts at the two different operating temperatures.

\subsection{SCS acquisition modes}

\begin{figure}
\centering
\includegraphics[width=\linewidth]{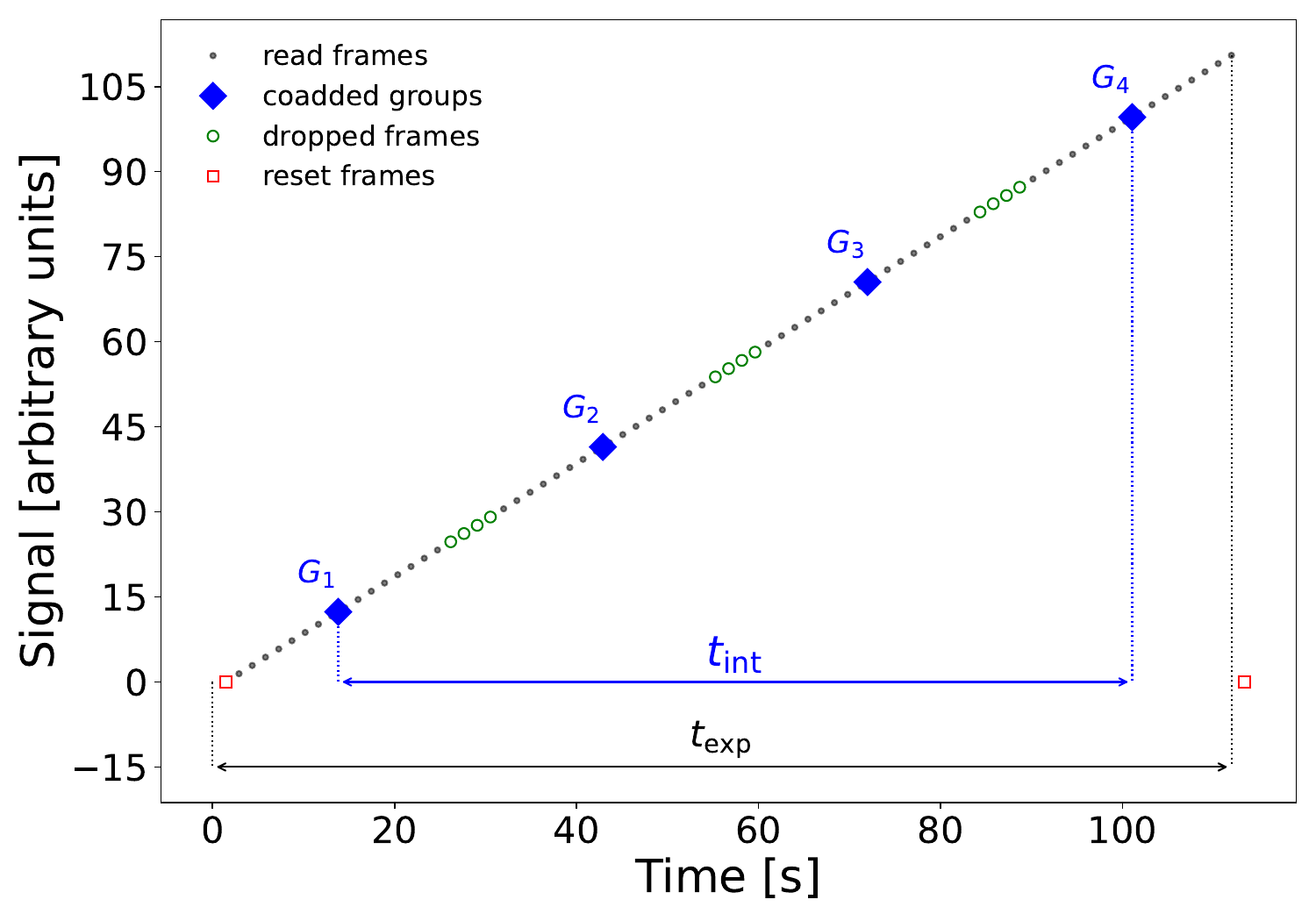}
\caption{\label{fig:scs_photo} Illustration of the multi-accumulation acquisition mode MACC($n_{\rm g},n_{\rm f},n_{\rm d}$) with $n_{\rm g}=4$, $n_{\rm f}=16$, $n_{\rm d}=4$ and one reset frame. Integration time $t_{\rm int}$ and exposure time $t_{\rm exp}$ defined in Eqs.~(\ref{eq:tint}) and (\ref{eq:texp}), respectively, are also indicated.} 
\end{figure}

The H2RG detectors can acquire signal in a so-called  multi-accumulation (MACC) acquisition mode \citep{Rauscher:2007gc} sketched in Fig.~\ref{fig:scs_photo}. The beginning of each exposure is defined by a reset frame. During the reset frame, each pixel is reset in single pixel reset mode or sequentially one by one. Immediately after reset electric charge accumulates at each pixel as generated by the incident absorbed photons.  Frames following the reset frame are read frames. During a read frame, each pixel is read non destructively.


A frame is the unit of data that results from sequentially clocking through and reading out a rectangular area of  $2048\times 2048$ pixels. Some of the frames may be dropped\footnote{The ROIC and SIDECAR ASIC clock all frames and transmit all data to the warm readout electronics in order to maintain thermal stability of the SCAs and SCEs. Selection of frames to coadd or drop is perfomed in the warm electronics.} while the integration of signal continues. At the end of the exposure, the signal per pixel is composed of $n_{\rm g}$ equally spaced groups. Each group contains $n_{\rm f}$ consecutive frames sampled up-the-ramp (UTR) and the groups are separated by $n_{\rm d}$ dropped frames. This non-destructive acquisition with UTR sampled data allows for a more  precise  signal  estimate  and  for the detection of anomalies that can occur during signal integration, such as cosmic ray hits, electronic jumps or deviations from linearity. 

The reset after each exposure is performed pixel by pixel in parallel in the 32 output channels. The time needed to reset all the pixels is equal to the time to read the entire detector. This reset scheme tends to reduce as much as possible the transient effect on the first read.

The NISP instrument during nominal observations of the sky acquires data in two predefined acquisition modes \citep{EuclidSkyNISP}:
\begin{itemize}
    \item The spectroscopic mode with 15 groups of 16 averaged frames and 11 dropped frames between each of two groups, MACC(15,16,11), with the total exposure time of about \SI{574.4}{s}.
    \item The photometric mode with 4 groups of 16 averaged frames and 4 dropped frames between each of two groups, MACC(4,16,4), with the total exposure time of about 112\,s.
\end{itemize}
This choice was a compromise between the signal to noise ratio (SNR) performance that varies depending on MACC schemes \citep{Kubik2015JATIS}, the limitations of the onboard computational resources, and of bandwidth  allocation to downlink the data to ground. It is not planned to vary the acquisition mode depending on the brightness of the observed sources, but specific acquisition modes are foreseen for in-flight calibrations, diagnostics, or for sanity checks of the detectors.

Table \ref{tab:modes} summarizes the main characteristics of the NISP acquisition modes for which detector performance is measured and presented in this paper. Integration time 
\begin{equation}\label{eq:tint}
   t_{\rm int}=(n_{\rm f} + n_{\rm d})(n_{\rm g}-1)t_{\rm fr}\,,
\end{equation}
relevant for science, and total exposure time 
\begin{equation}\label{eq:texp}
    t_{\rm exp}=\left[ n_{\rm f} n_{\rm g} + n_{\rm d}(n_{\rm g}-1) + n_{\rm r}\right]t_{\rm fr}\,,
\end{equation}
relevant for observation planning, are also reported. The exposure time includes the time of $n_{\rm r}=1$ reset frames at the beginning of each exposure.
In the table we also give the number of frames $n$ for up-the-ramp UTR($n$) acquisition with integration time equivalent to the given MACC, or explicitely $n = n_{\rm d}(n_{\rm g}-1) + n_{\rm f} n_{\rm g}$.

\begin{table}
\caption{NISP acquisition modes: MACC($n_{\rm g},n_{\rm f},n_{\rm d}$) and corresponding UTR($n$) used during ground characterisations with $n = n_{\rm f} n_{\rm g} + n_{\rm d}(n_{\rm g}-1)$. Integration time $t_{\rm int}$ and exposure time $t_{\rm exp}$ defined in Eqs.~(\ref{eq:tint}) and (\ref{eq:texp}), respectively, are also indicated.}
\label{tab:modes}
\begin{center}       
\begin{tabular}{l|cccccc}
\toprule
   NISP          & \multicolumn{3}{c}{MACC}               & UTR & $t_{\rm int}\,$[s] & $t_{\rm exp}\,$[s] \\
   exposure mode & $n_{\rm g}$ & $n_{\rm f}$ & $n_{\rm d}$ & $n$ &                    &                 \\
\midrule
   Spectrometer  & 15          & 16          & 11          & 394 &  549.6             & 574.4  \\
   Photometer    & 4           & 16          & 4           & 76  &  87.2              & 112.0  \\
\bottomrule
\end{tabular}
\end{center}
\end{table} 
 
\subsection{Removal of common modes using reference pixels}
\label{sec:ref_pix}

The sensitivity of integrating infrared detectors is limited by dark current and electronic readout noise. The dark current can be lowered down below the natural background level, such as zodiacal light described in \cite{Scaramella-EP1}, in the high-quality HgCdTe detectors by cooling them to temperatures below 100~K. Then the sensitivity of the array is essentially limited by the readout noise. This read noise is basically the noise of the SIDECAR ASIC and the noise of the Field-Effect Transistors (FET) used as a source-follower employed in the charge-to-voltage conversion in each pixel or "unit cell", including not only the statistical noise of this FET but also any noise associated with bias supplies and clocks. While the statistical noise of each unit cell FET is independent, any noise arising from common FETs in the signal chain or from common biases and clocks will be in general correlated.

The reference pixels provided within the detector allow us to reduce this common mode noise at least partially. Reference pixels do not respond to light but contain a simple capacitor $C_{\rm pix}$ with capacitance similar to that of the active pixels, which is connected to the polarisation voltage of the detector substrate. They are designed to electronically mimic a photosensitive pixel, therefore they are important for tracking polarisation and temperature changes during long exposures \citep{2010SPIE.7742E..1BM,Rauscher_2017}.

\begin{figure}
    \begin{center}
        \includegraphics[width=\linewidth]{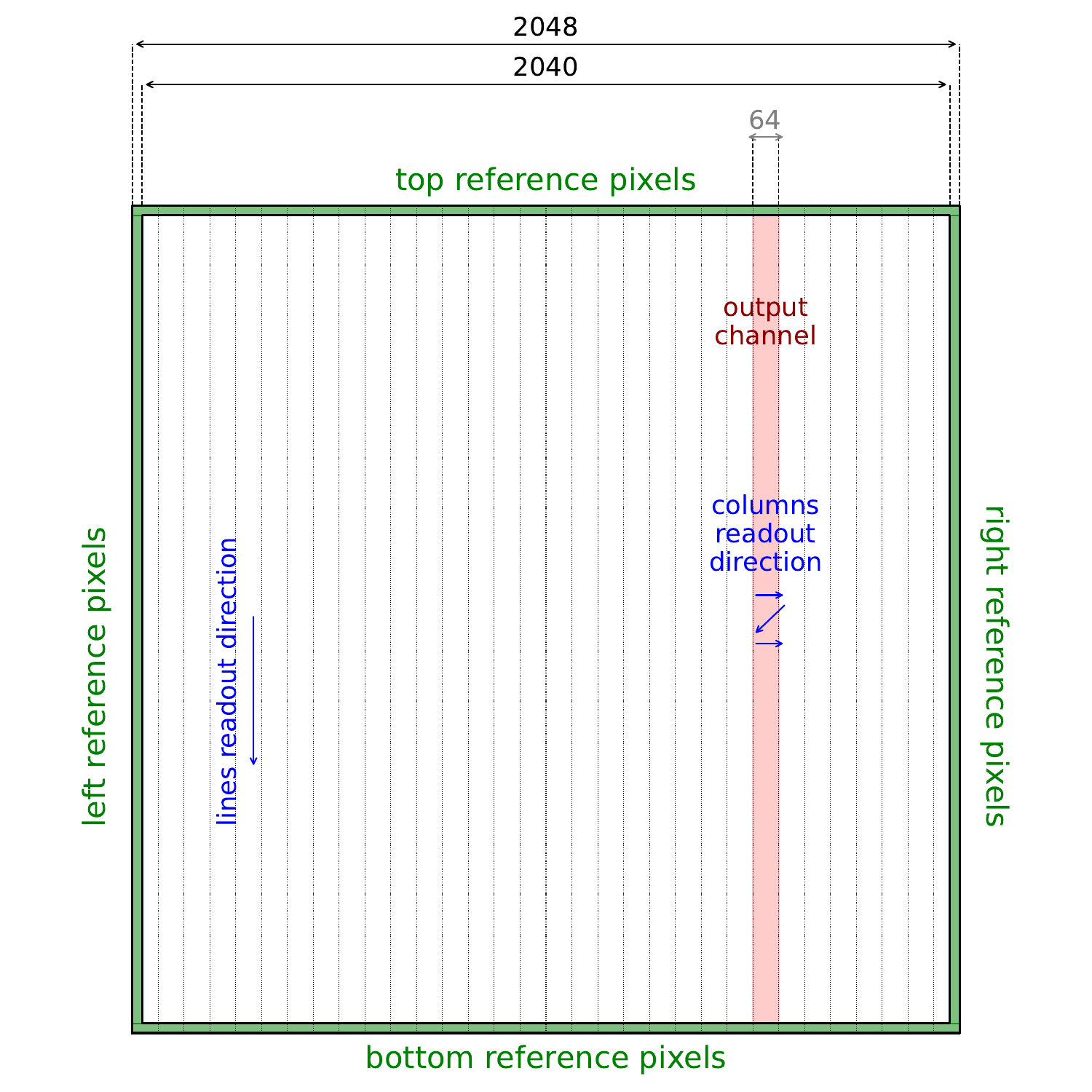}
        \caption{\label{fig:frame_geometry} H2RG frame geometry definition with reference pixels on the edges of the array in the 32-output channel mode. The read directions are indicated by the blue arrows. The widths are not in scale. }
    \end{center}
\end{figure}

Each of the detector channels has four rows of reference pixels at the top and at the bottom of the array as presented in Fig.~\ref{fig:frame_geometry}. Additionally the two outside channels include four columns of reference pixels at the outer edges of the array providing a reference for the output at the beginning and at the end of each 2040-pixel row (left and right references). 

Various possible corrections using reference pixels were defined and tested in \cite{KubikSPIE2014}. Possibilities included using only top and bottom references, using left and right references, or a combination of both. The impact of interpolation between top-to-bottom and left-to-right references and the use of sliding averages in the reading direction for side pixels was also examined. An indicator of the optimal correction was the minimum Correlated Double Sampling\footnote{CDS involves resetting the detector array and then reading it at least twice, generating two frames at an interval of at least one frame.} (CDS) noise level as it is the noise easiest to measure and most sensitive to common modes.

The analysis showed that the optimal correction was to subtract the average of top and bottom reference pixels per output channel to minimize the channel-to-channel noise variations and temporal fluctuations with periods of several frames.  The references on the left and right remove common modes on time scales of the readout of one line, suggesting the use of a sliding average centred on the selected line with 4 to 5~pixels on the sides. The effect of interpolation was found to be negligible when optimising noise in photometric and spectrometric exposures.  
For the ground characterisation and for the flight operations the optimal correction was used, corresponding to $c^{(ch)}_{3mn}(x, y)$ defined in Eq.~(2) in \cite{KubikSPIE2014}. We recall this correction below:

\begin{equation}\label{eq:refpixcorr} 
    c_{\rm{y}}^{\rm{(ch)}}(m) = \frac{1}{2}\left({\rm{T}}^{\rm{(ch)}} + {\rm{B}}^{\rm{(ch)}}\right)  + \frac{1}{2}\left( {\rm{L}}_{\rm{y}}(m) + {\rm{R}}_{\rm{y}}(m) \right) \, .
\end{equation}
This equation indicates that for a pixel located in column $x\in [0,64)$ and in line $y\in [4,2044)$ in the output channel $o\in [0,32)$, the correction value $c_{\rm{y}}^{\rm{(ch)}}(m)$ is calculated as the average of all top ${\rm{T}}^{\rm{(ch)}}$ and bottom ${\rm{B}}^{\rm{(ch)}}$ reference pixels in the same output channel $\rm{ch}$ and the average of the left ${\rm{L}}_{\rm{y}}(m)$ and right ${\rm{R}}_{\rm{y}}(m)$ reference pixels located in the window centred on line $y$ and of width $2m+1$.

The impact of the reference pixel correction on the noise performance is shown in Sect.~\ref{sec:noise}.

\subsection{Signal estimator}
\label{sec:llk}

One of the main aspects from the point of view of evaluating detector performance and defining methods for their calibration and correction is that the operation of NISP requires two different exposure times and acquisition modes in order to obtain the best SNR for the targeted scientific objects.
The limited daily bandwidth offered by the spacecraft requires sending only the slope calculated on-board from data points sampled up the ramp and the associated quality factor (QF) of the fit \citep{2016SPIE.9904E..5RB,2020SPIE11443E..59M}.

The signal estimator and QF computed on board is based on a likelihood estimator built on the group differences \citep{kubik2016}. 
The choice of this estimator was driven by two facts. First, it  is  a  more  efficient  estimator than  the  commonly  used  least square fit, i.e.\   its  variance  is  lower.  This translates directly to a higher SNR -- a vital parameter for the scientific outcome of the mission. Secondly, the QF can be computed at the same time as the signal without the need to reprocess the data.  This accelerates the computations and minimises the power consumption.  The QF allows to monitor the quality of the signal estimate and the linearity of the pixel response which otherwise would be impossible to track in the presence of non-destructive readout and signal fitted on-board.  



\section{Ground characterisation campaigns\label{sec:ground_charac}}
This section describes the objectives, workflow and tools developed for the characterisation campaign of the \Euclid infrared detectors. We provide a detailed description and motivation of the testing process and we describe the software developed for the data acquisition and verification in near real-time.

\subsection{Objectives}

The production of flight parts is a lengthy process that begins with acceptance tests and the associated requirements. The SCS triplets and their individual components (SCA and SCE) have undergone extensive testing throughout the production process. This includes acceptance testing and ranking at the Detector Characterisation Laboratory (DCL) at NASA/GSFC \citep{Waczynski2016,Bai2018}, SCE testing at NASA Jet Propulsion Lab \citep[JPL,][]{Holmes2022}, individual component tests of the SCAs at the CNRS-IN2P3 Center for Particle Physics in Marseille (CPPM), and thermal vacuum (TV) tests of the FPA in its final configuration at the CNRS-INSU Laboratory of Astrophysics in Marseille (LAM) facility. The ground characterisation was carried out in three stages:
\begin{enumerate}
\item  Acceptance tests at NASA allowed the selection of the best 20 detectors from a set of 60 that were available for \Euclid and provided overall average reference of the dark current, readout noise, and QE for subsequent detailed characterisation and performance measurements,
\item characterisation tests at CPPM provided per-pixel performance of individual detectors taken in a cross-validated and controlled environment, and  
\item tests of all detectors integrated onto the FPA of the NISP instrument.
\end{enumerate}
A complete overview of characterisation campaigns carried out by the NISP detector team can be found in \cite{barbier2018}.

The goals of the ground characterisation campaigns were to:
\begin{enumerate}
    \item Tune the operating parameters (such as the polarisation voltage, the gain, or the baseline) for all the detectors to optimise their performance,
    \item produce detailed pixel maps of detector performance for use by the science ground segment (SGS) as references for flight calibrations,
    \item produce readout chain correction functions with a relative accuracy of 1\%,
    \item estimate the accuracy of the in-flight calibration procedure of the readout chain through tests mimicking flight conditions,
    \item create models of how detector properties and performance are modified under varying environments experienced during flight, in order to monitor and quantify their impact on the detector chain error compared to optimal performance determined in ground tests, and 
    \item study the behaviour of the flight detectors in order to anticipate their evolution and the degradation of their performance during the mission.
\end{enumerate}
Handling the large volumes of data generated during these tests, which can reach up to 0.5\,TB per day per SCS, is crucial. Ensuring stability and reproducibility throughout the testing process is also essential. This requires a well-controlled and continuously monitored cryogenic environment, along with careful management of both the optical and the electrical equipments.

This paper focuses on the first and second goals and it presents the performance of flight detectors in their final configuration. Therefore most of the results presented in this paper are derived from the TV (thermal vacuum test) dataset. Only the QE (Sect. \ref{sec:qe}) and IPC (Sect. \ref{sec:ipc}) maps were derived from tests at DCL and CPPM tests respectively. 

The pixel effects reported in the sections below impact the signal measurements and introduce systematic errors. Based on these measurements, per-pixel correction functions can be derived with a requirement of 1\% accuracy on the relative response of the detector chain \citep{secroun2016,barbier2018}. The principal difficulty in the detector chain response calibration is to fulfil the correction at 1\% accuracy over the full dynamic range of the detector. This translates into a compliance to the test specifications ranging from very low dark signal and zodiacal background to the highest calibration fluence, knowing that most of the fluence (total integrated signal) of the sources of interest will lay in the range of a few thousands of electrons per pixel.

\subsection{Setup}
The Euclid Consortium used two facilities to carry out the detectors’ characterisation.
Firstly, the CPPM benches (Euclid Collaboration: Secroun et al. in prep) have been specifically designed to meet the needs of pixel-by-pixel characterisation of the NISP detectors’ performance, taking into account the 1\% accuracy objective. Significant effort has therefore been devoted to minimising and controlling the systematic errors introduced by the benches themselves.
In summary, two twin cryostats (to ensure redundancy) can each accommodate two detectors being read in parallel, using Markury electronics (see Sect. \ref{sec:SCStunning}). These cryostats consist of an outer stainless steel layer that maintains the vacuum, an internal copper layer coupled to a 90\,W cryocooler that ensures cooling within the cryostat and at the focal plane, and a second internal aluminium layer that simultaneously provides a deep dark environment and a very homogeneous flat field (Euclid Collaboration: Secroun et al. in prep). The detectors’ temperature can be controlled with a precision better than 10\,mK and a stability of 1\,mK. The flux exhibits homogeneity better than 1\% across the detector and a flux stability well below 1\% (Euclid Collaboration: Secroun et al. in prep).  These benches enable measurements of detector performance dependencies on temperature (typically between 70\,K and 120\,K), wavelength (across the entire sensitivity range of the detectors), and flux (from dark level to several times saturation).


Secondly, the ERIOS space simulation chamber at LAM \citep{Costille2016} has been designed to test full instruments like NISP. This chamber, measuring 4\,m in diameter and 6\,m in length, maintains a secondary vacuum ($\sim$\,10$^{-6}$\,mbar) and a cold ambient environment throughout its entire chamber (45\,m$^3$) at a temperature close to that of liquid nitrogen. To ensure mechanical stability, and hence accurate measurements, the instrument’s supporting table is connected to a 100tone concrete block under the floor and rests on pillars via spring boxes. The Thermal and Mechanical Ground Support Equipment (TMVS) developped for NISP TV provide thermal and mechanical interfaces of the instrument and simulate the PLM thermal environment with a stability of 4\,mK.

\subsection{TV test flow}

\begin{figure}
\centering
\includegraphics[width=\linewidth]{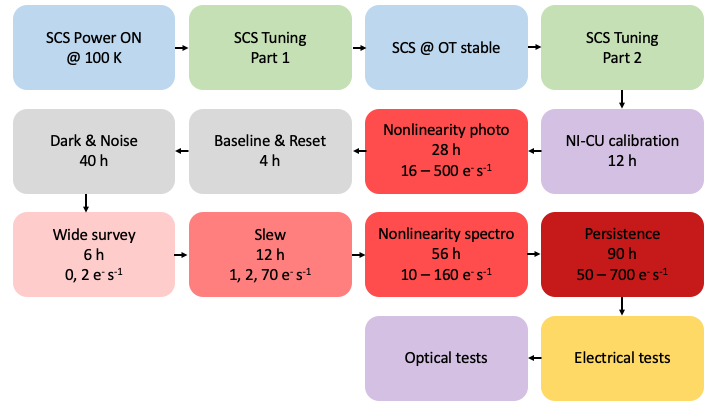}
\caption{\label{fig:TV1_workflow}The TV workflow as implemented and executed during the 14-day testing period. In the boxes we specify the type of test, the test duration in hours and the range of fluxes that were used. No indication of flux values means that the test was performed in dark conditions. All tests were performed with a 99.73\% time efficiency.} 
\end{figure}

The characterisation workflow, or test flow, is designed to produce high-quality data from the flight SCS within a time frame compatible with the NISP testing schedule. Data is collected continuously, 24 hours a day, over a 14-day period, with 3 days specifically allocated for tuning of the SCS system parameters. The definition of the test flow is the result of a trade-off between various constraints and requirements.
Key considerations while defining the test flow include
\begin{enumerate}
    \item Schedule management: Ensuring the schedule remains on track with built-in margins to accommodate potential setup failures.
    \item Thermal stability: Minimising the number of cooling cycles and large temperature variations across different tests.
    \item Mitigating long exposure issues: Avoiding long test runs that might be affected by spurious exposures.
    \item Latency mitigation: Controlling latency effects by ordering tests based on
increasing levels of illumination.
\end{enumerate}
Detector maps obtained during acceptance tests are used as cross-references. While these maps are not intended for direct verification or validation, they help ensure data coherence between different test facilities. 
The test flow, presented in Fig. \ref{fig:TV1_workflow}, includes the following runs,
\begin{enumerate}
    \item system temperature stabilisation,
    \item SCE register tuning,
    \item near-infrared calibration unit \citep[NI-CU;][]{EuclidSkyNISPCU} calibration and validation (stability and accuracy),
    \item baseline and reset noise measurements,
    \item dark measurements (includes the measurement of single frame readout noise and the slope noise),
    \item nonlinearity measurements, 
    \item latency measurements, and
    \item additional optical and electrical tests.
\end{enumerate}

\subsection{Data acquisition software}

The measurement of any detector characteristic is defined as a coherent entity consisting of pre-programmed acquisition cycles and environmental configurations. The number of cycles is typically determined by the required measurement accuracy, and their order is based on the physics of the detector and the specific parameters being measured. These measurements must be performed according to a predefined scenario that ensures the stability of both the environmental conditions.
This is crucial for meeting the statistical requirements necessary for the characterisation of individual pixels and mitigate contamination from persistence signal between exposures.

\begin{figure}
\centering
\includegraphics[width=\linewidth]{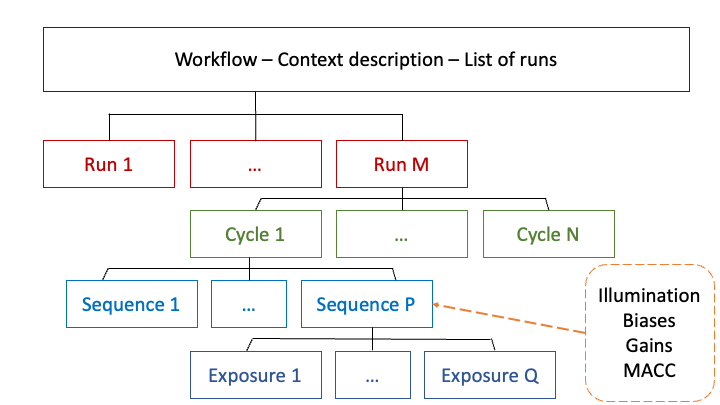}
\caption{\label{fig:das_dm}Hierarchical data structure defined for the SCS characterisation campaign.} 
\end{figure}

The runs specified in the test flow are executed by the data acquisition software (DAS). The scheme of the runs executed automatically according to the test plan is shown in Fig.~\ref{fig:das_dm}.

The building block of each run is an exposure. It consists of a ramp of non-destructive reads and the accompanying environmental settings, including SCS (gains, bias voltages) and NI-CU settings (e.g., LED selection and its current and duty cycle). 


The CPU time is used to synchronise the environment and illumination settings with the SCS readouts with a precision better than 0.5 seconds. 
The lighting setup is controlled by the DAS, which triggers key events such as switching the LEDs on and off. 

For ground characterisation, up-the-ramp acquisitions UTR($n$) were used with the number of frames $n$ equivalent to the specified MACC (Table~\ref{tab:modes}). The frames corresponding to the MACC readout were selected, if needed, during the post-acquisition analysis of data. 

The overall workflow is managed by a Super Scheduler, i.e.\ pseudo-code of specifications that generates a scenario of the test flow covering the 14 days of TV tests. The scheduler supports asynchronous tasks, tasks of different types can be executed in parallel \citep{Williams2019}, one of the most challenging aspects of implementation: there may be a few hundred milliseconds of inaccuracy in time-stamping at the CPU level, depending on the operating system used for acquisition. 

Thanks to the automated data acquisition system, the tests were conducted continuously, 24 hours a day, achieving a time efficiency of 99.73\%. This indicates that less than 0.3\% of the total time was lost due to failures.

\subsection{Data quality quick check}

The implemented automated quality check enables efficient and straightforward monitoring of data quality during the data acquisition process. It allows for almost real-time control and evaluation of system performance through comparison of CDS signal and noise levels against validated references. Statistical analysis carried out at the level of individual lines, columns, channels, or frames, enables the observation of both temporal and spatial characteristics of the data. Reactivity, including the management of warnings/alerts, is crucial as the system needs to validate the run in less than one hour after a two-day run period. This highlights the need for clear figures of merit or key performance indicators to guide the evaluation of quality check data efficiently.


\section{Detector configuration settings\label{sec:tuning}}
The performance of pixel sensors involves a delicate trade-off between noise characteristics and the accessible dynamic range. For photometric applications, a minimum dynamic range of 60\,ke$^-$ is required and there is a requirement on the maximum acceptable readout noise value of 13 and 9 electrons in photometric and spectroscopic readout modes respectively (see also Sect. \ref{sec:noise}). 

The accessible dynamic range of the signal is primarily determined by the gain of the SCE and is constrained by several factors, including the nonlinearity of the SCE at both the upper and lower limits, the inhomogeneity of the baseline and the minimum pixel full well capacity which is defined by the applied polarisation voltage. On the other hand, both the position of the baseline, the polarisation and the SCE gain impact the detector noise performance.

\subsection{Baseline settings\label{sec:baseline_tunning}}
First, the level of the baseline in the 65\,535 counts range of the 16\,bits ADC is of course a direct driver of the maximal science signal achievable in both acquisition modes. The limiting factor to the lowering of the baseline is the low-end of the differential non linearity (DNL) of the ADC, which is on the order of 1000 ADU. If the baseline is set at the lowest possible level not exceeding the lower DNL threshold, good stability is obtained for dark current and for weak signals, but with readout noise slightly higher than in the middle of the ADC's dynamic range. We therefore set the baseline 5000 ADU higher than the lower DNL limit allows, in order to minimize the readout noise while maintaining sufficient dynamic range for scientific signals. 

\subsection{Detector polarisation voltage settings}
The detector polarisation voltage, which is the difference in voltage between the backside substrate and the diode reset voltage, was set to 500\,mV. This value was found to be a good compromise between noise performance, the dark current value and full well capacity. The compromise was based on the following considerations: firstly, setting the highest possible polarisation, while keeping the 95th percentile of total noise below specification; secondly, achieving a maximum integrated signal of 60\,ke$^-$ in photometric acquisition mode; and thirdly, shifting the full well capacity (130\,ke$^-$) above the ADC saturation (65535\,ADU) to minimise diode nonlinearity.

\subsection{SCE gain settings\label{sec:gain_tunning}}
Various preamplifier gains\footnote{The preamplifier provides control over the gain and bandwidth of the analogue signal using programmable capacitors and resistors. Preamplifier gain is not the same quantity as a the conversion gain, decribed in Sect. \ref{sec:convgain}.} were tested and discussed during the Non-Recurring Engineering (NRE) phase\footnote{This was a phase that lasted from 2012 to 2015. During this phase, pre-development versions of SCS parts were designed, manufactured, qualified, and evaluated for performance. The phase ended with approval for production of the flight parts.} and acceptance tests. Advantages and disadvantages were raised for three possible preamplifier gain values, 15\,dB, 18\,dB and 21\,dB, consistent with the target SNR. The final choice was to use the lowest SCE gain of 15 dB to improve the dynamic range from 60\,ke$^-$ to 115\,ke$^-$. This mitigates the nonlinear effects close to pixel full well and increases the range of the measurable signal, which is very important when calculating the persistence contribution for subsequent exposures. 

\subsection{Tuning of flight SCS parameters\label{sec:SCStunning}}
To speed up the tuning process of the 16 SCSs and of the SIDECAR ASICs, the data acquisition is done in parallel, but some parameters are adjusted one by one within a certain predefined range, already optimised during the acceptance tests carried out by NASA with the expertise and supervision of Markury Scientific\footnote{\url{https://www.markury-scientific.com}}. At the end, after optimising the different functions, each SCS has its own set of parameters. The tuning was carried out in the following steps.

Firstly, the dynamic response of the pixels was optimised by analysing the unit cell current controlled by the ROIC voltage settings. The minimised value in this case is the residual signal that can be observed on the first pixel adjacent to the stimulated pixel with respect to the reading direction, if the cell current is too weak to drive the signal within a time slot of one clock (10\,$\muup$s). Obviously, this current must be kept as low as possible to reduce the power consumption of the ROIC during the readout sequence.

Secondly, the CDS noise of the groups taken from the MACC(4,16,4) photometric acquisition mode, which represents the most demanding scenario in terms of noise performance, was minimised. The values of median noise in each of the 32 channels and the value of the noisiest channel were used as the references for this process. 

Thirdly, baseline adjustment was performed through the reference voltage of the capacitive trans-impedance preamplifier before the 15\,dB gain setting. For each detector, a maximum value was selected from the 32 DNL limits measured during the acceptance test for each output channel and used as the lowest threshold for the median baseline of the reference pixels. The baseline setting is driven by the reference pixels, because they exhibit lower baseline values compared to the photosensitive pixels and too low baseline of reference pixels could introduce nonlinearites in the signal during reference pixel correction. A closed-loop algorithm was developed to rapidly converge to the target baseline values by adjusting the bias for the 16 SCS in parallel.
At the end of the tuning process, the average dynamic range of each SCS is around 115\,ke$^-$.

\section{Pre-launch SCS properties and performance\label{sec:properties}}
This section is the core section of the paper. Here we present the main properties of the \Euclid near-infrared detectors as measured during the ground characterisation campaign. For each property, we describe its physical background and its role in the overall performance of the detector system. We then outline the measurement protocols used to evaluate each property, presenting the steps taken to ensure accuracy and consistency. Finally, we present the measurement results, highlighting their potential implications for detector performance.

In the sections below, we identify the detectors either by their position in the focal plane (e.g., DET 11) or by their SCA serial number (a five-digit identifier in the form 18***). The table below provides the explicit correspondence between these two naming conventions:

\begin{center}
\begin{tabular}{|l|c|c|c|c|}
\hline
FPA position & 41    & 42    & 43    & 44    \\
SCA number   & 18458 & 18249 & 18221 & 18628 \\
\hline
FPA position & 31    & 32    & 33    & 34    \\
SCA number   & 18280 & 18284 & 18278 & 18269 \\
\hline
FPA position & 21    & 22    & 23    & 24 \\
SCA number   & 18268 & 18285 & 18548 & 18452 \\
\hline
FPA position & 11    & 12    & 13    & 14 \\
SCA number   &  18453 & 18272 & 18632 & 18267 \\
\hline
\end{tabular}
\end{center}

\subsection{Disconnected pixels\label{sec:disconnected}}
Pixels with missing or not fully connected indium bumps between the P-diode and the metal pad of the ROIC occur infrequently and are an issue in the detector system that needs to be characterised. These pixels are permanently inoperable. To efficiently discriminate between disconnected and operational pixels, a strategy was adopted to examine the pixel output response to substrate bias ($D_{\rm sub}$) at room temperature.

Connectivity tests were performed before the entire setup was cooled down. Two sequences of 64 ramps were acquired in UTR(1) with $D_{\rm sub}$ set to \SI{500}{mV} and \SI{550}{mV}, respectively. The estimator used to identify the disconnected pixels is the difference between the pixel response measurements $b_1$ and $b_2$ at two different polarisation values (changing $D_{\rm sub}$) of the substrate. Specifically, the estimator is calculated as \[d = \frac{b_2-b_1}{\langle b_2-b_1 \rangle}-1\,,\] where $\langle b_2-b_1 \rangle$ is the spatial mean of the difference of two images $b_1$ and $b_2$. For operating pixels this value is close to 0, whereas for disconnected pixels it is $-1$. In practice, a threshold of $-0.7$ was set to discriminate between connected and disconnected pixels. Pixels with estimator values ranging from $-0.7$ to $-0.2$ were considered potentially disconnected and were recorded for further investigation. 

Depending on the SCA, the number of disconnected pixels ranges from a few hundred to a few thousand. The precise numbers for each detector are reported in Table~\ref{tab:PFA_main_properties}. The results show that disconnected pixels are randomly distributed, with no noticeable clusters. Reference pixels formally have a disconnectivity estimator $d$ close to $-1$ due to their capacitive behaviour, but should not be considered inoperable. The temporal stability of disconnected pixels was tested in three independent tests, at DCL, at CPPM and at LAM, in part with different bias voltages. The two populations of pixels, connected and disconnected, were shown to be stable over time and thermal cycling.

\subsection{Baseline and dynamical range\label{sec:baseline}}
Since in-flight acquisitions are based on MACC ramps, the baseline is defined as the average value of the first 16 frames after a pixel reset. It represents the pedestal value for sampling up the ramp. The tuning of the baseline value, described in Sect.~\ref{sec:tuning}, is critical since it defines the maximal signal that can be detected and it prevents entry into the nonlinear zone of ADC if the baseline setting is correct.

\begin{figure}
\centering
\includegraphics[width=\linewidth]{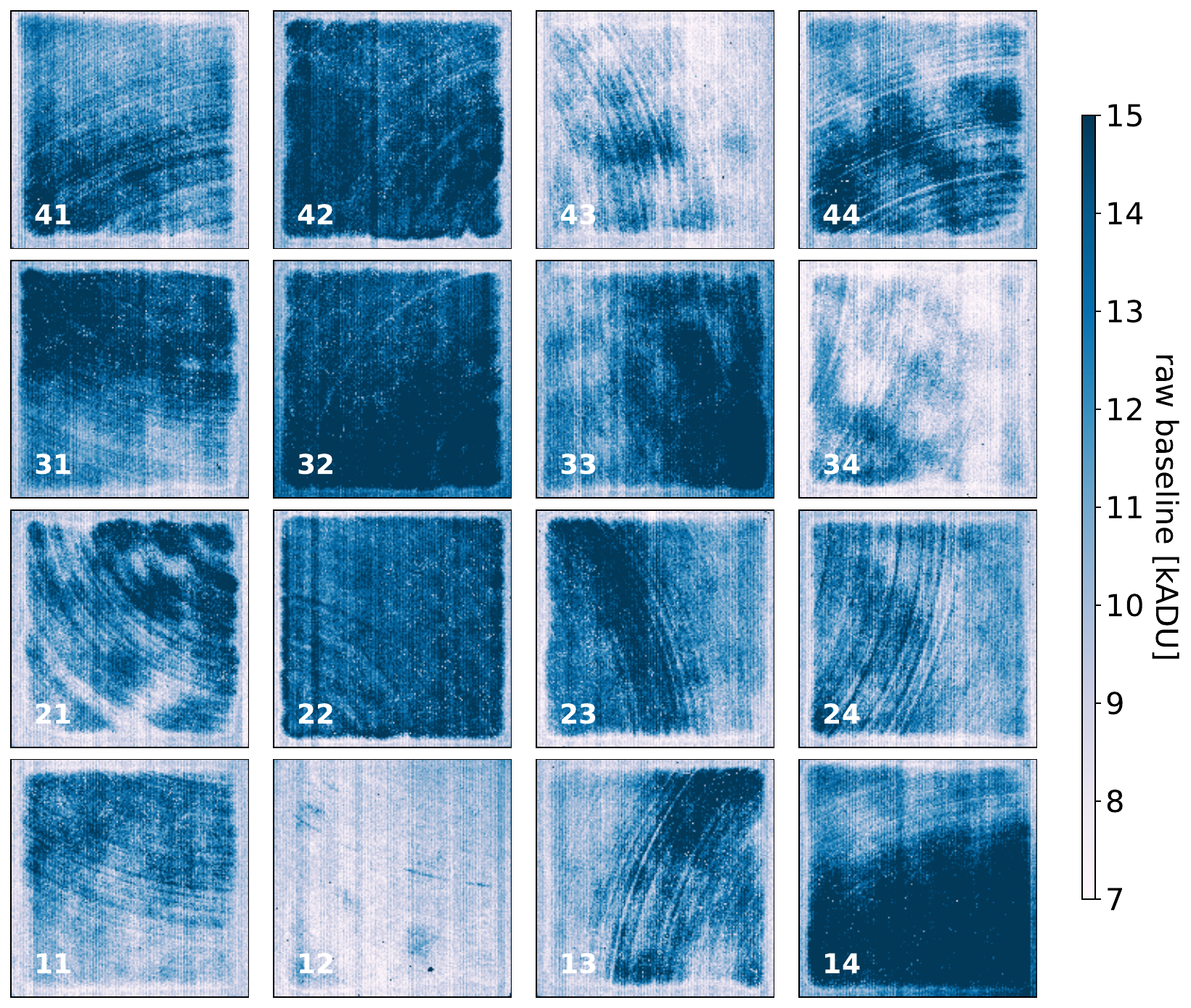}
\caption{\label{fig:baseline-img}Baseline image of the whole FPA before reference pixel correction. This and all subsequent detector maps are displayed in the R-MOSAIC coordinate system \citep{EuclidSkyNISP}.}
\end{figure}

\begin{figure}
\centering
\includegraphics[width=\linewidth]{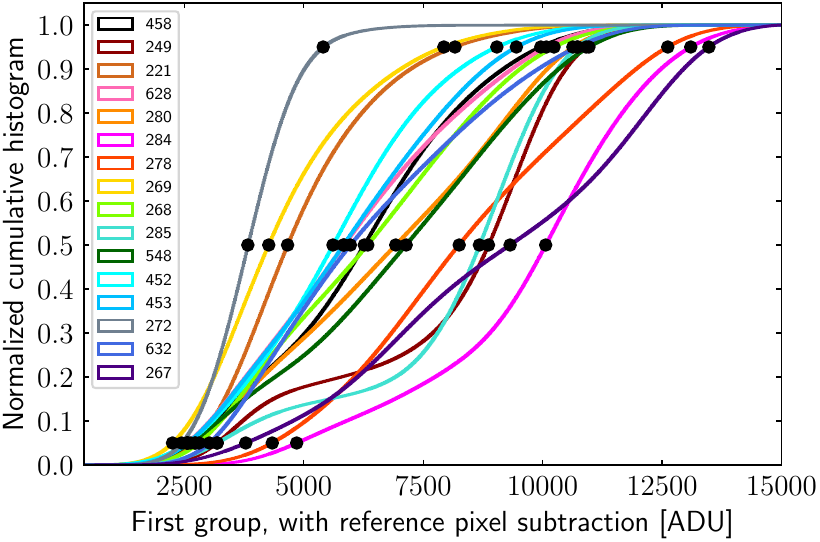}
\caption{\label{fig:baseline-dist}Distribution of the baseline values after reference pixel correction. Black dots represent the 5th, 50th and 95th percentile of the science pixels baseline distribution.}
\end{figure}

The `master' baseline B is measured as the average baseline over 500 ramps of 16 frames. For \Euclid H2RG detectors, the baseline typically ranges from 7000 to \SI{15000}{ADU} before reference pixel subtraction, as shown in the baseline image in Fig.~\ref{fig:baseline-img}, and is reduced to 4000 to \SI{10000}{ADU} after reference pixel correction as indicated in Fig.~\ref{fig:baseline-dist}. Table~\ref{tab:PFA_main_properties} reports the median baseline values of science pixels after reference pixel correction. Additionally, the median baseline values of reference pixels are provided for each detector.

For photosensitive pixels, the difference between the 5th and 95th percentiles (spatially over an array) of the baseline values is about 7000~ADU and this spread remains unchanged by the reference pixel correction. The shape of the distribution of baseline values per detector deviates significantly from a typical Gaussian function, which may be related to the manufacturing process.

The baseline of the reference pixels differs significantly from that of the photosensitive pixels; it is lower and more uniform than the one of photosensitive pixels. For reference pixels, the per detector median baseline values, with the settings described in Sect.~\ref{sec:tuning}, range from about 5000 to 6000\,ADU. The difference between the 5th and 95th percentiles is slightly less than 3000\,ADU. 

The value of the baseline is directly related to the dynamic range of the pixels of 115\,ke$^-$. This corresponds to a flux of \SI{1000}{e^{-}.s^{-1}} in MACC(4,16,4) and \SI{200}{e^{-}.s^{-1}} in MACC(15,16,11). The spatial spread of the baseline values also represents the spatial spread of the pixel dynamic range. It means that not all pixels will be able to observe an equally strong signal, some will saturate earlier or, on the other hand, the same signal falling on pixels with different baseline values will be subjected to different sources of nonlinear behaviour.

Photosensitive pixels with a baseline significantly higher than \SI{60}{kADU} do not have sufficient dynamic range for scientific applications and should be considered as unusable. Pixels with baseline values lower or higher than the acceptable DNL range shall be also be flagged as unusable. These pixels are included in the bad pixel budget described in Sect. \ref{sec:bad_pixels}.

\subsection{Bad pixels\label{sec:bad_pixels}}
\begin{figure}
\centering
\includegraphics[width=\linewidth]{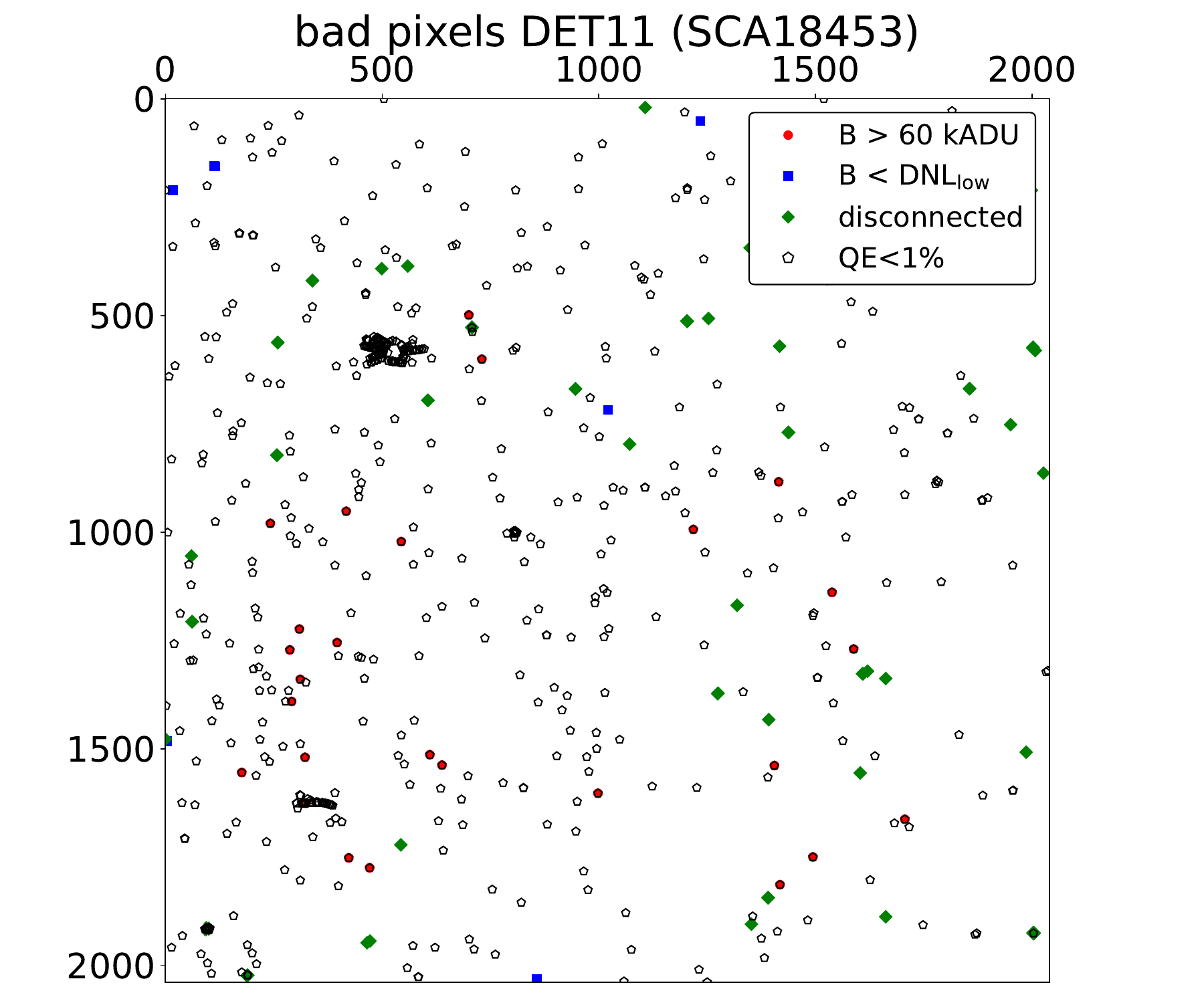}
\caption{\label{fig:bad}Example of the spatial distribution of the bad pixels across an array.}
\end{figure}

The survey efficiency requirement translates into a 95th percentile requirement of operational pixels for science.
Pixels can be inoperable permanently or for some amount of time, for example if they are saturated by an energetic particle hit and the signal cannot be recovered. Pixels are considered permanently inoperable or `bad' if they meet any of the following criteria: 
\begin{enumerate} 
    \item Pixels that are disconnected from the circuit are non-functional for obvious reasons.
    \item Pixels that have the QE of less than 1\% are non-functional due to lack of light response.
    \item Pixels whose baseline values are outside the acceptable DNL limits ($B<\text{DNL}_\text{low}$) are non-functional as they may exhibit nonlinear behaviour that is difficult to correct.
    \item Pixels with $B> 60\,\text{kADU}$ are non-functional because they lack sufficient dynamic range for scientific purposes.
\end{enumerate}
The budget of the inoperable pixels in each of the above-mentioned categories, as well as their combined total, excluding double-counting, is shown in Table~\ref{tab:bad_pix}. The number of non-operational pixels does not exceed 0.2\% per detector which is far below the requirement. The spatial distribution of bad pixels is typically sparse across the detector, as illustrated in Fig.~\ref{fig:bad}. Clustering is observed only in pixels with low QE.

\begin{table*}
\caption{Bad pixels budget per detector as measures during ground tests. In the last raw the combined total, excluding double-counting, is given.} 
\label{tab:bad_pix}
\begin{center}       
\begin{tabular}{@{}l|rrrrrrrrrrrrrrrr@{}}
\toprule
\toprule
FPA position & 11 & 12 & 13 & 14 & 21 & 22 & 23 & 24 & 31& 32 & 33 & 34 & 41 & 42 & 43 & 44 \\
SCA 18*** & 453 & 272 & 632 & 267
    & 268 & 285 & 548 & 452
    & 280 & 284 & 278 & 269
    & 458 & 249 & 221 & 628 \\
\midrule
QE<1\%
&  630 & 3913 &  792 & 1640
&  969 & 3024 &  869 &  572 
& 1431 &  635 & 1058 & 1952 
& 1712 &  656 &  945 & 7181 \\
disconnected
&  135 & 3223 &  589 & 252
&  313 & 2580 & 1512 & 146 
& 1420 &  269 &  262 & 970
&  198 &  334 &  222 & 255 \\
$B<\text{DNL}_\text{low}$
& 13 &   13 & 15 &   18
& 21 & 2040 & 16 &    4 
&  7 &   11 &  1 &   61 
& 10 &    7 &  6 & 1077 \\
$B> 60\,\text{kADU}$
&  26 &  10 & 70 & 178
& 136 & 108 & 77 &  14 
&  52 &  41 & 97 &  11 
&  49 & 107 & 10 &  34 \\
\midrule
Total
&  764 & 5526 & 1266 & 1906
& 1295 & 3525 & 2157 &  694 
& 2066 &  890 & 1289 & 2117 
& 1864 &  927 & 1128 & 8483 \\
\bottomrule
\end{tabular}
\end{center}
\end{table*}

\subsection{Inter pixel capacitance\label{sec:ipc}}
In most infrared pixel detectors, which are based on CMOS hybrid readout technology, there is a phenomenon known as electrostatic cross-talk between pixels. This is because the fields coming from the edges of the node capacitors in neighbouring pixels affect the voltage readings in the central pixel. This results in a dependence of the signal in the central pixel on the charges in adjacent pixels. This effect is typically modelled by introducing a coupling capacitance between pixels, effectively connecting every pixel to its neighbours. Naturally, the inter pixel capacitance (IPC) is becoming more important in near-infrared detectors as pixel size decreases \citep{Moore2004,Moore2006,Fox:2009mk}. It is crucial to understand that this IPC is not the same as charge diffusion. The latter involves the physical movement of charge carriers between neighbouring pixels before charge collection and is a stochastic process. Poisson noise resulting from charge diffusion in neighbouring pixels is uncorrelated. IPC cross-talk occurs after charge collection and is deterministic resulting in correlated Poisson noise in neighbouring pixels.

The presence of IPC needs to be accounted for in applications like photometry or astrometry, as it blurs the PSF, modifying both the size and shape of the sources. Moreover, as it correlates the Poisson noise of the signal in adjacent pixels, 
 it can lead to an overestimation of the gain (ADU/e$^-$) and consequently to an overestimation of the QE while lowering the effective SNR in each pixel \citep{Moore2006,Fox:2009mk}.

Furthermore, pixels frequently display a nonlinear response, whereby the pixel capacitance depends on the charge. This deviation from the nominal capacitance value is typically modelled separately, leaving the IPC to be treated as a linear effect. However, accounting for IPC is essential, particularly when calculating the conversion gain of the system \citep{2018SPIE10709E..21S,legraet2022,Graet:2024ytq}.

\begin{figure}
\centering
\includegraphics[width=\linewidth]{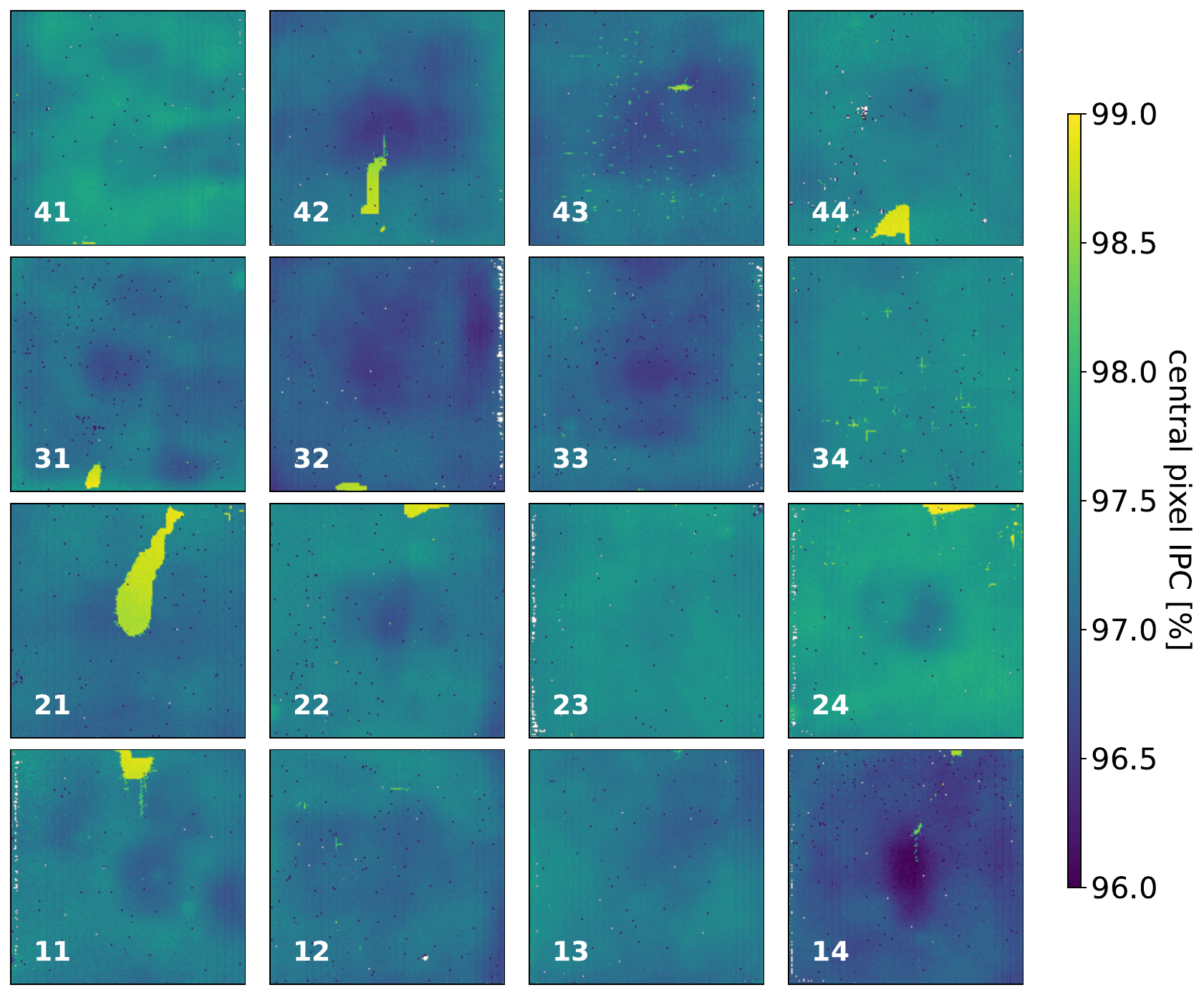}
\caption{\label{fig:ipc_map}Map of the central pixel value  of the inter-pixel capacitance kernel. Void regions are visible as yellow shapes.}
\end{figure}

The IPC for the \Euclid SCAs was measured using the single pixel reset (SPR) technique enabled by the guide mode\footnote{H2RG can interleave the readout of the guiding window with the readout of the entire array, so that the
science array can perform the guiding function required to accurately stabilize the telescope \citep{BeleticSPIE2008}.} of H2RGs. SPR enables the direct characterisation of IPC, eliminating the necessity for an illumination source. This SPR characterization mode is incorporated into the command structure of the Euclid firmware and can be used as part of on orbit calibration. This method is useful for isolating IPC, since the charge is not generated in the photosensitive material, and is therefore not susceptible to the effects of charge diffusion \citep{FingerSPIE2006,2012ApOpt..51.2877D}. In SPR, following the setting of all pixels in the SCA to a single voltage and the initial readout, a grid of widely spaced single pixels is reset to a second voltage level. Following this reset, all pixels are again read out. The difference of these two images will reveal any IPC as a signal in pixels adjacent to the reseted pixels. The 3\,$\times$\,3\,pixel IPC kernels were obtained for all SCA through SPR testing. These couplings are uniform across the detector arrays with the observed dispersion of order of 1\% if the  specific regions called `voids' are excluded. The 'void' regions, presumably with a missing epoxy layer between the HgCdTe substrate and the ROIC, exhibit a significant deviation from the average value (yellow regions in Fig.~\ref{fig:ipc_map}) and affect the uniformity of the detector response but the flat field correction can remove this effect. 

\begin{figure}
\centering
\includegraphics[width=\linewidth]{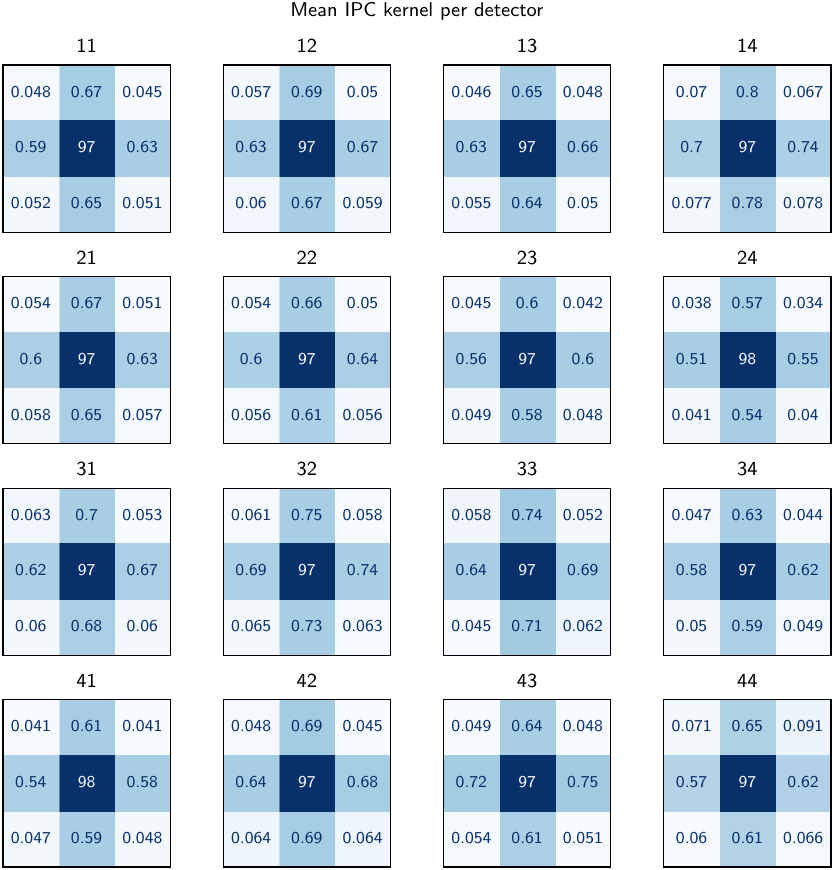}
\caption{\label{fig:ipc_mean}Mean IPC kernel per detector with sub-percent accuracy.}
\end{figure}

To achieve an accuracy below 1\% for the central pixel, we provide in Fig.~\ref{fig:ipc_mean} the average IPC value per detector. The parasitic coupling introduced by IPC contributes to less than 3\% of the pixel's nodal capacitance and produces less than 1\% of cross-talk between nearest neighbours. The median values of the IPC kernel are reported in Table~\ref{tab:PFA_main_properties}.

\subsection{Quantum efficiency\label{sec:qe}}
QE is a measure of the fraction of photons incident on a detector that are converted into electrons and detected as a signal. A high QE is therefore essential for a sensitive and efficient detection system. Measuring QE generally involves comparing the signal recorded by a characterised sensor with the signal recorded by an independent calibrated signal sensor (such as a photodiode). The measurement is also sensitive to the efficiency of charge collection in the detector and to the effective charge-to-voltage conversion. By its nature, QE depends on the incident photons' wavelength.

\begin{figure}
\centering
\includegraphics[width=\linewidth]{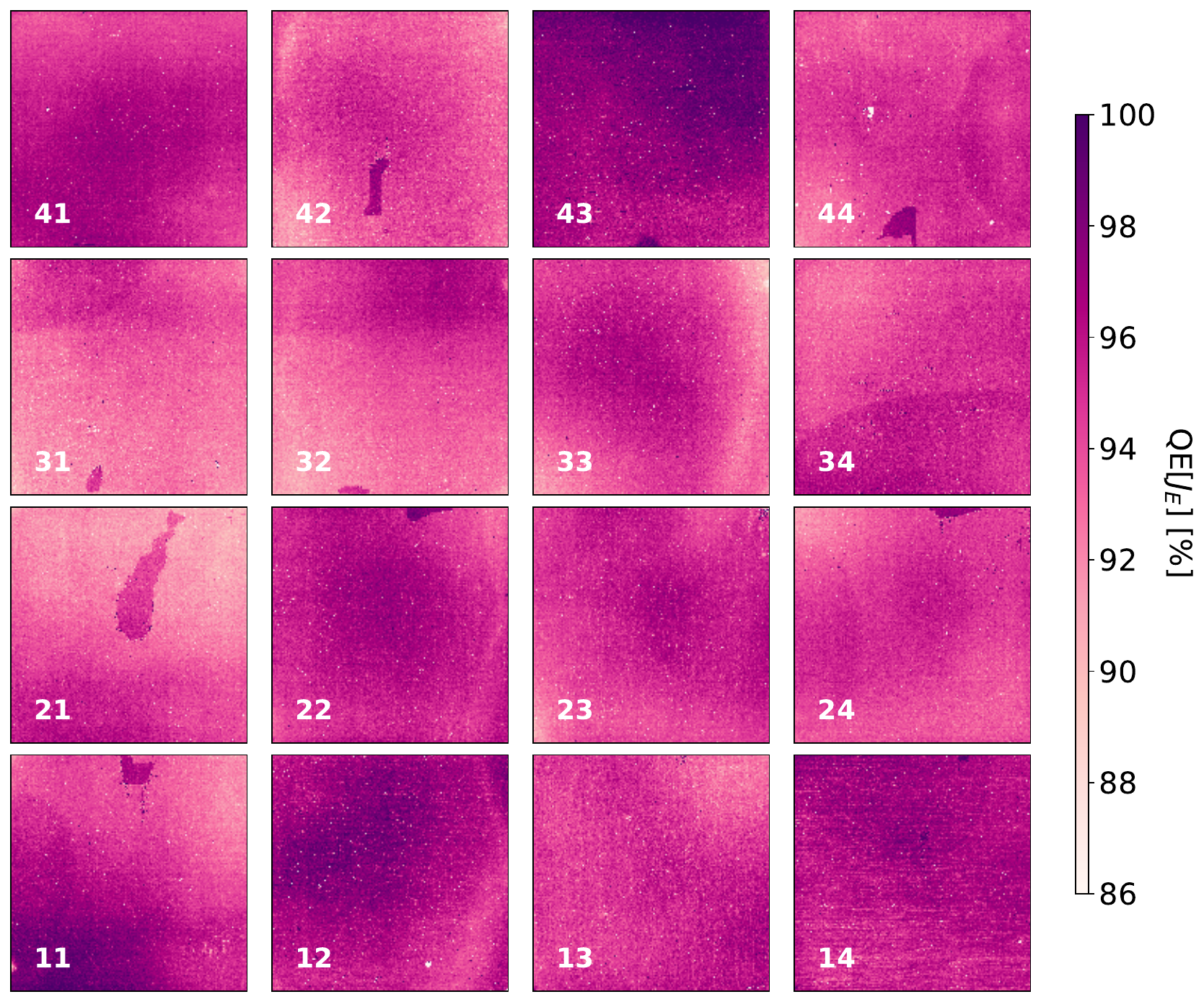}
\caption{\label{fig:qe_map}Map of the average QE in the NISP \JE  band (1168--1567\,nm).}
\end{figure}

\begin{figure}
\centering
\includegraphics[width=\linewidth]{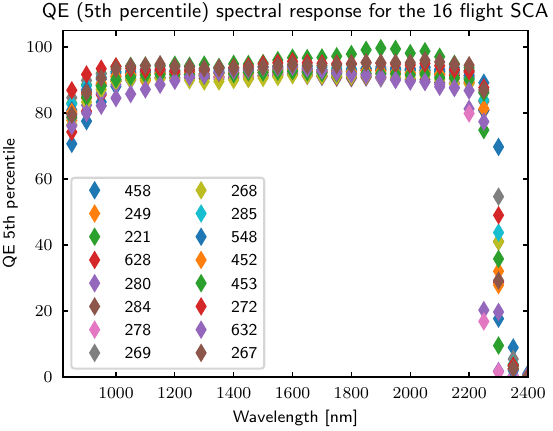}
\caption{\label{fig:qe_16scs_5perc} QE as function of wavelength. The points indicate the lowest 5th percentile per detector. 95\% of pixels have QE values higher than 80\% over most of the wavelength range.}
\end{figure}

For the NISP detectors, QE data were obtained at DCL during acceptance tests. The DCL data set includes pixel QE maps measured at 40 different wavelengths ranging from 0.6 to 2.6\,\micron, with a resolution of 50\,nm steps and an absolute accuracy of the QE measurements of 5\%. 
A map of the mean QE in the NISP \JE band, calculated as the average QE from eight measurements in the wavelength range of (1168--1567\,nm), is shown in Fig. \ref{fig:qe_map}. 
The reported median QE values are above 90\% and the 95\% of pixles have QE above 80\% across the entire spectral range of \Euclid as shown in Fig.~\ref{fig:qe_16scs_5perc}. Additionally we provide the explicit values for the 5th, 50th, and 95th percentiles of the average QE in the NISP photometric bands \YE, \JE and \HE. in Table \ref{tab:qe}. For the QE measurements, an independent detector-averaged gain was used, which may account for QE values exceeding 100\%. Additional factors affecting the measurement precision include the relative accuracy of the photodiode calibration, the uniformity of the illumination, and the exact definition of the pixel area.

\begin{table*}[ht!]
\caption{5th, 50th, and 95th percentiles of the average QE in the NISP photometric bands \YE, \JE and \HE.} 
\label{tab:qe}
\begin{center}       
\begin{tabular}{@{}l|rrrrrrrrrrrrrrrr@{}}
\toprule
\toprule
FPA position & 11 & 12 & 13 & 14 & 21 & 22 & 23 & 24 & 31& 32 & 33 & 34 & 41 & 42 & 43 & 44 \\
SCA 18*** & 453 & 272 & 632 & 267
    & 268 & 285 & 548 & 452
    & 280 & 284 & 278 & 269
    & 458 & 249 & 221 & 628 \\
\midrule
QE[\YE]-5\%    & 90 & 93 & 86 & 93 & 89 & 92 & 92 & 91 & 89 & 90 & 91 & 92 & 91 & 91 & 94 & 90 \\
QE[\YE]-50\%   & 93 & 96 & 88 & 96 & 92 & 94 & 94 & 93 & 91 & 92 & 94 & 94 & 93 & 93 & 96 & 92 \\
QE[\YE]-95\%   & 97 & 98 & 90 & 97 & 96 & 96 & 96 & 94 & 94 & 95 & 96 & 96 & 94 & 95 & 99 & 94 \\
\midrule
QE[\JE]-5\%    & 92 & 94 & 92 & 94 & 90 & 93 & 92 & 92 & 91 & 91 & 91 & 92 & 93 & 91 & 95 & 92 \\
QE[\JE]-50\%   & 95 & 96 & 94 & 96 & 92 & 95 & 95 & 94 & 92 & 93 & 94 & 94 & 95 & 93 & 97 & 94 \\
QE[\JE]-95\%   & 98 & 98 & 96 & 97 & 95 & 97 & 96 & 95 & 95 & 96 & 96 & 96 & 97 & 95 & 99 & 96 \\
\midrule
QE[\HE]-5\%    & 92 & 95 & 92 & 95 & 91 & 94 & 94 & 93 & 91 & 91 & 92 & 94 & 93 & 92 & 98 & 93 \\
QE[\HE]-50\%   & 95 & 98 & 95 & 97 & 94 & 96 & 96 & 95 & 93 & 94 & 95 & 96 & 95 & 94 & 100 & 95 \\
QE[\HE]-95\%   & 99 & 100 & 96 & 99 & 97 & 98 & 97 & 96 & 96 & 97 & 97 & 98 & 97 & 96 & 103 & 97\\
\bottomrule
\end{tabular}
\end{center}
\end{table*}

\subsection{Conversion gain\label{sec:convgain}}
The conversion gain is essential for measuring a variety of detector properties, including readout noise, dark current, and QE. Since it defines the relationship between digitized counts from the ADC and the photogenerated electrons detected by the system, it allows for accurate calculations of detector performance parameters in physical units. 

The gain is generally viewed as a combination of three distinct processes: (1) Charge-to-voltage conversion from electrons to volts in the photodiode, (2) voltage amplification and buffering in the ROIC, and (3) conversion from volts to ADU in the external electronics. A detailed description of these transfer functions can be found in \citet{barbier2018}.
A common method for determining the conversion gain is the mean-variance method, also known as the photon transfer curve (PTC) described in \citet{JanesicPTC}. In this approach, a series of exposures with varying fluences are taken under constant flux to generate data. The signal variance is plotted against the mean signal of each exposure, and a linear fit is applied to this data. The inverse slope of this line represents the total conversion gain from ADU to electrons. 

\begin{figure*}
\centering
\includegraphics[width=\linewidth]{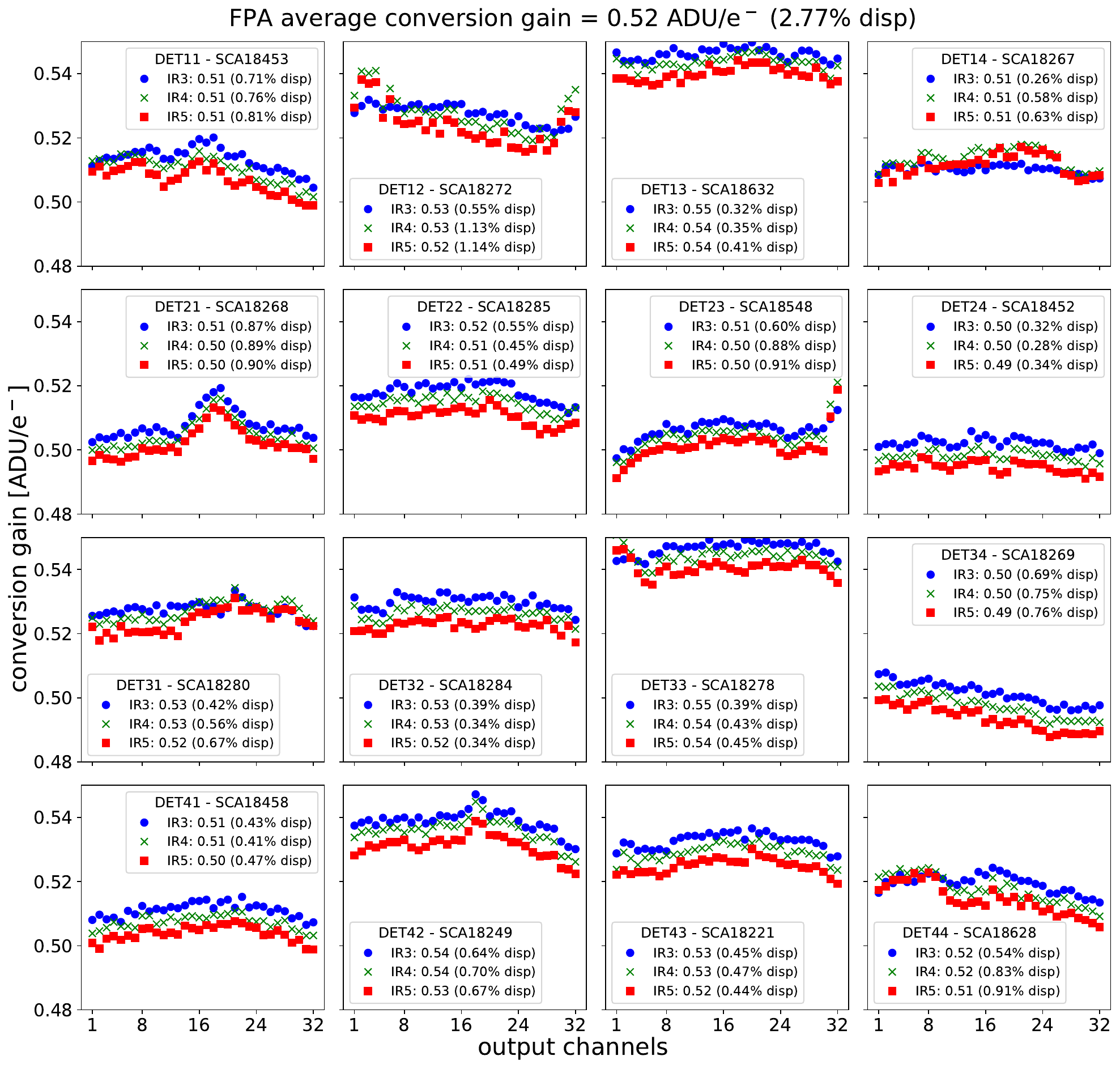}
\caption{\label{fig:cg}Conversion gain per channel computed using flat field ramps with increasing (IR3 $<$ IR4 $<$ IR5) illumination levels. In the legends for each detector the average gain over the 32 channels and the dispersion between channels are given. For all the three illuminations the mean conversion gain for the whole FPA is 0.52\,ADU/e$^-$ with a 3\% variation between arrays. The channel-to-channel gain variation inside a detector is less than 1\%. The gain variation with signal is described in \cite{Graet:2024ytq}.}
\end{figure*}

The mean-variance curves were constructed using 15 flat-field acquisitions, where the variance and mean signal were computed for each pixel across 15 ramps taken under the same flux conditions. Next, the average spatial variance was calculated for each of the 32 readout channels of $64\times 2040$\,pixels after baseline subtraction, which removes fixed pattern noise and excluding bad pixels. Spatial correlations between pixels were neglected. Spatial dispersion in the gain predominantly arises from the output buffer of the multiplexer (MUX), the channel amplifier before the ADC and the ADC itself -- all these contributions are shared across the same readout channel. Per-pixel contributions, such as the pixel source follower gain and transimpedance gain (charge-to-voltage conversion factor), are not included in the spatial variance calculation. Average conversion gain values for each channel are computed and shown in Fig.~\ref{fig:cg}. The dispersion of the conversion gain per channel is less than 1\%. In Table~\ref{tab:PFA_main_properties} the average values of conversion gain per detector are reported. The average conversion gain of the focal plane array (FPA) is 0.52\,ADU/e$^-$, with about 3\% variation between different arrays. Average conversion gain values for each channel are used to convert ADU to electrons in subsequent analyses. 

The influence of the dynamic range used to compute the conversion gain was studied in \cite{2018SPIE10709E..21S}. A method for evaluating gain with higher spatial resolution, using $16 \times 16$ super-pixels, was derived in \cite{Graet:2024ytq}. Using the conversion gain measured per super-pixel, the effect of gain and QE can be decorrelated, thus obtaining a more precise measurement of QE. \cite{Graet:2024ytq} also discusses the influence of IPC and signal nonlinearity and proposes a method for estimating gain corrected for these factors. 

\subsection{Noise performance\label{sec:noise}}
The source of noise can be the signal source itself, but in the absence of illumination, the dominant sources of noise are in the readout electronics and in the light-sensitive material of the pixel. The resulting noise from different sources in the light-sensitive material and in the electronics is called readout noise.  It describes the typical variability of recorded counts from one reading to another in the absence of illumination and, in the case of sky observations, adds to the noise from the source itself. Which noise sources are dominant in a given case depends on the exposure time, acquisition mode, and signal estimation method. 
Reset noise is the noise of the pixel's zero level generated by the diodes when connecting to the power supply. It is also called kTC noise, as it varies proportionally with temperature $T$ and pixel capacitance $C$ along with the Boltzmann's constant $k_\mathrm{B}$ as the proportionality factor. Fortunately, the reset noise is completely removed by sampling up-the-ramp, but it is interesting to quantify it to see by how much the offset of signal integration can vary from exposure to exposure. It is also a reference for checking the state of the detectors. The current baseline can be compared to the reference baseline and the kTC noise, and in this way it can be checked if it is within the expected range, or if there is too much deviation, which signals that additional tuning is required. 

In the case of very bright sources, which saturate the detector in a very short time, a typical solution is to use the CDS technique, which makes it possible to easily adapt the exposure time while avoiding saturation, thus increasing the accuracy of the measurement. In such a case, it is necessary to estimate the readout noise of a single readout. In the case of \Euclid, no acquisition mode change is anticipated for bright sources. However, the noise value of a single readout is required as input for the algorithm of signal estimation using multiple frames up-the-ramp \citep{kubik2016}. Finally, the noise characteristics will also depend on how the flux is estimated from up-the-ramp samples. 

The following sections present an overview of the noise characteristics of the \Euclid H2RG detectors. First, typical reset noise values are discussed, followed by an overview of single frame readout noise and finally the readout noise of the signal estimator, when multiple signal readouts are taken up the ramp and the slope is calculated using the algorithm described in \cite{kubik2016}, is presented.

\subsubsection{Reset noise\label{sec:kTC}}

\begin{figure*}
\centering
\includegraphics[width=\linewidth]{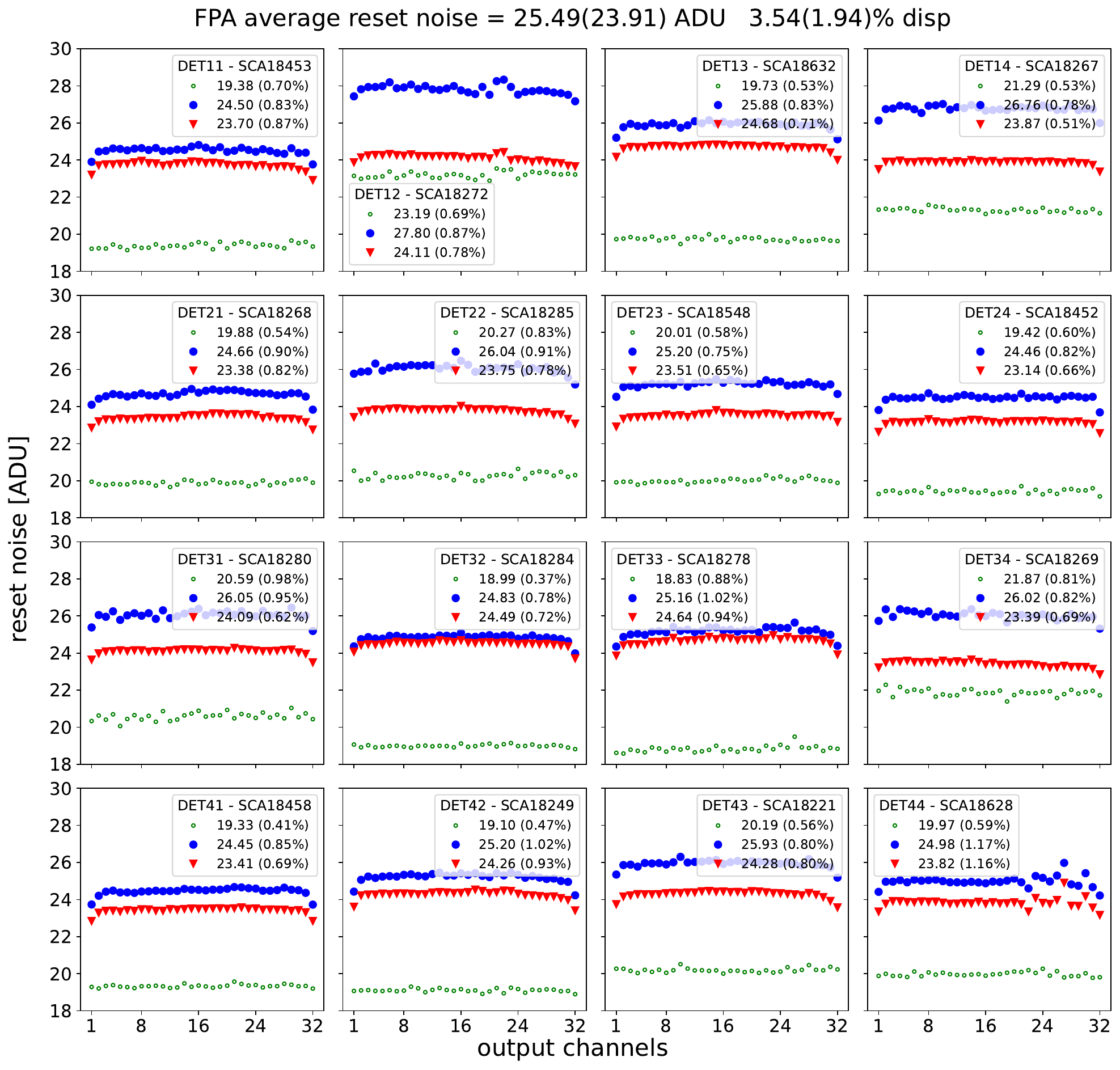}
\caption{\label{fig:ktc_noise}Reset noise of science pixels before (blue circles) and after (red triangles) reference pixel correction. The reset noise of reference pixels is shown as green circles. For each detector in the legend the average reset noise over the 32~channels is indicated and in parentheses the relative dispersion of the average reset noise between channels is given. In the top main title we indicate the average reset noise over the whole FPA and the dispersion between detectors before reference pixels correction. In parentheses the values after reference pixel correction are given.}
\end{figure*}

To estimate the reset noise of the NISP detectors, 500 dark exposures of 16 frames up-the-ramp were taken. The 16 frames were averaged per pixel for each of the exposures, forming 500 groups. The standard deviation of the groups per pixel was computed and this represents the reset noise of the first group in each of the science acquisition modes of NISP detectors. These calculations were performed twice: first prior to the reference pixels correction, and then after the frames were corrected using reference pixels. In Fig.~\ref{fig:ktc_noise} we show the obtained values as well as the reset noise of reference pixels.

For photosensitive pixels, the reset noise is typically in the range of 25\,ADU (about 50\,e$^-$), while reference pixels have a reset noise of around 20\,ADU.  The noise is fairly homogeneous within each of the 32 channels of a detector, but each channel has a different average noise level. The reference pixel correction effectively reduces the baseline noise of photosensitive pixels to approximately 23\,ADU and reduces the spatial dispersion of noise levels across channels. The FPA average reset noise before reference pixels correction is equal to 25.5\,ADU with a 3.5\% dispersion between arrays. After reference pixels correction the average FPA reset noise is reduced to 23.9\,ADU and the dispersion between detectors is lowered to 2\%. The median values of the reset noise after reference pixel correction are reported in Table~\ref{tab:PFA_main_properties}. There is a correlation between the reset noise of the reference pixels and the correction amplitude: higher reset noise in the reference pixels results in a stronger correction.

\subsubsection{Single frame readout noise\label{sec:ron}}

To estimate the detection chain noise of a single readout we used 90 exposures of 394 frames, UTR(394), acquired with no illumination. After reference pixel correction, the non-overlapping consecutive pairs of frames are subtracted to form 197 CDS frames per exposure. The single frame readout noise is computed as the standard deviation per pixel across these 197 CDS frames and divided by $\sqrt 2$ to account for two frames in one CDS pair. The shot noise contribution of the dark current is negligible and is not subtracted.  The median per pixel is then taken across 90 values of noise frames to reject glitches and to form the final single frame readout noise map. This map is converted to electrons using the average conversion gain per channel. 

\begin{figure}
\centering
\includegraphics[width=\linewidth]{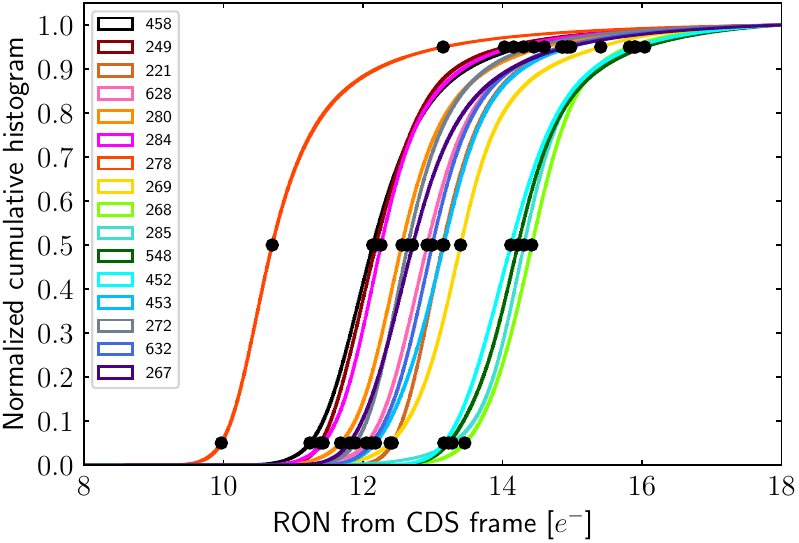}
\caption{\label{fig:ron_cds}Cumulative distributions of single frame readout noise for the 16 detector arrays after reference pixel correction. Black dots represent the 5th, 50th, and 95th percentiles of the distribution.}
\end{figure}

The values of median noise per detector are reported in Table~\ref{tab:PFA_main_properties}. The median values of the single frame readout noise of \Euclid H2RG detectors are less than \SI{15}{e^-}. The noise distribution is log-normal, the 95th percentile does not exceed \SI{16}{e^-} and the 5th percentile lies between \SI{10}{e^-} and \SI{14}{e^-} depending on the SCA as shown in Fig.~\ref{fig:ron_cds}.


\subsubsection{Readout noise\label{sec:dcn}}

The readout noise (RON) of the detecting system is defined as the standard deviation per pixel of the signal estimates. It depends on the total exposure time and on the readout mode. 

To estimate the readout noise in photometric and spectroscopic readout modes, the same slopes used for the dark current computation (see next section) were used. The slopes were multiplied by the integration time characteristic for each of the readout modes, i.e.\ by \SI{87.2}{s} and by \SI{549.6}{s}, respectively, to obtain the total integrated signal in electrons. After that, the standard deviation per pixel was computed from the signal variation between the 90 available ramps. The median values per detector are reported in Table~\ref{tab:PFA_main_properties}. The median photometric noise $\sigma_{\rm ph}$ is below \SI{7}{e^-}. The median spectroscopic noise $\sigma_{\rm sp}$ is below \SI{9}{e^-}. The noise does not scale with exposure time, because different readout modes are used. In spectroscopic mode 15 groups are taken, effectively lowering the noise in signal estimation compared to when fewer groups would be acquired along the ramp \citep{Kubik2015JATIS}.

\subsection{Dark current\label{sec:dc}}
Pixel dark current is another major factor affecting the performance of pixel detectors. Its cause is the leakage current of the photodiode, which is mainly thermally generated. However, the measurement is also sensitive to the differential voltage drift between reference and science pixels. Hence, an important factor to consider when assessing the dark current is the whole SCS -- comprising the SCA and the SCE components. 
Moreover, it is recommended to use matching dark and science exposures for most astronomical infrared array detectors to remove any residual bias \citep{Rauscher:2007gc}. This is achieved by ensuring that the dark exposures are taken using the same acquisition mode as the science exposures.
This is why in this section we report values of `photometric' and `spectrometric' darks that are evaluated using the MACC(4,16,4) or MACC(15,16,11) acquisition modes, respectively. 

At the operating temperature of 87\,K, the pixel leakage current of H2RG detectors is very low, typically below \SI{e-3}{e^{-}.s^{-1}}. This is generally good for science, but is also making it difficult to measure with high accuracy in science acquisition mode. Accurate dark current measurements typically require many repeated exposures. For example, assuming a dark current of \SI{e-3}{e^{-}.s^{-1}} we would need to take more than 7000 ramps (requiring more than a week of dark measurement) to achieve 1\% accuracy on the dark in photometric acquisition mode for pixels with the readout noise close to the median value. Therefore, in our measurements, we did not aim for an accuracy of 1\% when measuring dark current, but were interested in measuring dark current under ‘operational’ conditions with fixed MACC modes and exposure times.

Dark current values were computed from 90 exposures acquired with UTR(394), each with a total exposure time of about 10 minutes. The reference pixel correction was applied to all exposures as the first step of data reduction. The signal estimation algorithm used was the one implemented in the flight electronics on board \Euclid to obtain comparable values \citep{kubik2016}. To calculate the signal, among 394 acquired frames only the frames corresponding to the in flight acquisition modes, i.e.\ MACC(4,16,4) or MACC(15,16,11), were selected as input to the algorithm. Thus, two different values of dark current were obtained: `photometric' dark $d_{\rm ph}$ and `spectrometric' dark $d_{\rm sp}$ respectively. 
After removing exposures with detected anomalies at the pixel level, a median filter was applied to obtain dark current maps.

The average conversion gain per channel was applied to convert ADU to electrons. The results for dark current measurements $d_{\rm ph}$ and $d_{\rm sp}$, given in Table~\ref{tab:PFA_main_properties}, are expressed in electrons per second allowing easier comparison between the two. A higher value of $d_{\rm sp}$  compared to $d_{\rm ph}$ could be due to the imperfect measurement conditions with e.g.\ low-level residual stray-light in the cryostat. Also, a different readout mode can affect the measured value of the dark current, but it is difficult to estimate the relative influence of different factors on the measurement.

\subsection{Nonlinearity\label{sec:nl}}
The response of H2RG detectors, like that of many other pixel-based devices, can deviate from the ideal linear behaviour \citep{Vacca2004,Plazas2017,Mosby2020JATIS}. This analysis concentrates on `classical' nonlinearity, namely the nonlinearity associated with the conversion from charge to voltage domain. In contrast to CCD devices, the nonlinearity in HgCdTe detectors varies from pixel to pixel. Attaining sub-percent accuracy in correcting for this nonlinearity is crucial for the mission's scientific objectives. In this section, we examine the extent to which the pixel response deviates from ideal linear behaviour following the tuning of the electronic registers. 

The nonlinearity is measured separately in photometric and spectroscopic acquisition modes at several fluence levels ranging from 1000 up to 80\,ke$^-$ which corresponds to about 60\% of the typical pixel full well (assumed here to be of 130\,ke$^-$ on average). Typically, nonlinearity is a measure of how much the signal recorded at a pixel in the maximum dynamic range deviates from the signal measured as a linear fit to the start of the ramp of that pixel. The response of the pixel is then modelled by a polynomial function, the inverse of which is used as the linearising function of the measured signal. In the case of \Euclid, the inherently nonlinear signal sampled up-the-ramp is estimated on board using a linear function and the slope result is sent to the ground. This signal estimate is subject to an error due to the nonlinearity of the pixel response. We therefore need to quantify the deviation of the signal measured as a fit of the linear function to the nonlinear response with respect to the expected linear signal.

To determine the degree of nonlinearity in spectroscopic acquisition mode a typical acquisition pattern involves a sequence of 30 flat-field ramps of UTR(394). The number of exposures per flux is calculated in order to maintain a relative flux error below 1\%, provided that the number of exposures does not exceed 30. Beyond this threshold, the measurement was constrained by time limitations. The ramps after reference pixels correction are modelled one by one using a third-order polynomial per pixel. The use of higher-order polynomials has been verified as unnecessary for the purpose of describing the nonlinearity of the ramps. 

What we call the reference linear signal $S_{\rm lin}$ -- see also discussion in Sect.~\ref{sec:refsignal} -- is defined as the 1st-order coefficient of the polynomial expansion, described in Appendix~\ref{app:ortho}, multiplied by the exposure integration time. The zeroth-order coefficient represents the baseline estimator and can be compared to the independent baseline measurements for the purpose of verification. In our analysis, such a comparison was made, and the measured coefficients were found to be consistent with the independent baseline estimate.

The nonlinear integrated signal $S_{\rm nl}$ is calculated using the nominal NISP signal fitting algorithm described in Sect.~\ref{sec:llk} with the spectroscopic acquisition mode. 
At each illumination level the average of $S_{\rm lin}$ and $S_{\rm nl}$ per pixel is computed over 30 ramps and gives average $\langle S_{\rm lin}\rangle$ and $\langle S_{\rm nl}\rangle$ maps.
The degree of nonlinearity is quantified by measuring the deviation of the nonlinear signal from the expected linear signal, that is by taking the difference between $\langle  S_{\rm lin}\rangle$ and $\langle S_{\rm nl}\rangle$ per pixel and is expressed as percentage of $\langle S_{\rm lin}\rangle$ per pixel.

In photometric acquisition mode, the nonlinearity was measured using exactly the same analysis and very similar statistics. The only difference was that the ramps were acquired with shorter integration time and the estimate of the nonlinear signal $S_{\rm nl}$ was calculated using photometric readout mode. 

\begin{table*}[ht!]
\caption{5th, 50th, and 95th percentiles of nonlinearity at $80\,000$ integrated electrons in photometric and spectroscopic exposure time.} 
\label{tab:nl}
\begin{center}       
\begin{tabular}{@{}l|rrrrrrrrrrrrrrrr@{}}
\toprule
\toprule
FPA position & 11 & 12 & 13 & 14 & 21 & 22 & 23 & 24 & 31& 32 & 33 & 34 & 41 & 42 & 43 & 44 \\
SCA 18*** & 453 & 272 & 632 & 267
    & 268 & 285 & 548 & 452
    & 280 & 284 & 278 & 269
    & 458 & 249 & 221 & 628 \\
\midrule
ph-5\%  & 2.65 & 2.06 & 1.05 & 0.60 
                 & 0.15 & 2.67 & 1.48 & 1.69 
                 & 1.36 & 1.15 &$-$0.45 & 2.37 
                 & 2.02 & 2.37 & 2.03 & 1.81 \\
ph-50\% & 3.46 & 2.94 & 2.45 & 2.54 
                 & 1.79 & 3.52 & 2.49 & 2.56 
                 & 2.61 & 2.49 & 2.04 & 3.31 
                 & 2.78 & 3.09 & 2.96 & 2.90 \\
ph-95\% & 4.30 & 3.82 & 3.31 & 4.03 
                 & 2.84 & 4.40 & 3.35 & 3.32
                 & 3.58 & 3.42 & 3.36 & 4.37
                 & 3.48 & 3.78 & 3.81 & 3.77 \\
\midrule
sp-5\%  & 2.70 & 2.01 & 0.08 & 0.17 
                 &$-$0.65 & 2.65 &$-$1.11 & 1.52 
                 & 0.66 & 0.59 &$-$1.11 & 2.34 
                 & 1.95 & 2.33 &$-$1.84 & 1.50 \\
sp-50\% & 3.72 & 3.10 & 2.35 & 2.51 
                 & 1.46 & 3.72 & 2.46 & 2.64
                 & 2.39 & 2.31 & 1.59 & 3.49 
                 & 2.92 & 3.21 & 3.07 & 2.92 \\
sp-95\% & 4.70 & 4.18 & 3.57 & 4.27 
                 & 2.79 & 4.77 & 3.53 & 3.56
                 & 3.66 & 3.56 & 3.29 & 4.73
                 & 3.75 & 4.04 & 4.11 & 3.95 \\
\bottomrule
\end{tabular}
\end{center}
\end{table*}

\begin{figure}
\centering
\includegraphics[width=\linewidth]{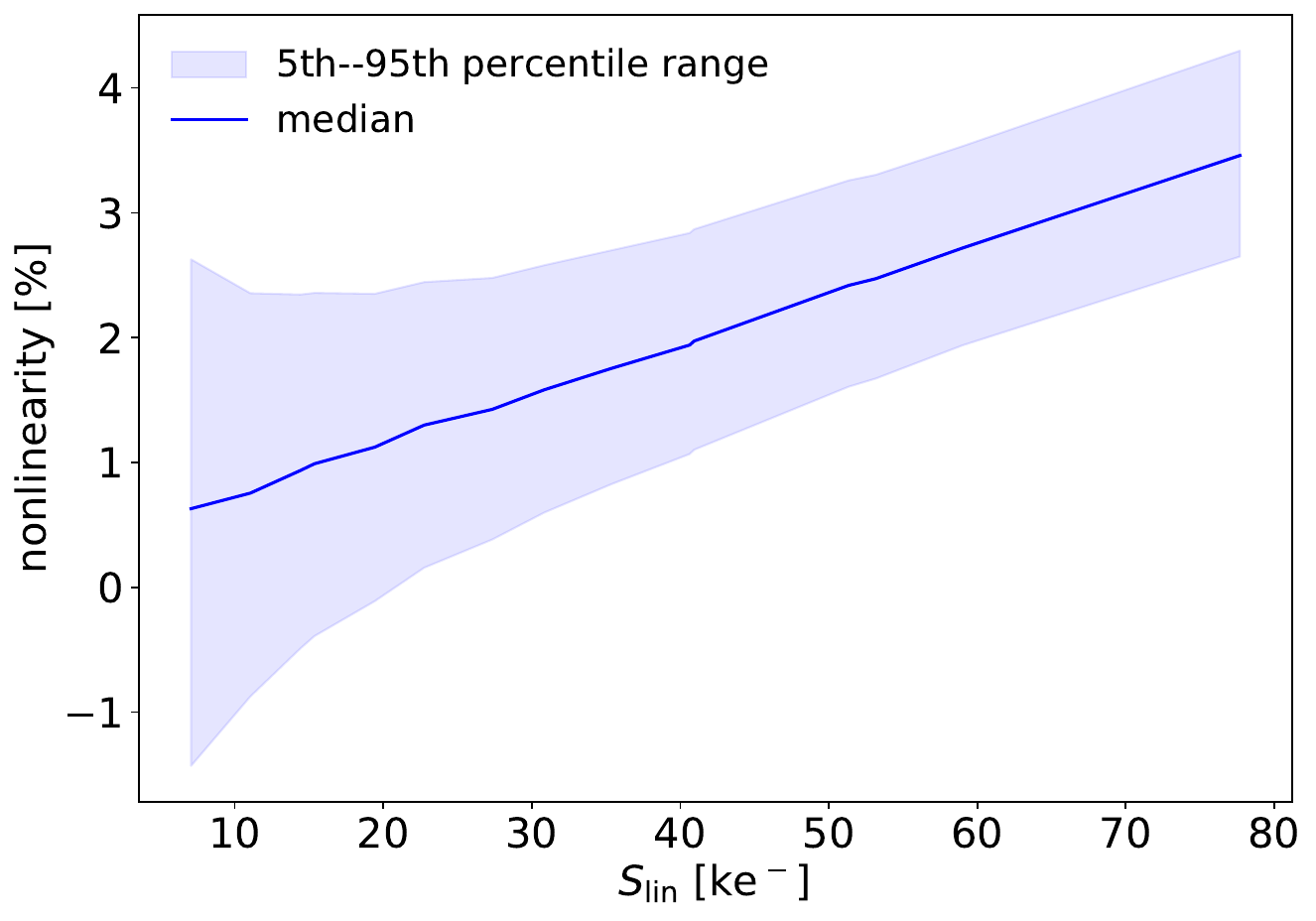 }
\caption{\label{fig:nl_ff}Median nonlinearity of the pixel response in photometric readout mode for SCA 18453 over all dynamical range measured during ground tests. The shadowed band extends from the 5th to the 95th percentiles of the per-pixel distribution.}
\end{figure}

In Table~\ref{tab:nl} we report the median nonlinearity at maximum measured signal of \num{80000}~electrons in photometric and spectroscopic acquisition modes. 50\% of pixels have nonlinearity lower than 3.5\% in photometric and 3.7\% in spectroscopic acquisition modes at 60\% of the full well. 95\% of pixels do not exceed 4.8\% deviation from linear integration. 
Typical nonlinear behaviour is shown in Fig.~\ref{fig:nl_ff} for one detector in photometric acquisition mode. This response is observed for all detectors, the exact median values and the scatter between the 5th and 95th percentile can be read in Table~\ref{tab:nl}.

\subsubsection{Reference signal definition problem\label{sec:refsignal}}

Since the definition of the reference signal, as the linear coefficient of a polynomial function fitted to the ramp, could also have a number of other definitions, it deserves a few words of commentary. 

For the same constant incident flux and assuming a constant baseline level, the resulting value of the reference signal depends on (1) the number and spacing of samples up the ramp, i.e.\ on the acquisition mode, (2) the dynamic range in which the signal is measured, i.e.\ on the exposure time, and (3) the initial state of the detector, which may be more or less susceptible to other sources of linearity distortion, such as capture and release of charges. These differences were measured for each of the cases we analysed.

Least important is the impact of the number and spacing between UTR samples used in the polynomial fit. Whether the ramp is fitted using all the frames UTR or using several groups of 16 averaged frames, as for example in the nominal acquisition modes used by NISP, the difference does not exceed 0.1\% and is usually less than 0.05\% provided that the exposure time is kept constant. Between these two options we choose fitting using all the sampled frames, as it is less prone to anomalies. 

\begin{figure}
\centering
\includegraphics[width=\linewidth]{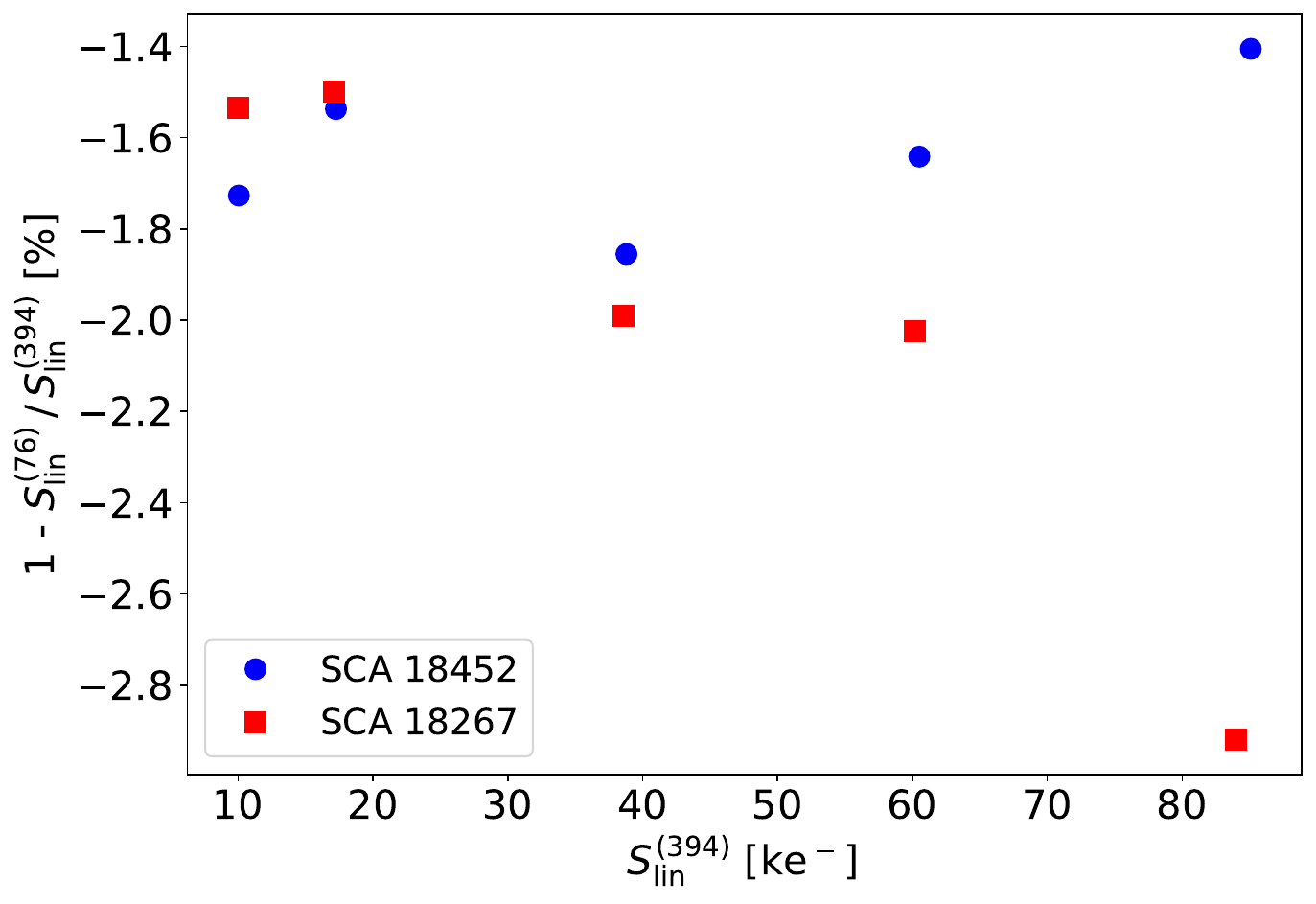}
\caption{\label{fig:nl_ref_lin} Example of linear flux definition using ramps acquired in UTR(394). Fit performed using 76 frames $S_\textrm{lin}^{(76)}$ and 394 frames $S_\textrm{lin}^{(394)}$. The difference per pixel of first order coefficients was computed and the relative differences between the median signals per detector are shown for two detectors: low persistence SCA18452 (blue points) a higher persistence SCA18267 (red squares).}
\end{figure}

More important is the dynamic range that is used for the fit. In the case where the polynomial fit uses only a part of the ramp, naturally the linear coefficient will be sensitive to higher order corrections. In our acquisition scheme for measuring nonlinearity, in which several flat-field ramps are taken one after the other, fitting a polynomial to the first part of the ramp gives a higher flux estimate than an estimate based on all frames up the ramp and the same order of polynomial function. For example, the linear coefficient of 3rd order polynomial function using 76 first frames $S_\textrm{lin}^{(76)}$ can differ by 3\% with respect to the 3rd order polynomial function using 394 frames of the same ramp $S_\textrm{lin}^{(394)}$. Interestingly, this difference is in general constant with flux as illustrated in Fig.~\ref{fig:nl_ref_lin} by blue circles. 

\begin{figure}
\centering
\includegraphics[width=\linewidth]{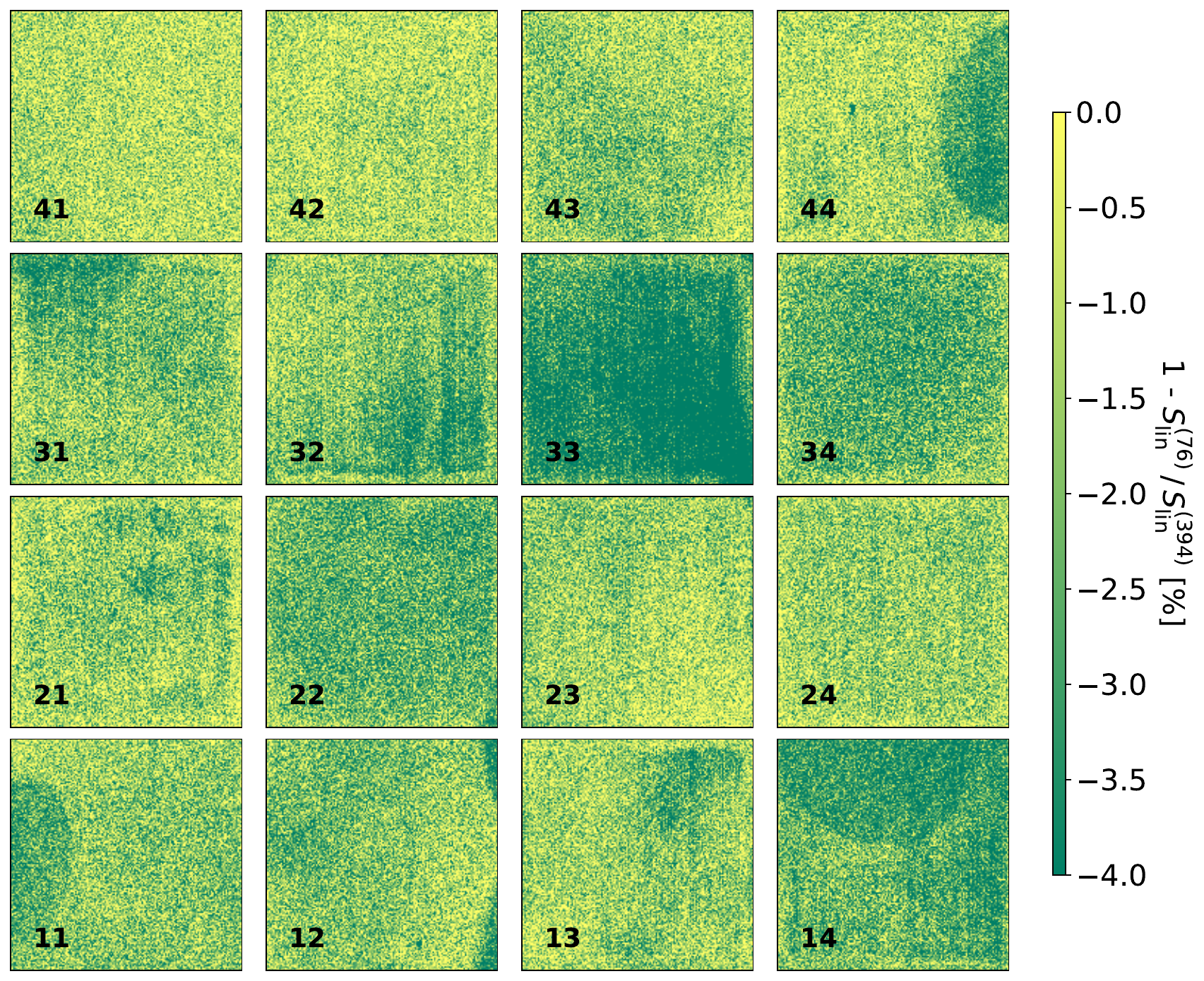}
\caption{\label{fig:Slin_diff_img} The difference between the linear coefficient of the polynomial to the first 76 frames of the ramp with respect to the linear coefficient of the polynomial to all the 396~frames that were acquired. A correlation with the persistence map (Fig. \ref{fig:pers_det_structures}) is evident. The difference in the linear flux estimate for pixels with high persistence can reach 4\%.}
\end{figure}

The only cases in which this difference increases with flux are detectors with a large amplitude of persistence (red squares in Fig.~\ref{fig:nl_ref_lin}). 
It is the continuous capture and trapping of charges, the rate of which varies depending on the state of the detector, that has the greatest impact on the linear flux estimation. A clear correlation between the nonlinear reference definition and the persistence amplitude is evident while comparing the structures seen in Figs.~\ref{fig:Slin_diff_img} and \ref{fig:pers_det_structures}. An accurate estimate of this effect in all possible acquisition schemes and in all the range of fluxes is beyond the scope of this paper. We will limit ourselves here only to point out that the measurement of nonlinearity is systematically subject to error due to other physical effects present in the pixel, of which capture and emission of charges  is, in our case, one of the most significant. For example, at an illumination level of about \num{12000}\,e$^-$, the difference in the estimation of the linear signal using a fit of a 3rd-order polynomial function to the exposure of \SI{100}{s} can reach about 10\% depending on the measurement scheme for detectors characterised by high persistence currents. 

Therefore, to estimate the degree of nonlinearity, we use a fixed measurement scheme in which the detector is in a fixed illumination state and average the measurements over a set of exposures measured in a similar state to reduce other systematic effects. The nonlinearity is measured independently for photometric and spectrometric exposures and the expected linear signals are calculated independently in these two schemes using the maximum number of points through the ADC saturation in each case.

\subsection{Persistence\label{sec:latency}}
Image persistence is a significant problem in infrared detectors, severely limiting data quality if its amplitude is large compared to the background level. This problem manifests itself as residual images, which can persist for several hours after observing a very bright source. 

The phenomenological model explaining image persistence, proposed in \cite{Smith_2008a,Smith_2008b}, attributes it to charge trapping and release in the diode's depletion region. Subsequent studies characterised persistence amplitudes and decay based on flux, fluence, and exposure time. Some models describe persistence decay exponentially \citep{Serra_2015,Tulloch_2019,Mosby2020JATIS}, while others use a power-law \citep{Long_2012,Long_2015}, linking it to a wide range of trapping time constants. 

A comprehensive description of the persistence properties of the NISP detectors, as characterised during ground tests, is provided in \cite{Kubik_SPIE_2024}. Here, we summarize the typical persistence amplitudes observed after non-saturating fluences and the characteristic timescale for persistence decay.

We measured the persistence current as a function of stimulus amplitudes ranging from 5000 to \SI{95000}{e^-}. For each stimulus level, the persistence current was recorded during a dark exposure UTR(276), following a flat-field exposure of UTR(76). At each fluence level, the measurement was repeated 15 consecutive times. 

\begin{figure}
\begin{center}
\begin{tabular}{c} 
\includegraphics[width=\linewidth]{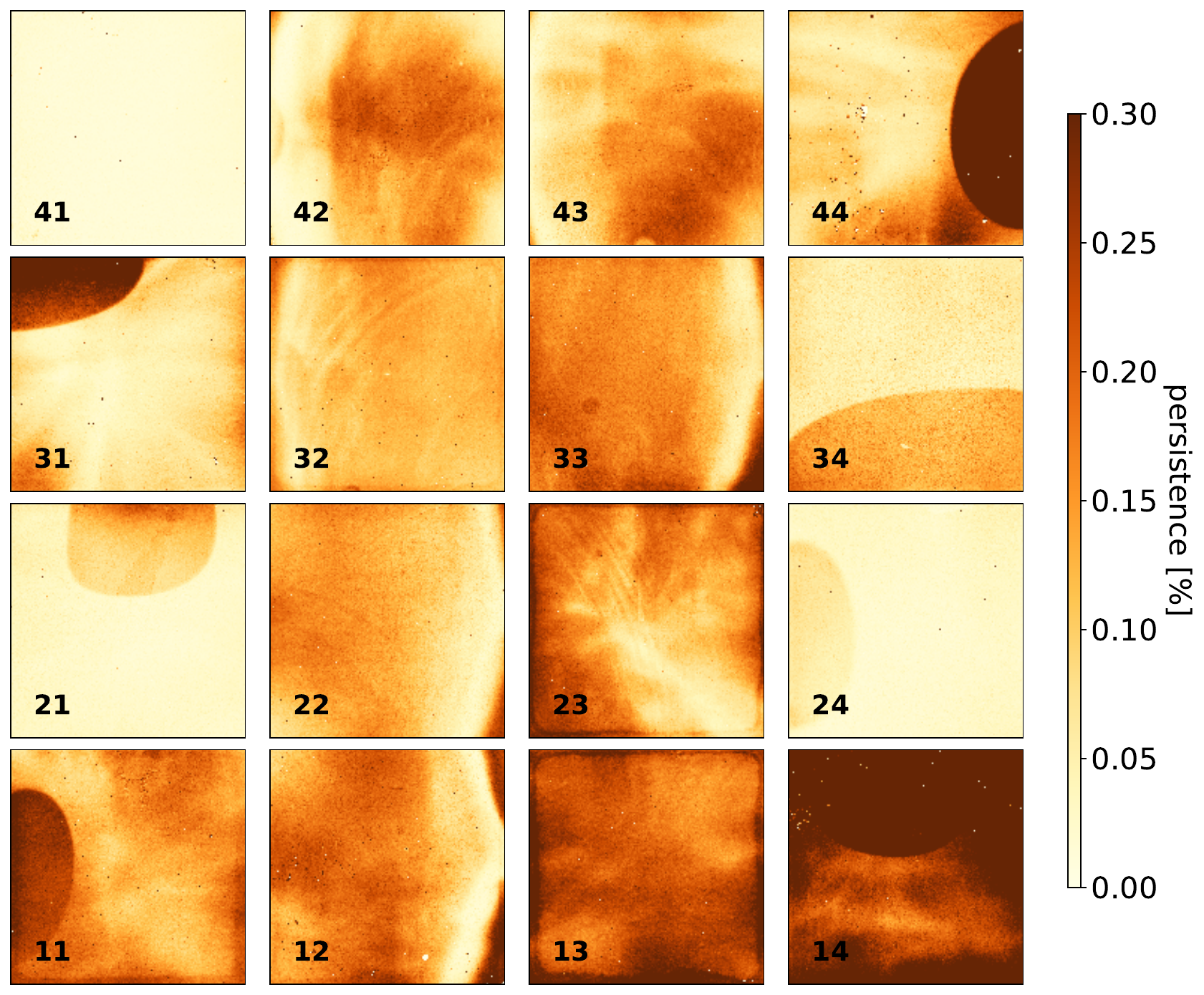}
\end{tabular}
\end{center}
\caption{\label{fig:pers_det_structures}Persistence amplitudes in percentage of the previous flat-field illumination below saturation.}
\end{figure}

The integrated persistence charge per pixel was determined by performing a linear fit to the dark ramp in UTR(276), multiplying the resulting slope by the typical NISP photometric exposure time, and averaging over the 15 exposures. Following a 5000\,e$^-$ stimulus, the median persistence per detector ranges from 1 to 45\,e$^-$, while after a \SI{95000}{e^-} stimulus (near the pixel full-well capacity of \SI{130000}{e^-}), it spans from 10 to 300\,e$^-$. On average, the median persistence is below 0.3\% of the stimulus for sub-saturation levels for an exposure UTR(276) directly following the stimulus (Table~\ref{tab:PFA_main_properties}). However, the persistence signal varies across pixels within individual detectors as shown in Fig.~\ref{fig:pers_det_structures}. The dependence of the persistence amplitudes on the stimulus intensity is not linear and the detailed description can be found in \cite{Kubik_SPIE_2024}.

\begin{figure}
\begin{center}
\begin{tabular}{c} 
\includegraphics[width=\linewidth]{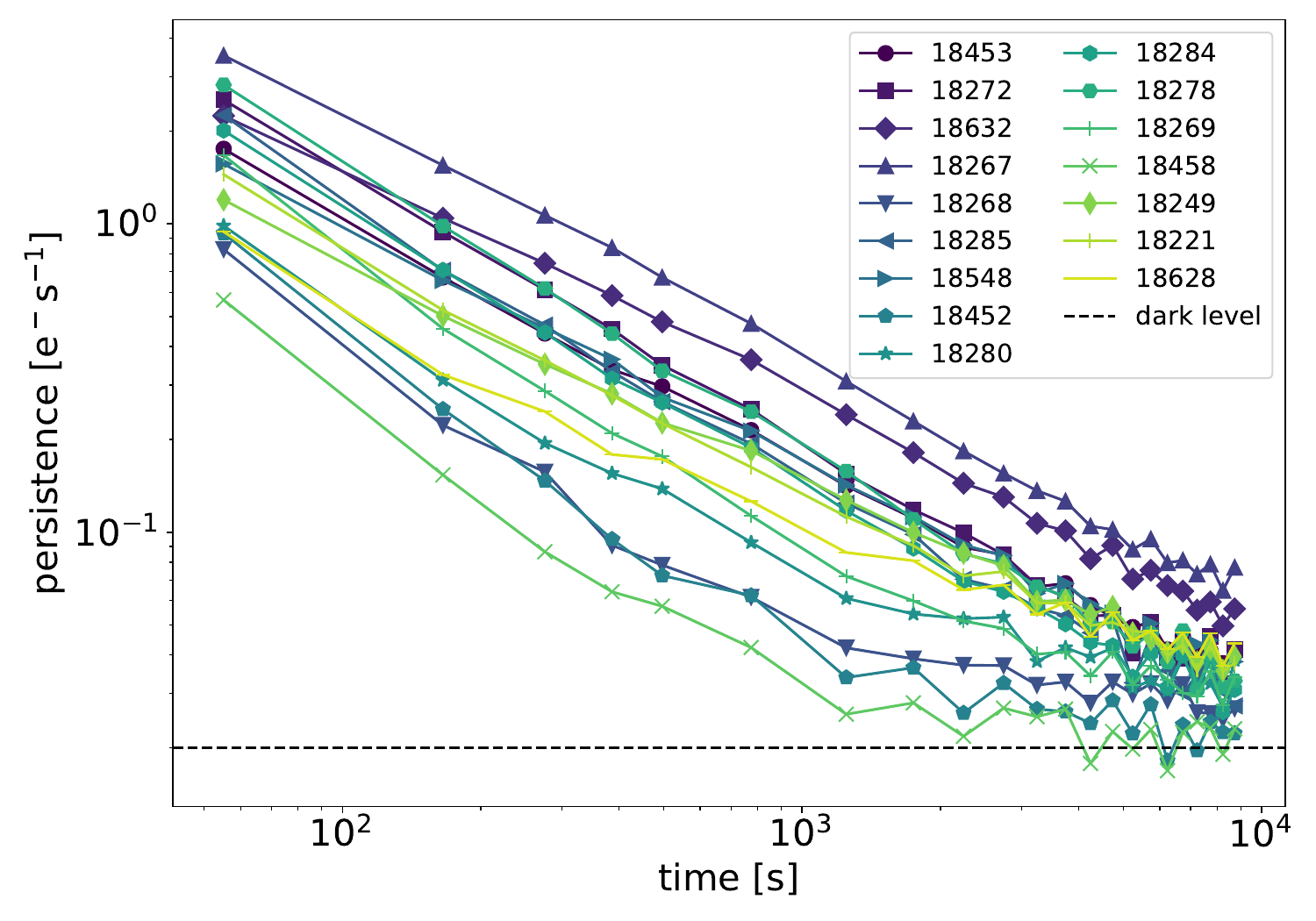}
\end{tabular}
\end{center}
\caption{\label{fig:pers_decay}Persistence decay over two hours after a flat-field of 80\,000\,e$^-$. The lines between the points were added to guide the eye, it is not the result of fitting the model to the data.}
\end{figure}

In Fig.~\ref{fig:pers_decay} we show the decay of the persistence current measured over two hours in dark conditions. The persistence current follows a power-law decay 
\begin{equation}
    I(t) = I_0 t^{-\beta}
\end{equation}
for all the detectors with the average power index $\beta = 0.72$ with about 15\% of dispersion between detectors. The average decay power per detector are given in Table~\ref{tab:PFA_main_properties}. The detailed per-pixel analysis of the power law decay is presented in \cite{Kubik_SPIE_2024}.

\section{Conclusions\label{sec:conclusions}}
The infrared detectors designed for the NISP instrument underwent a comprehensive characterisation in 2019 in their flight configuration to ensure optimal performance for space mission. During this process, a detailed evaluation of all key detector properties was carried out. This included identification of defective or `bad' pixels, evaluation of IPC, determination of gain and noise levels, measurement of dark current, and analysis of response nonlinearity and persistence. As a result of these rigorous measurements, detailed detector maps were created, providing a comprehensive representation of pixel-specific properties. 

The maps derived from the ground characterisation campaign serve as a reference for in-flight calibration efforts, enabling continuous monitoring and adjustment of instrument performance under operational conditions \citep{Cogato24}. In addition, they are an essential tool for simulating instrument behaviour \citep{EP-Serrano}, facilitating the development and refinement of advanced data processing techniques essential for correcting data collected during instrument operation. By integrating these maps with both calibration and simulation processes, the accuracy and reliability of scientific observations is greatly enhanced, providing the high-quality data necessary to achieve the mission's scientific goals.

\paragraph{Ground characterisation workflow:}
The test results were achieved through the automated execution of a tightly designed test plan to reach the required 1\% measurement precision, using advanced data acquisition software to efficiently execute the process. This automation ensured nearly 100\% operational efficiency, with an extremely low failure rate, underscoring the robustness of the test structure. The tests were conducted continuously, operating 24 hours a day, 7 days a week, with minimal downtime demonstrating the reliability of the system and its ability to maintain high performance over extended operational periods.

\paragraph{Detector system parameter tuning:}
An automatic tuning procedure for all of the SCS parameters was implemented and executed, significantly reducing the testing time and improving the overall efficiency of the process. The detector system operating parameters were optimised as follows: the gain was set to 15\,dB, the polarisation voltage to 500\,mV, and the baseline for the science pixels after reference pixel correction is in the range from \SI{4000} to \SI{10000}\,ADU. These settings allowed for the maximisation of the dynamic range (excluding the DNL range) while maintaining low readout noise. The resulting dynamic range was approximately 115\,ke$^-$ , corresponding to a maximum measurable flux of \SI{1000}\,e$^-$s$^{-1}$ in MACC(4,16,4) and 200\,e$^-$s$^{-1}$ in MACC(15,16,11).

\paragraph{Detector system performance:}
Regarding performance, the number of disconnected pixels was found minimal -- fewer than 1000 per detector -- and this number remained stable throughout thermal cycles, demonstrating the detectors' reliability under varying thermal conditions. Pixels classified as inoperable accounted for less than 0.2\% of the total science pixels. 

The IPC coupling is minimal, with less than 1\% crosstalk between neighbouring pixels. Furthermore, 95\% of the pixels exhibit a QE greater than 80\% across the full spectral range of the \Euclid mission, indicating high sensitivity and efficient light capture.

The conversion gain is approximately 0.52\,ADU/e$^-$, with a 3\% variation between detectors and less than 1\% variation between channels of the same detector, ensuring consistent system response. The reset noise is approximately equal to 25\,ADU and is effectively reduced to 23\,ADU through reference pixel correction.

The single frame readout noise is approximately 13\,e$^-$, with minimal variation between detectors, ensuring consistent noise characteristics across the system. The signal estimator noise is measured at 7\,e$^-$ in photometric mode and 9\,e$^-$ in spectroscopic acquisition mode. The dark current is found to be below the measurement accuracy threshold. 

The deviation from linear response is less than 5\% for 95\% of the pixels, even at signal levels up to 80\,ke$^-$ , in both photometric and spectroscopic MACC modes, reflecting the high linearity of the detector's response. 

Median persistence amplitude is less than 0.3\% of the signal, though it exhibits significant spatial variation and differences between detectors. 

\paragraph{Issues:}
Achieving 1\% precision per pixel necessitates exceptionally stable measurement conditions, particularly for dark current measurements, as well as a substantial volume of data to reduce statistical errors. Defining and measuring a linear signal reference is particularly challenging in the absence of absolute calibration. The measurement duration is a crucial factor, as continuous charge capture and release affect the flux stability. As such, the charge trapping and release has an impact on all the measurements, including the characterisation of persistence itself, adding further complexity to the analysis.

\begin{acknowledgements}
    \AckEC
\end{acknowledgements}

\bibliography{Euclid,paper}

\begin{appendix}
 \onecolumn 
 
\section{Detector properties summary table \label{app:det_prop}}
\begin{sidewaystable}[ht!] 
\caption{H2RG median properties from ground characterisations across the NISP focal plane. dpix -- number of disconnected pixels, B$_\text{R}$ -- baseline of reference pixels, B -- baseline of science pixels, RN$^c$ -- reset noise after reference pixel subtraction, \textbf{QE}[\YE,\JE,\HE] -- quantum efficiency in \% in three photometric bands, \textbf{IPC} -- central pixel value  of the inter-pixel capacitance kernel [\%], \textbf{CG} -- conversion gain, $\sigma_R$ -- single frame readout noise after reference pixels correction, $d_{\rm ph}$, $d_{\rm sp}$ -- dark currents measured in photomeric and spectroscopic integration times at 80\,000 integrated electrons, $\sigma_{ph}$ -- noise of signal estimator in MACC(4,16,4), $\sigma_{sp}$ -- noise of signal estimator in MACC(15,16,11), \textbf{nl} -- departure from linearity at 80\,000 integrated electrons in MACC(4,16,4), \textbf{pers} -- persistence amplitude as percentage of 80\,ke$^-$ flat-field intensity, $\beta$ -- persistence decay power index.} 
\label{tab:PFA_main_properties}
\begin{center}       
\begin{tabular}{@{}ccrrrrrrrrrrrrrrrrr@{}}
\toprule
\toprule 
    \textbf{FPA} 
 &  \textbf{ID}  
 &  \textbf{dpix} 
 &  \textbf{B$_\text{R}$} 
 &  \textbf{B} 
 &  \textbf{RN$^c$}
 &  \textbf{QE}
 &  \textbf{QE}
 &  \textbf{QE}
 &  \textbf{IPC}
 &  \textbf{CG} 
 &  \textbf{$\sigma_{\rm R}$}
 &  \textbf{$d_{\rm ph}$} 
 &  \textbf{$d_{\rm sp}$} 
 &  \textbf{$\sigma_{\rm ph}$}
 &  \textbf{$\sigma_{\rm sp}$} 
 &  \textbf{nl} 
 &  \textbf{pers} 
 &  \textbf{$\beta$} \\
   \textbf{position} 
 & \textbf{18***}
 & 
 & ADU
 & ADU
 & ADU
 & [\YE]\%
 & [\JE]\%
 & [\HE]\%
 & \%
 & ADU/e$^-$
 & e$^-$
 & e$^-$s$^{-1}$
 & e$^-$s$^{-1}$
 & e$^-$
 & e$^-$
 & \%
 & \% 
 & \\
\midrule
11 & 18453 &  135 &  5665 &  5842 & 23.6 & 94 & 95 & 96  & 97.26 & 0.51 & 13.2 &  0.006 & 0.024 & 6.0 & 8.2 & 3.46 & 0.16 & 0.74 \\
12 & 18272 & 3223 &  5977 &  3846 & 23.9 & 96 & 97 & 99  & 97.11 & 0.53 & 12.7 & $-0.001$ & 0.022 & 6.2 & 9.0 & 2.94 & 0.17 & 0.85 \\
13 & 18632 &  589 &  5744 &  5988 & 24.6 & 89 & 95 & 95  & 97.21 & 0.55 & 13.0 &  0.009 & 0.031 & 5.9 & 7.8 & 2.45 & 0.22 & 0.76 \\
14 & 18267 &  252 &  5687 &  9344 & 23.7 & 96 & 96 & 98  & 96.69 & 0.51 & 12.7 &  0.024 & 0.043 & 6.4 & 9.2 & 2.54 & 0.32 & 0.80 \\
21 & 18268 &  313 &  5778 &  6344 & 23.3 & 93 & 93 & 94  & 97.23 & 0.51 & 14.4 &  0.013 & 0.023 & 6.3 & 8.2 & 1.79 & 0.03 & 0.58 \\
22 & 18285 & 2580 &  5687 &  8692 & 23.6 & 95 & 96 & 96  & 97.27 & 0.52 & 14.4 &  0.001 & 0.021 & 6.7 & 8.9 & 3.52 & 0.13 & 0.83 \\
23 & 18548 & 1512 &  5657 &  7149 & 23.4 & 95 & 95 & 96  & 97.47 & 0.51 & 14.3 &  0.004 & 0.022 & 6.4 & 8.3 & 2.49 & 0.15 & 0.74 \\
24 & 18452 &  146 &  6079 &  5624 & 23.1 & 93 & 95 & 95  & 97.67 & 0.50 & 14.1 & $-0.003$ & 0.019 & 6.4 & 8.4 & 2.56 & 0.03 & 0.65 \\
31 & 18280 & 1420 &  6389 &  6936 & 24.0 & 92 & 93 & 93  & 97.10 & 0.53 & 12.6 &  0.002 & 0.022 & 5.9 & 8.1 & 2.61 & 0.07 & 0.62 \\
32 & 18284 &  269 &  5751 & 10081 & 24.4 & 93 & 94 & 94  & 96.84 & 0.53 & 12.3 &  0.004 & 0.021 & 5.8 & 7.9 & 2.49 & 0.12 & 0.82 \\
33 & 18278 &  262 &  5775 &  8258 & 24.5 & 94 & 95 & 95  & 97.00 & 0.54 & 10.7 & $-0.003$ & 0.021 & 5.6 & 8.0 & 2.04 & 0.16 & 0.87 \\
34 & 18269 &  970 &  5631 &  4275 & 23.3 & 95 & 95 & 96  & 97.39 & 0.50 & 13.4 &  0.001 & 0.023 & 6.4 & 8.7 & 3.31 & 0.07 & 0.74 \\
41 & 18458 &  198 &  5767 &  6279 & 23.3 & 93 & 96 & 96  & 97.50 & 0.51 & 12.2 & $-0.002$ & 0.018 & 5.7 & 7.7 & 2.78 & 0.01 & 0.54 \\
42 & 18249 &  334 &  6234 &  8878 & 24.2 & 94 & 94 & 95  & 97.08 & 0.54 & 12.2 &  0.006 & 0.025 & 5.6 & 7.9 & 3.09 & 0.13 & 0.68 \\
43 & 18221 &  222 &  5534 &  4673 & 24.2 & 97 & 98 & 101 & 97.07 & 0.53 & 13.2 &  0.001 & 0.024 & 6.1 & 8.1 & 2.96 & 0.12 & 0.70 \\
44 & 18628 &  255 &  6359 &  5860 & 23.7 & 93 & 95 & 96  & 97.25 & 0.52 & 12.9 &  0.012 & 0.027 & 6.1 & 8.7 & 2.90 & 0.10 & 0.57 \\
\bottomrule
\end{tabular}
\end{center}
\end{sidewaystable} 

\clearpage
\section{Data-based orthogonal polynomials formalism \label{app:ortho}}

We have a vector of dimension $D$ of measurement points $x_i = \{ x_1, x_2, \dots, x_D \}$ and the vector of measured values $y_i(x_i) = \{ y_1, y_2, \dots, y_D \}$, e.g.\ in an UTR(400) exposure $x_i$ would be the frame number $x_i = \{ 1, 2, \dots, 400  \}$ and $y_i$ would correspond to the ADU read in the pixel at each frame, $y_i(x_i) = \{ 100, 200, \dots, 40\,000 \}$. 

The pixel response is nonlinear and we assume that it can be modeled by a polynomial of order $N$,
\begin{equation}\label{eq:y_x_beta}
    y(x) = \sum_{i=0}^{N} \beta_i x^i\;.
\end{equation}
Then we choose to decompose the function $y(x)$ on the basis of orthogonal polynomials $P_i(x)$
\begin{equation}\label{eq:y_x_P_i}
    y(x) = \sum_{i=0}^N \alpha_i P_i(x)\;,
\end{equation}
where $P_i(x)$ is a polynomial of order $i$, defined as
\begin{equation}
    P_{i}(x) = \sum_{j=0}^{i} a_{ij}x^i\;.
\end{equation}

By identifying the coefficients in front of the $j$-th power of $x$ in Eqns.~(\ref{eq:y_x_beta}) and (\ref{eq:y_x_P_i}) we get the expression of $\beta_j$ as function of $\alpha_i$ and $a_{ij}$,

\begin{equation}\label{eq:beta_j}
    \beta_j = \sum_{i=j}^{N} \alpha_i a_{ij}\;.
\end{equation}
Physically $\beta_0$ corresponds to the pixel baseline [ADU], $\beta_1$ is the expected linear flux value [ADU/frame] and $\beta_{i>1}$ are the terms that describe the deviations from the ideal linear response of the pixel to a constant incident flux $[\beta_j] = $ADU/frame${}^{j}$.

The problem is to find the set of coefficients $\beta_j$ that fits the pixel response, so to find $\alpha_{i}$s and $a_{ij}$s with $i, j\in {1\dots N}$. Below we present the formulas to compute $\alpha_{i}$s and $a_{ij}$s given measurement $x_i = \{ x_1, x_2, \dots, x_D \}$ and $y_i(x_i) = \{ y_1, y_2, \dots, y_D \}$.
The coefficients $a_{ij}$ of each polynomial $P_i(x)$ are computed using the orthogonality and normalization conditions on $x_i = \{ x_1, x_2, \dots, x_D \}$
\begin{equation}
    \langle P_m, P_n \rangle \equiv \sum_{i=1}^{D} P_m(x_i) P_n(x_i) = \delta_{mn}\;.
\end{equation}
For $m > n$ this gives

\begin{equation}
    \langle P_m, P_n \rangle =  \sum_{i=1}^{D} P_m(x_i) P_n(x_i) = \sum_{k=0}^m a_{mk} \gamma_{nk}\,.
\end{equation}

%
Dividing the above equation by $a_{m0}$ and defining $a_{mj}^{\prime} = \frac{a_{mj}}{a_{m0}}$ we get
\begin{equation}
    0 = \gamma_{n0} + a^{\prime}_{m1} \gamma_{n1} + a^{\prime}_{m2} \gamma_{n2} + \dots + a^{\prime}_{mm} \gamma_{nm}\,.
\end{equation}
Considering the orthogonality of $P_m$ to all $P_n$ with $n<m$ ($P_n = \{ P_0, P_1, \dots, P_{m-1}\}$) we construct a system of $m$ linear equations with $m$ unknowns $a^{\prime}_{mj}$ with $j = 1\dots m$,

\begin{equation}
             \bordermatrix{\langle P_n, P_m     \rangle           ~ & j=1            & j=2      & \dots & j=m            &                  \cr
                           \langle P_0, P_m     \rangle \Rightarrow & \gamma_{01}    a^{\prime}_{m1} & \gamma_{02}    a^{\prime}_{m2} & \dots & \gamma_{0m}    a^{\prime}_{mm} & = \gamma_{00}    \cr
                           \langle P_1, P_m     \rangle \Rightarrow & \gamma_{11}    a^{\prime}_{m1} & \gamma_{12}    a^{\prime}_{m2} & \dots & \gamma_{1m}    a^{\prime}_{mm} & = \gamma_{10}    \cr
                           \langle P_2, P_m     \rangle \Rightarrow & \gamma_{21}    a^{\prime}_{m1} & \gamma_{22}    a^{\prime}_{m2} & \dots & \gamma_{2m}    a^{\prime}_{mm} & = \gamma_{20}    \cr
                           ~                                        & ~                         & ~              a^{\prime}_{m2} & \dots & ~                                 \cr
                           \langle P_{m-1}, P_m \rangle \Rightarrow & \gamma_{m-1,1} a^{\prime}_{m1} & \gamma_{m-1,2} a^{\prime}_{m2} & \dots & \gamma_{m-1,m} a^{\prime}_{mm} & = \gamma_{m-1,0} \cr}
 \end{equation}
The system can be easily solved by any well established method. We propose the determinant method. To find $a^{\prime}_{mj}$ we compute the determinant of the matrix $\Gamma$,

\begin{equation}
\Gamma = \begin{pmatrix}
                 \gamma_{01}    & \gamma_{02}    & \cdots & \gamma_{0j}    & \cdots & \gamma_{0m}    \\
                 \gamma_{11}    & \gamma_{12}    & \cdots & \gamma_{1j}    & \cdots & \gamma_{1m}    \\
                 \gamma_{21}    & \gamma_{22}    & \cdots & \gamma_{2j}    & \cdots & \gamma_{2m}    \\
                 \vdots         & \vdots         & \vdots & \vdots         & \ddots & \vdots         \\
                 \gamma_{m-1,1} & \gamma_{m-1,2} & \cdots & \gamma_{m-1,j} & \cdots & \gamma_{m-1,m} \\
         \end{pmatrix}
\end{equation}
 and the determinant of $\Gamma_j$ where the $j$-th column of $\Gamma$ was replaced by the vector of free terms $\{ \gamma_{00} \dots \gamma_{m-1, 0}\}$,
 \begin{equation}
\Gamma_j = \begin{pmatrix}
                 \gamma_{01}    & \gamma_{02}     & \cdots & \gamma_{00}    & \cdots &  \gamma_{0m}    \\
                 \gamma_{11}    & \gamma_{12}     & \cdots & \gamma_{10}    & \cdots & \gamma_{1m}    \\
                 \gamma_{21}    & \gamma_{22}     & \cdots & \gamma_{20}    & \cdots & \gamma_{2m}    \\
                 \vdots         & \vdots          & \vdots & \vdots         & \ddots & \vdots         \\
                 \gamma_{m-1,1} & \gamma_{m-1,2}  & \cdots & \gamma_{m-1,0} & \cdots & \gamma_{m-1,m} \\
           \end{pmatrix}
 \end{equation}
 The solution for $a^{\prime}_{mj}$ is then:
\begin{equation}\label{eq:a'_mn}
    a^{\prime}_{mj} = \frac{\det \Gamma_j}{\det \Gamma}\,.
\end{equation}

The absolute scale of the coefficients $a_{mi}$ is found by considering the normalization of $P_m(x)$
\begin{equation}
    \langle P_m, P_m \rangle = \sum x^{i+j}\,,
\end{equation}

%
%
from which we have
\begin{equation}\label{eq:a_m0}
    a_{m0} = \left( \sum_{i=0}^{N}\left( a_{mi}^{'2}\sum x^{2i} \right) + 2a^{\prime}_{mi}\sum_{j=i+1}^N a^{\prime}_{mj} \sum x^{i+j} \right)^{-1/2}
\end{equation}
and
\begin{equation}\label{eq:a_mn}
    a_{mj} = a_{mj}^{\prime} a_{m0}\;.
\end{equation}
The coefficients $\alpha_{i}$ are found by projecting the measurements $y_i$ on the basis of the orthogonal polynomials
\begin{equation}\label{eq:alpha_m}
    \alpha_{m} = \langle y, P_m \rangle = \sum_{i=1}^D y_i P_m(x_i) = \sum_{i=1}^D y_i \sum_{j=0}^m a_{mj} x_i^j \;.
\end{equation}

In summary, in order to fit the data using orthogonal polynomials one needs to 
\begin{enumerate}
    \item Compute $a_{ij}$ using Eqs.~(\ref{eq:a'_mn}), (\ref{eq:a_m0}), (\ref{eq:a_mn}). It is common for all the pixels, as it depends only on the time coordinates $x$.
    \item Compute $\alpha_{i}$ using Eq.~(\ref{eq:alpha_m}). This quantity is pixel dependent, as it depends linearly on the pixel signal values $y$.
    \item Compute $\beta_{j}$ using Eq.~(\ref{eq:beta_j}). 
\end{enumerate}

This method offers a significant advantage over traditional numerical fitting approaches, particularly when applied to a dataset of this magnitude (involving independent fits for $2040\times 2040$ pixels by 16 detectors across numerous fluence levels). The key benefit lies in the pre-computation of the $a_{ij}$ coefficients for the entire dataset. Once these coefficients are determined, the computation is greatly simplified to straightforward multiplications of per-pixel signal values to calculate $\alpha_{ij}$. This makes the method highly efficient and scalable for large datasets.

\end{appendix}
\label{LastPage}
\end{document}